\documentclass{emulateapj}

\begin{document}
\title{The Highly Dynamic Behavior of the Innermost Dust and Gas in the Transition Disk Variable LRLL 31}
\author{Flaherty, K.M. \altaffilmark{1}, Muzerolle, J. \altaffilmark{2}, Rieke, G. \altaffilmark{1}, Gutermuth, R. \altaffilmark{3}, Balog, Z.\altaffilmark{4}, Herbst, W. \altaffilmark{5}, Megeath, S.T. \altaffilmark{6}, Kun, M.\altaffilmark{7}}
\email{kflaherty@as.arizona.edu}

\altaffiltext{1}{Steward Observatory, University of Arizona, Tucson, AZ 85721}
\altaffiltext{2}{Space Telescope Science Institute, 3700 San Martin Dr., Baltimore, MD, 21218}
\altaffiltext{3}{Smith College, Northampton, MA 01063}
\altaffiltext{4}{Max-Planck Institut f\"{u}r Astronomie, K\"{o}nigstuhl 17, D-69117 Heidelberg, Germany}
\altaffiltext{5}{Department of Astronomy, Wesleyan College, Middletown, CT}
\altaffiltext{6}{Department of Physics and Astronomy, University of Toledo, Toledo, OH}
\altaffiltext{7}{Konkoly Observatory, H-1121 Budapest, Konkoly Thege \'{u}t 15-17, Hungary}

\begin{abstract}
We describe extensive synoptic multi-wavelength observations of the transition disk LRLL 31 in the young cluster IC 348. We combined four epochs of IRS spectra, nine epochs of MIPS photometry, seven epochs of cold-mission IRAC photometry and 36 epochs of warm-mission IRAC photometry along with multi-epoch near-infrared spectra, optical spectra and polarimetry to explore the nature of the rapid variability of this object. We find that the inner disk, as traced by the 2-5\micron\ excess stays at the dust sublimation radius while the strength of the excess changes by a factor of 8 on weekly timescales, and the 3.6 and 4.5\micron\ photometry show a drop of 0.35 magnitudes in one week followed by a slow 0.5 magnitude increase over the next three weeks. The accretion rate, as measured by Pa$\beta$ and Br$\gamma$ emission lines, varies by a factor of five with evidence for a correlation between the accretion rate and the infrared excess. While the gas and dust in the inner disk are fluctuating the central star stays relatively static. Our observations allow us to put constraints on the physical mechanism responsible for the variability. The variable accretion, and wind, are unlikely to be causes of the variability, but are both effects of the same physical process that disturbs the disk. The lack of periodicity in our infrared monitoring indicates that it is unlikely that there is a companion within $\sim0.4$ AU that is perturbing the disk. The most likely explanation is either a companion beyond $\sim0.4$ AU or a dynamic interface between the stellar magnetic field and the disk leading to a variable scale height and/or warping of the inner disk.
\end{abstract}

\section{Introduction}
The widely accepted view is that a pre-main sequence star slowly evolves from a deeply embedded core to a fully revealed star. After initial collapse of a dense core, the star enters the class 0/I phase in which it is rapidly accreting mass from its massive envelope and disk. As the envelope dissipates and the accretion rate decreases, the star enters the class II phase where it becomes optically revealed and is being fed slowly by a viscously-evolving irradiated accretion disk. The dust in the disk eventually coalesces into large grains and the gas dissipates from the system leaving a class III star. 

Confounding this picture of slow evolution, there is growing evidence that these systems are in fact highly dynamic. It has been known since the discovery of T Tauri stars, which are at the heart of class II systems, that the star is usually variable \citep{joy45}. Hot and cold spots rotating across the surface of the star, as well as variable accretion onto the star can create changes of a magnitude or more in only a few days in the optical \citep[e.g.][]{her94}. Recent observations have shown that the circumstellar disk, as traced by the infrared emission, can also change dramatically in only a few days. The shape of the disk continuum appears to change \citep{mor09,juh07,muz09} and the silicate feature, which traces the properties of the dust in the upper layers of the disk close to the star, also varies rapidly \citep{bar09,sit08,ske10}. The behavior may be related to large structural changes in the inner disk \citep{sit08}, large accretion events, such as in FU Ori outbursts \citep{har96} or to turbulence within the disk \citep{tur10}. The previous studies provide some insight into the source of the variability, but are limited by the lack of ancillary data and limited temporal or wavelength coverage. 

To understand better what could be causing these changes, we have studied one object, the highly variable star LRLL 31, in detail. This object shows dramatic mid-infrared variability \citep{muz09} that may be due to a non-axisymmetric structure in the inner disk with a changing scale height \citep{fla10}. We report continued monitoring in the mid-infrared, as well as multiple epochs of near-infrared spectra, optical spectra and polarimetry measurements to understand the dynamics of the system better. Using these observations, we study the changes in the dust, the gas and the central star, and any connections among them. The near simultaneity of these data sets allow us to rule out specific mechanisms that might be responsible for the observed variability.

\section{Data}
Our monitoring of LRLL 31 consists of multiple epochs of ground and space based observations covering the optical as well as the near and mid-infrared. 
We obtained 0.8-5\micron\ spectra with the SpeX instrument \citep{ray03} on the NASA Infrared Telescope Facility (IRTF) in fall 2008 and 2009. Table~\ref{obs_log} lists the observing log for all of our observations, and the spectra are shown in Figures~\ref{spectra} and ~\ref{spectra2}. On all but two nights we used both the SXD (0.8-2.2\micron) and LXD2.1 (2-4.8\micron) gratings, while on October 8,9 2009 we only obtained SXD spectra. On Oct 10,18,19 2008 we used the 0.5"x15" slit (R$\sim$1500) while on the other nights we used the 0.8"x15" (R$\sim$800) slit due to poorer seeing conditions. We also supplement our observations with SXD spectra taken in 2005 \citep{muz09} and 2006 \citep{dah08}. The data were reduced using Spextool \citep{cus04} and corrected for telluric absorption using {\it xtellcor} \citep{vac03}. Our telluric standards were the nearby A0V stars HD 19600 and HD 23441. The total exposure times were 30 minutes for the SXD and one hour for the LXD spectra. A set of calibration exposures, including flat field and argon arc spectra, were obtained for each target-standard star pair. 

We also observed the bright weak line T Tauri star (WTTS) HD 283572, which shown no sign of accretion or an infrared excess, with an exposure time of 10 minutes. It has the same spectral type (G6) as LRLL 31 and is used as a comparison star when measuring the veiling and strength of the emission lines. We took SXD spectra of this star with a slit of 0.8"x15" on 5 Jan 2010. The spectra were reduced in the same manner as the LRLL 31 spectra using the telluric standard HD 25152.

In addition to the spectra taken by Spex, we also obtained JHK photometry (Table~\ref{photometry}) from the slit-viewing near-infrared guide camera. LRLL 31 was observed with three dithers, standard reduction techniques were used and aperture photometry was performed on the two stars in the field of view. We did not observe any photometric standards, but instead derived the photometry of LRLL 31 relative to the other star in the field (LRLL 169, 2MASS03441776+3204476). This allows us to derive accurate relative photometry, but there could be large uncertainties in the absolute photometry. LRLL 169 is a diskless cluster member and is not marked as variable ($\sigma$=0.06 mag) in the optical monitoring of \citet{coh04}. Our Spitzer warm mission photometry (described below) of LRLL 169 shows variations of 1-2\% at 3.6 and 4.5\micron\, which is consistent with the uncertainty in the photometry. This suggests that the photometric variability is no more than a few percent, which is smaller than the absolute uncertainty (0.1 mag) associated with using only one standard and we ignore it.

We also include high-resolution H$\alpha$ spectra from Keck (kindly provided by Scott Dahm) and from the MMT. The MMT spectrum was taken with Hectochelle \citep{sze98} in 29 Feb 2008 with a resolving power of R$\sim30,000$. The Keck spectra were taken with HIRES \citep{vog94} on 28 Nov 2006 and 03 Dec 2008 with a resolving power of R$\sim33,000$. Observing details for the MMT spectra are reported in \citet{fla08} while the reduction process for the Keck spectra is reported in \citet{dah08}.

We include spectro-polarimetric observations taken on several nights using the SPOL imaging/spectropolarimeter \citep{sch92} on the Steward Observatory 2.3 m telescope. These low-resolution optical spectra, shown in Figure~\ref{opt_spec}, extend from 4000-7500$\AA$, and from them we can extract the strength of the H$\alpha$ line as well as the shape of the continuum, which is useful for measuring the reddening toward the source. Details on the observational mode and the data reduction can be found in \citet{smi03}. An intermediate resolution optical spectrum of LRLL 31 was obtained on 2009 October 11 using the CAFOS instrument on the 2.2-m telescope of the Calar Alto Observatory. The exposure time was 1880s. The R-100 grism covered the wavelength range 5800-9000 \AA. The resolving power, using a 1.5-arcsec slit, was R$\sim$3500 at 6600\AA. We observed the spectrum  of a He-Ne-Rb lamp for wavelength calibration. The spectrum was reduced using standard IRAF routines. 

We have observed LRLL 31 on multiple epochs with all three instruments onboard the Spitzer Space Telescope. IRS \citep{hou04} data were originally obtained on 09 Oct 2007, 16 Oct 2007, 24 Feb 2008 and 02 Mar 2008 and these spectra were presented in \citet{muz09}. The IRS data reduction pipeline has been improved and the updated spectra are presented here. The spectra were extracted using the Spectroscopic Modeling, Analysis and Reduction Tool (SMART;
\citet{hig04}) version 8.1.2, starting with the basic calibrated data products from the Spitzer Science Center reduction pipeline version S18.7. Rogue pixels were first removed using the "IRS\_CLEAN" program, constructing a pixel mask for each of the SL and LL modes using off-order images as a reference. SMART was then used with optimal manual point source extraction using local sky subtraction. The local sky background was calculated by fitting a polynomial to all sky pixels on either side of the source at each row in the spatial direction. The final calibrated spectra for each nod and order were then combined using the sigma-clipped averaging function in SMART; typical measurement uncertainties were about 2\% at most wavelengths. 

IRAC \citep{faz04} observations were taken during the cryo-mission, when all four channels were available, on 19-23 March 2009 (PID 50596) (Table~\ref{irac-cryo}) along with the original GTO \citep{lad06} and C2D \citep{jor06} data taken on 11 Feb 2004 and 08 Sep 2004. A warm-mission observing program (PID 60160) obtained repeated 3.6,4.5\micron\ photometry at a varying cadence with observations ranging from every four hours to every other day over most of the 40 day visibility window of IC 348 (Table~\ref{irac_wm}). The IRAC data reduction pipeline is described in \citet{gut09} with updates appropriate for the warm Spitzer mission described in Gutermuth et al. (in prep). MIPS \citep{rie04} observations were taken on 21 Feb 2004 \citep{lad06}, 19 Sep 2004 \citep{reb07} and 24-27 Sep 2007, 12 Mar 2008, 19 Mar 2008 \citep{muz09}. Details on the reduction of the MIPS data are in \citet{muz09}. Information on all of these observations is included in Table~\ref{obs_log}.

\section{Evidence for Fluctuating Gas and Dust Properties with Static Stellar Properties}

\subsection{Stellar Properties}
\begin{figure}
\center
\includegraphics[scale=.4]{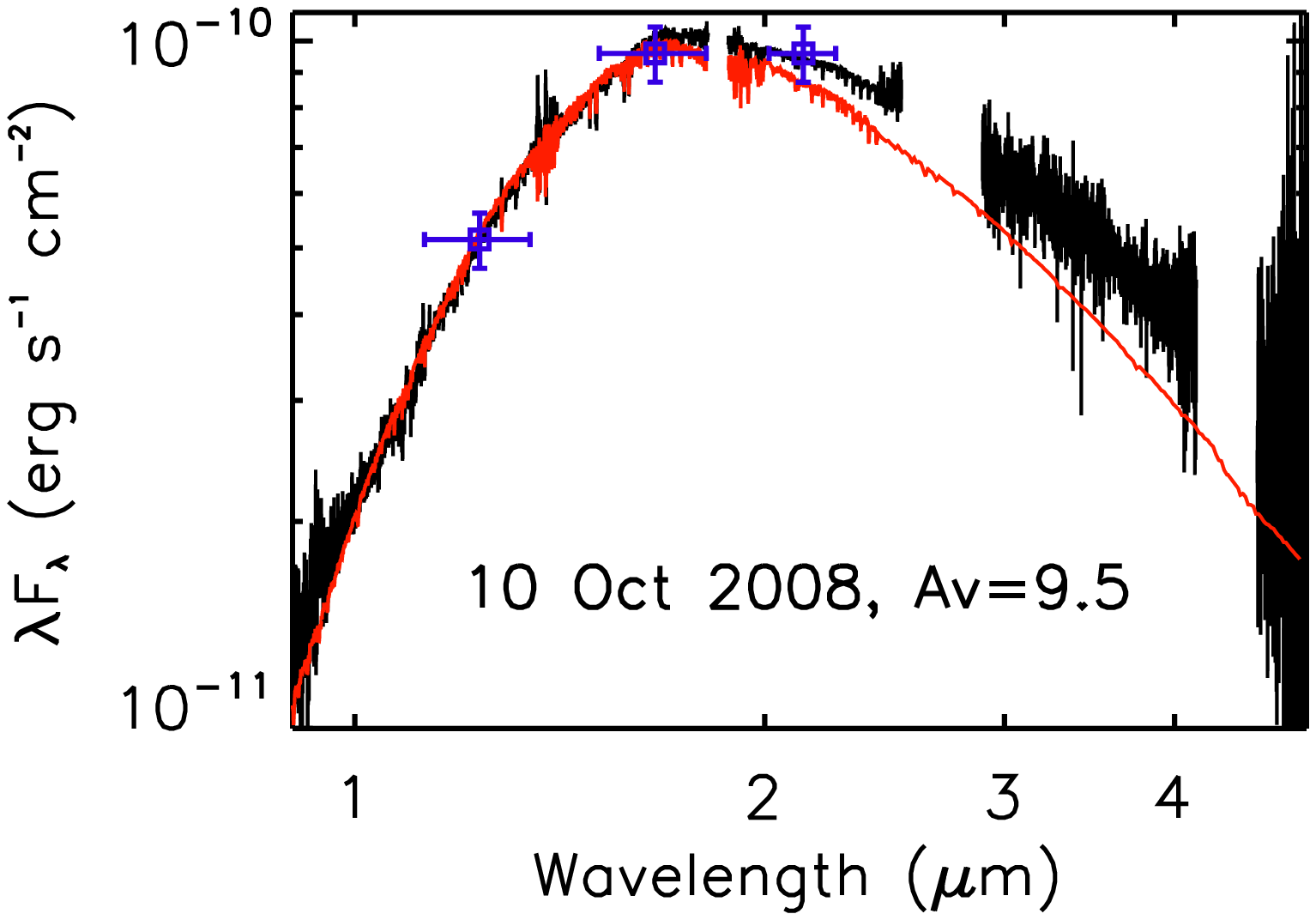}
\includegraphics[scale=.4]{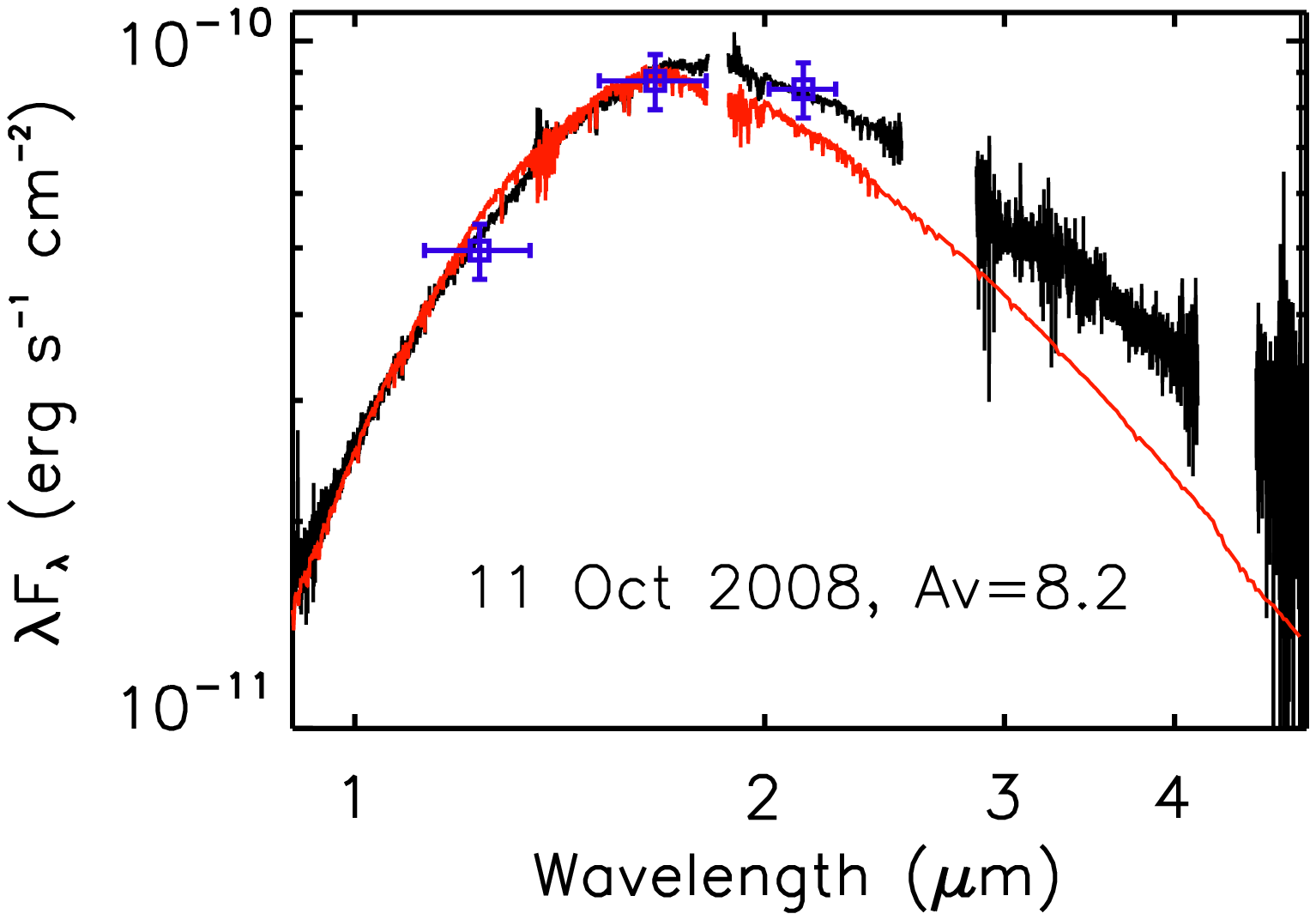}
\includegraphics[scale=.4]{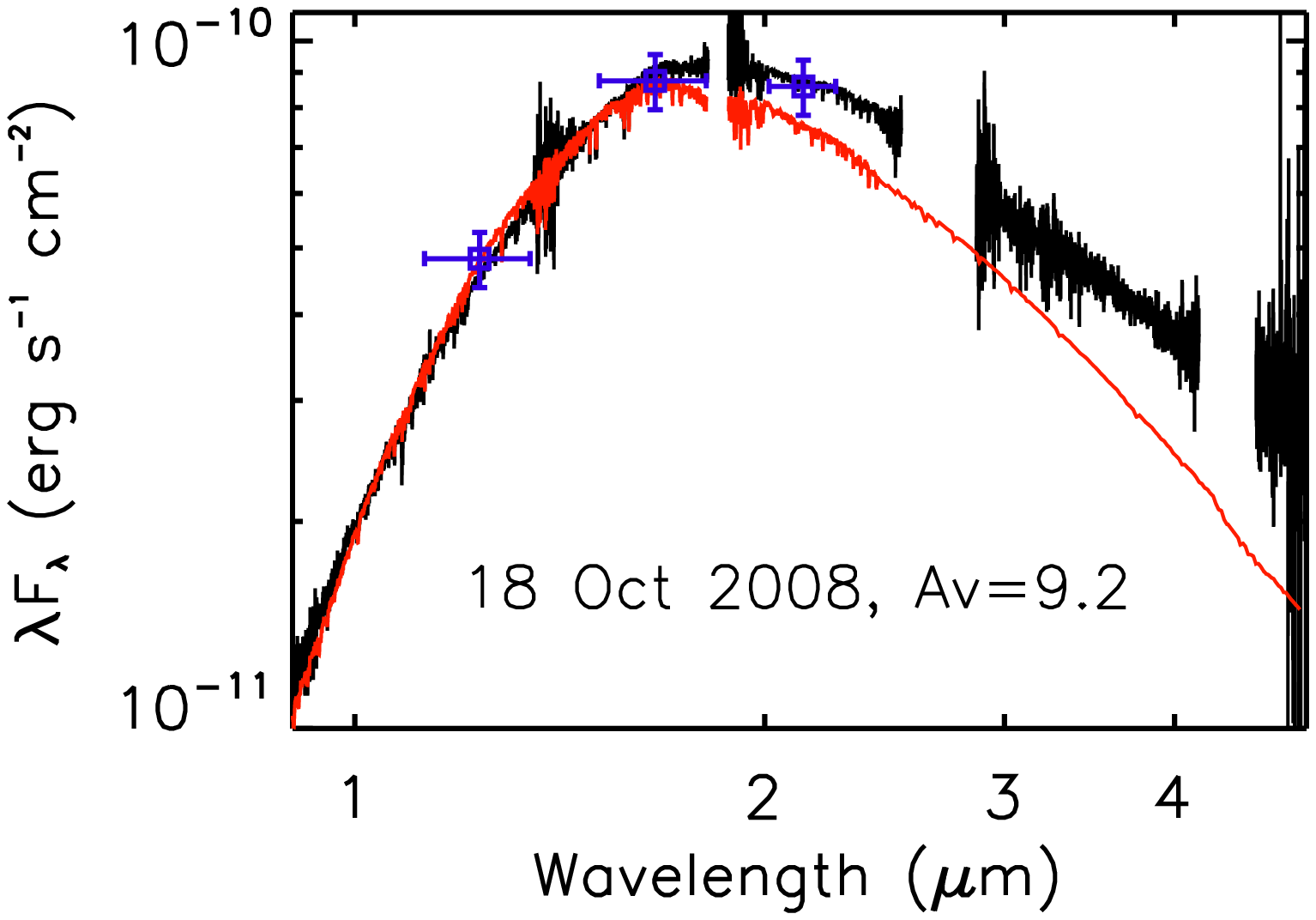}
\includegraphics[scale=.4]{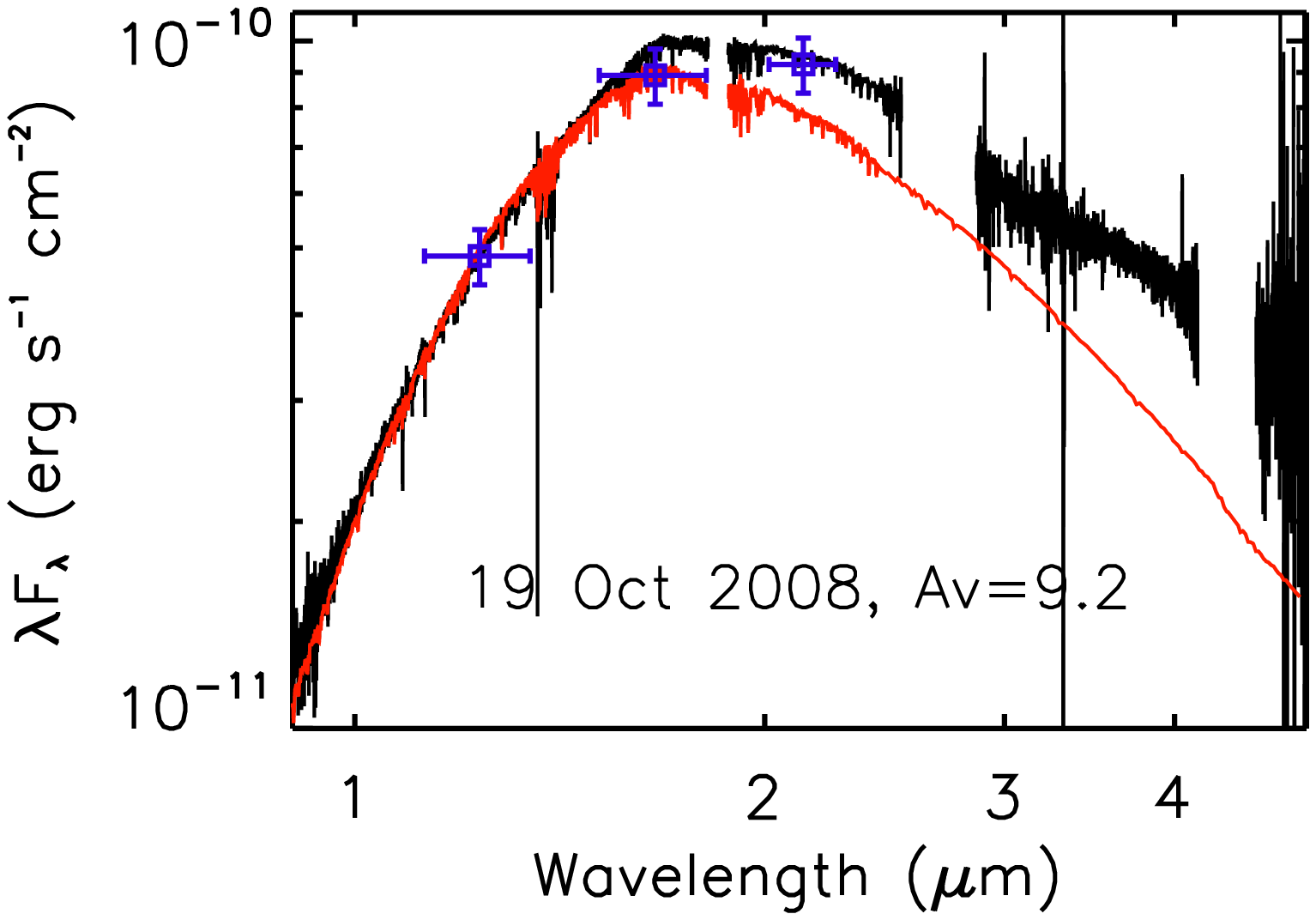}
\includegraphics[scale=.4]{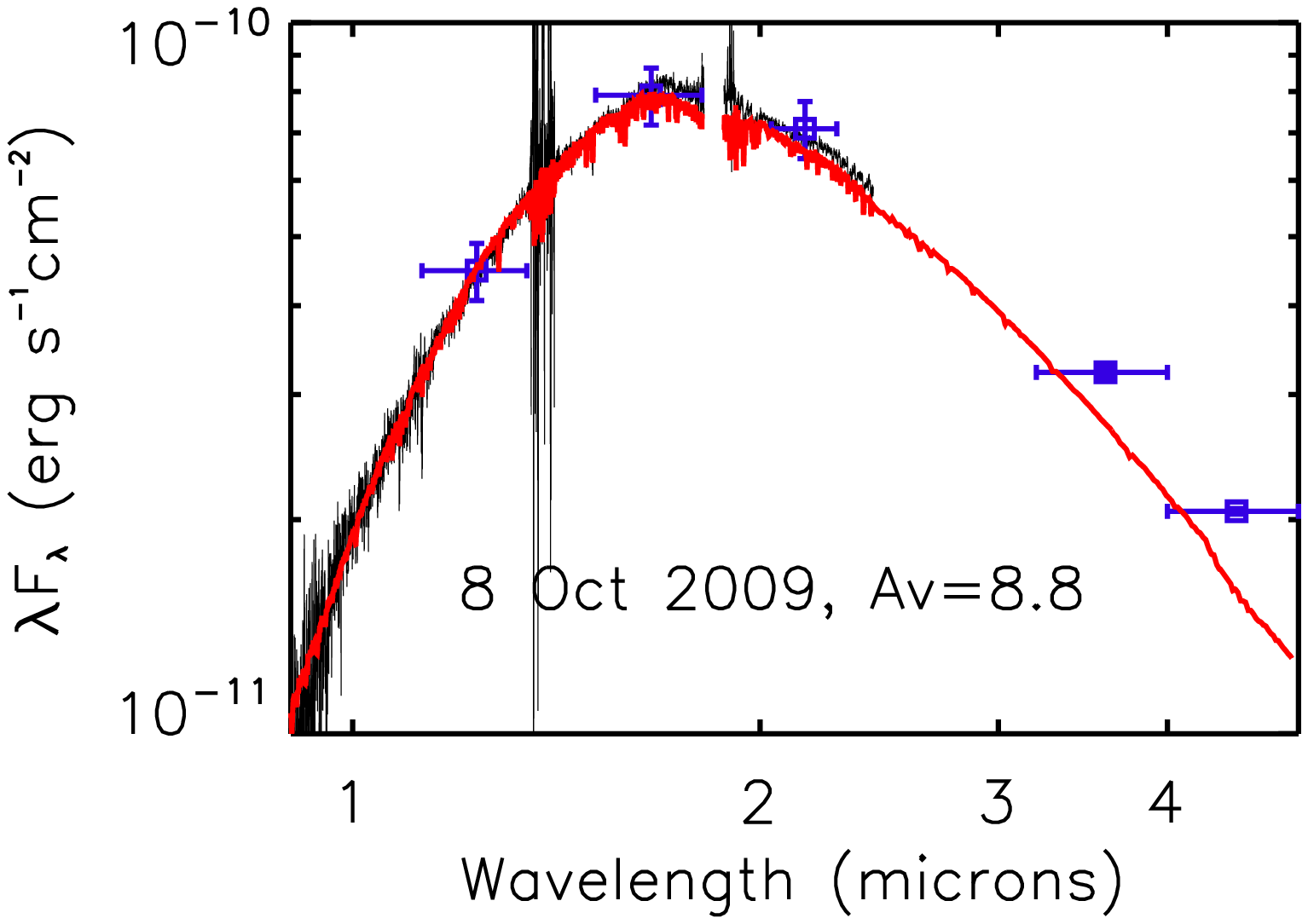}
\includegraphics[scale=.4]{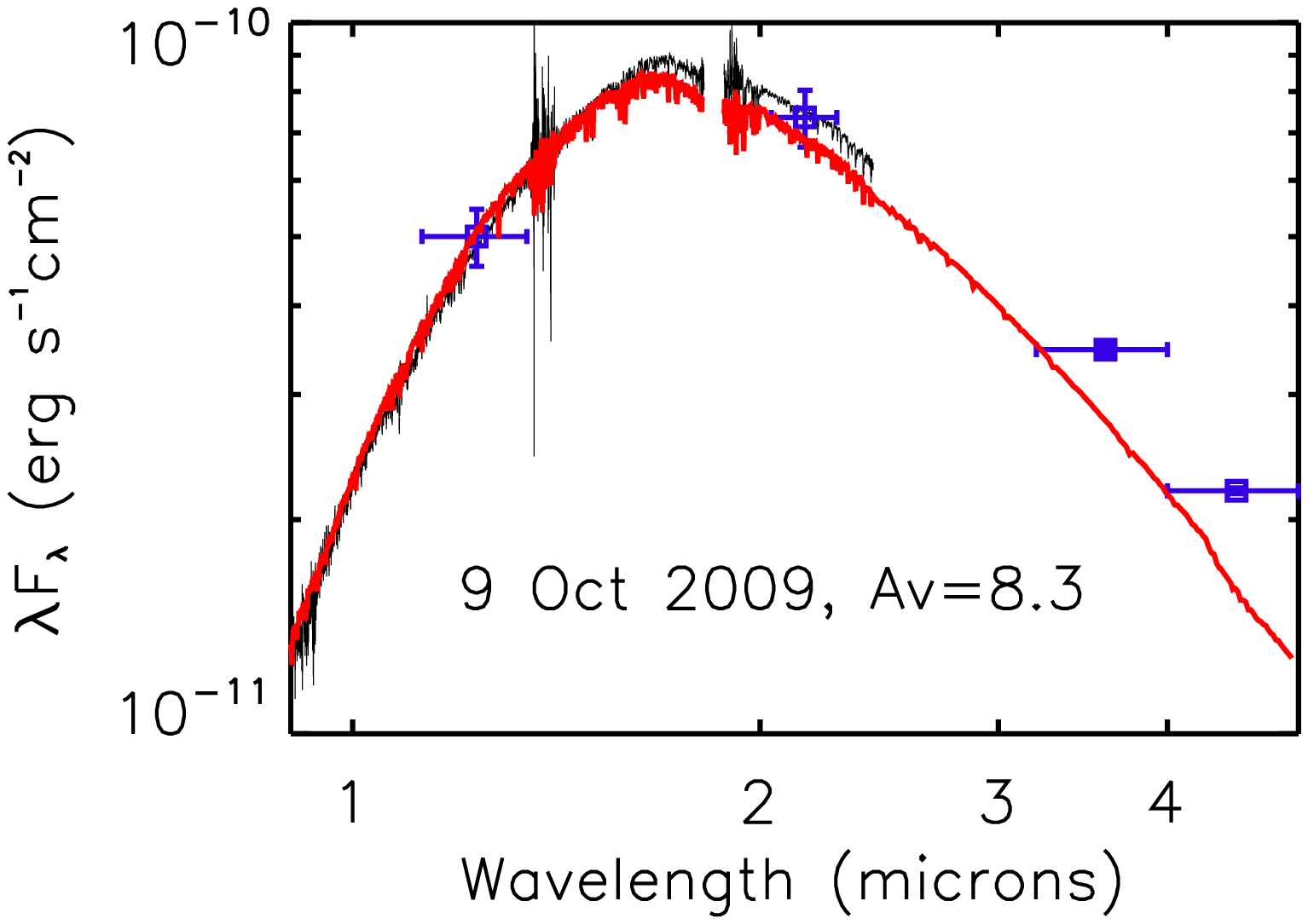}
\caption{Observed 0.8-5\micron\ spectra of LRLL 31 (black). Also included are simultaneous JHK and 3.6,4.5\micron\ photometry (blue), when available. The spectra are compared to the standard with our best fit reddening (red spectra). All spectra have been scaled to match the near-infrared photometry, except the 2005 and 2006 spectra for which we do not have contemporaneous photometry and have not been flux calibrated. For the photometric points the 'error bars' in the x-axis show the size of the passband.\label{spectra}}
\end{figure}

\begin{figure}
\center
\includegraphics[scale=.4]{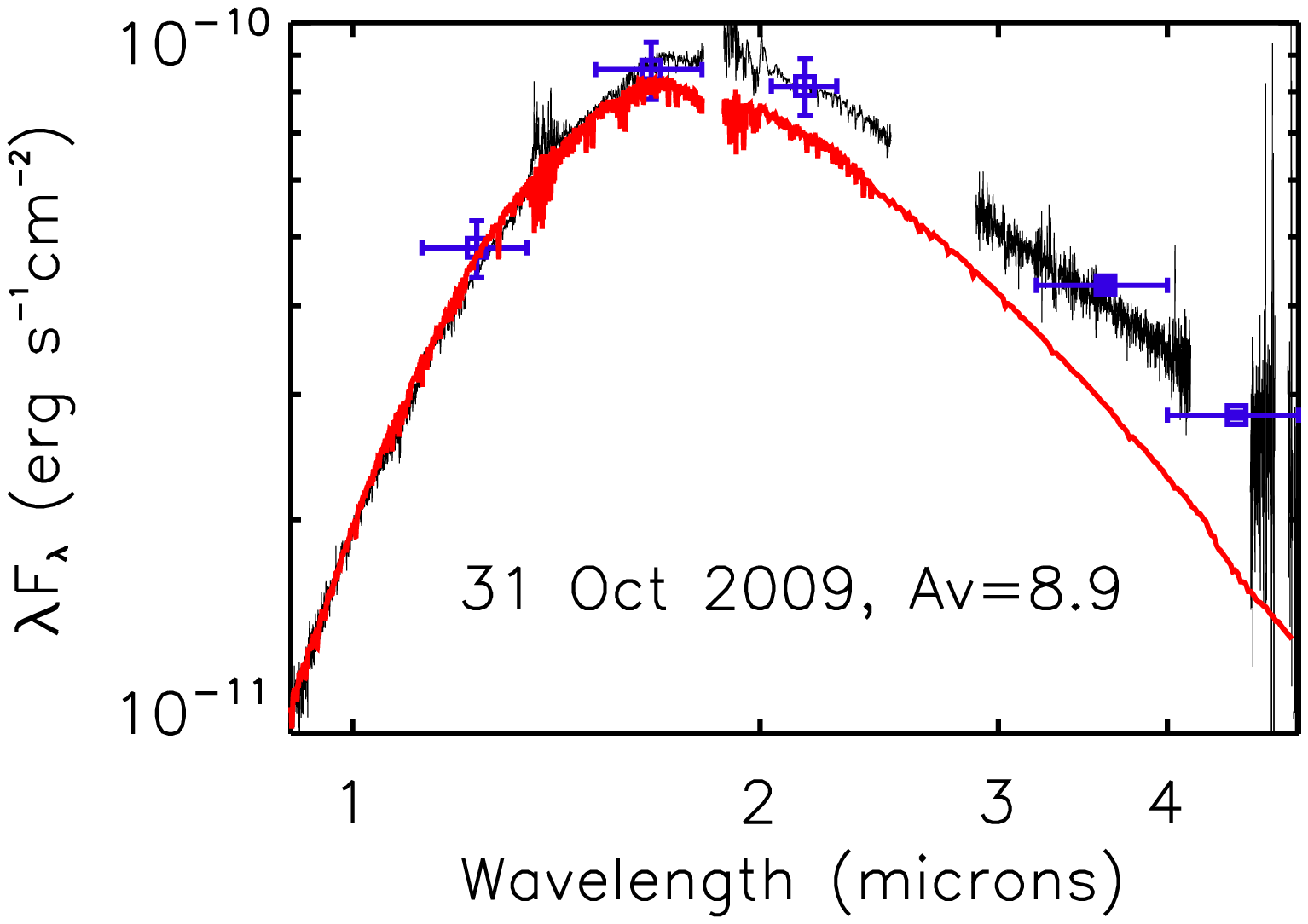}
\includegraphics[scale=.4]{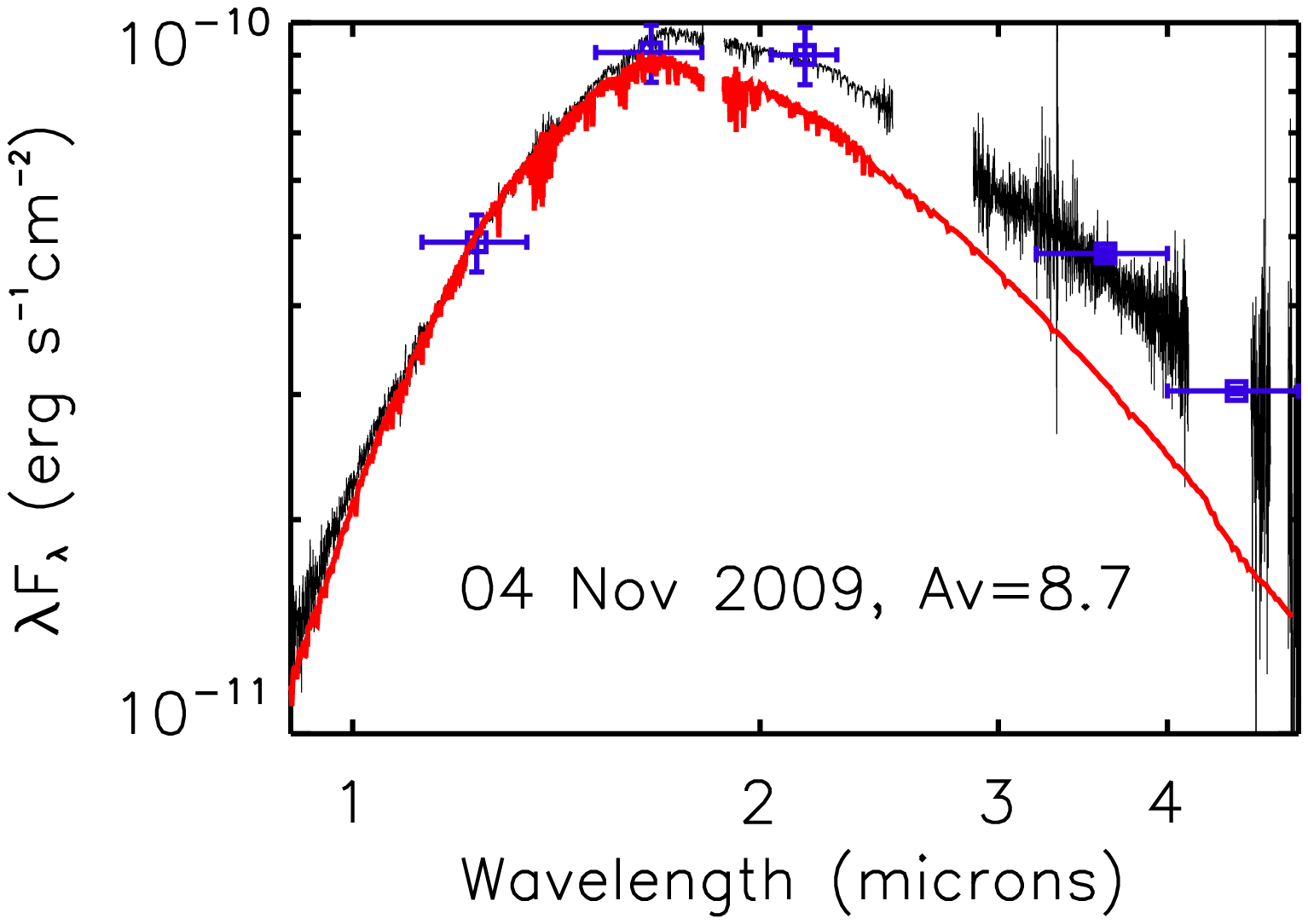}
\includegraphics[scale=.4]{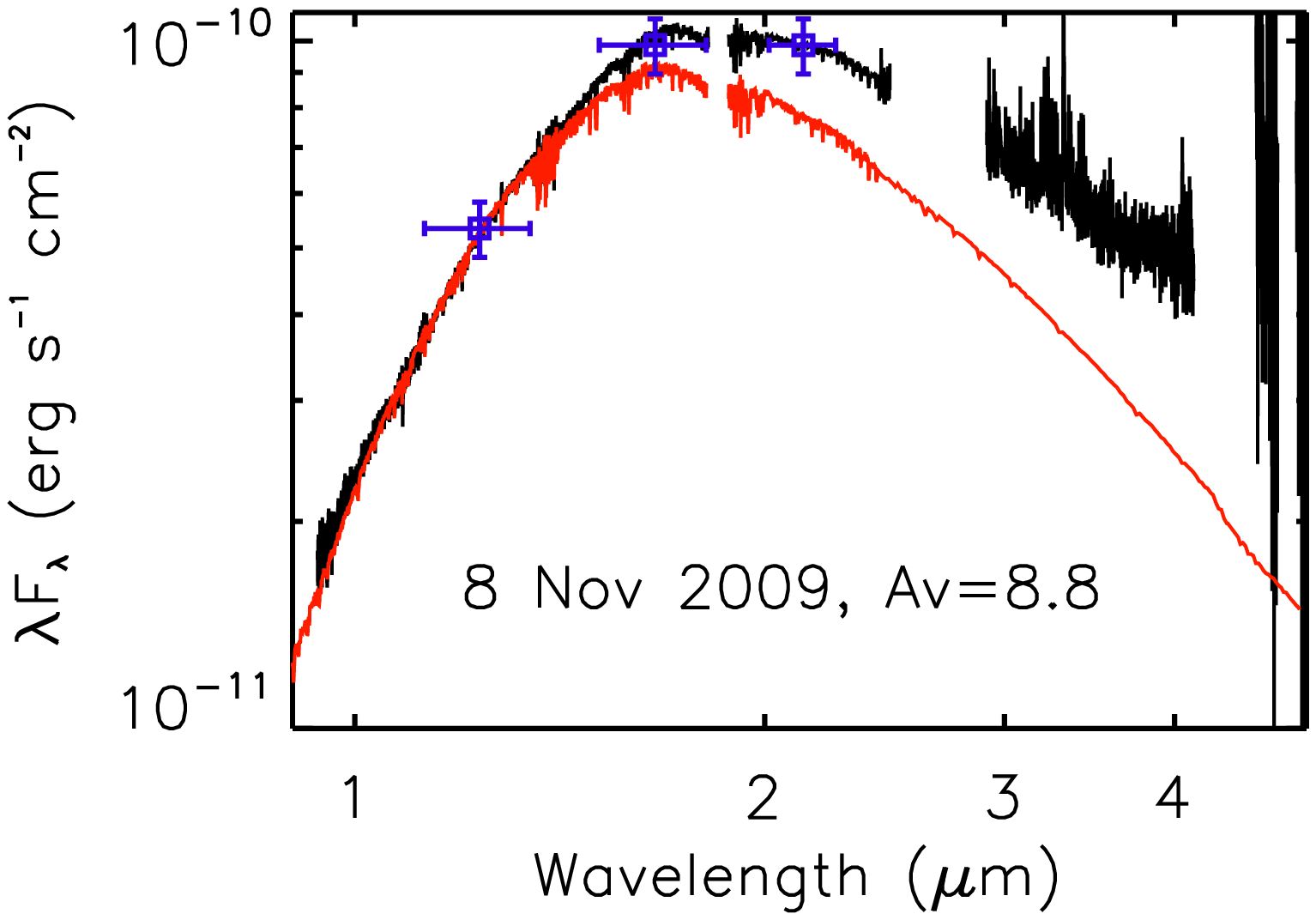}
\includegraphics[scale=.4]{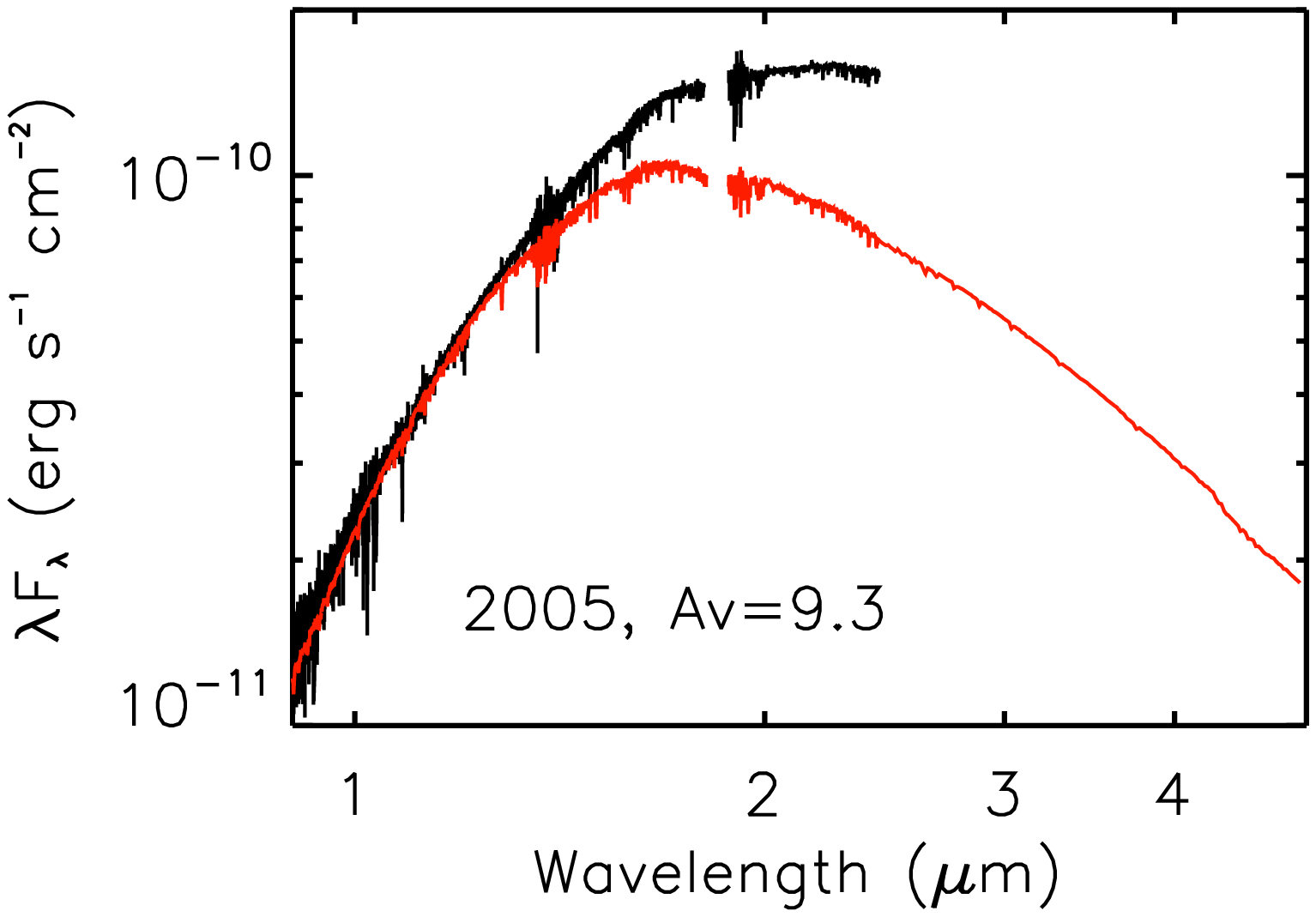}
\includegraphics[scale=.4]{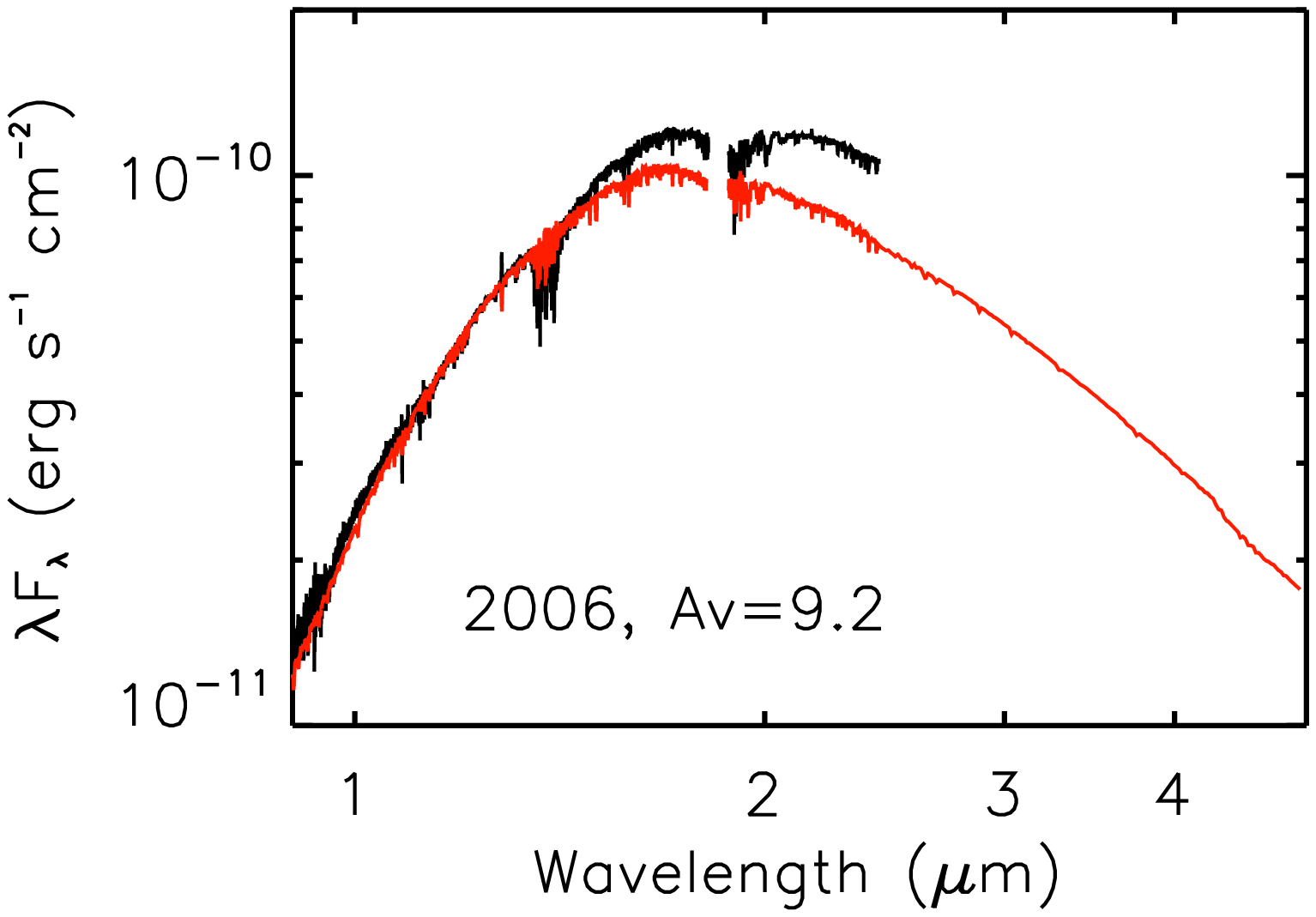}
\caption{Fig~\ref{spectra} continued. Curved shape of the H and K band near 1.8\micron\ in 2006, which is not seen in the other spectra, is a sign of imperfect telluric correction.\label{spectra2}}
\end{figure}

\begin{figure}
\includegraphics[scale=.4]{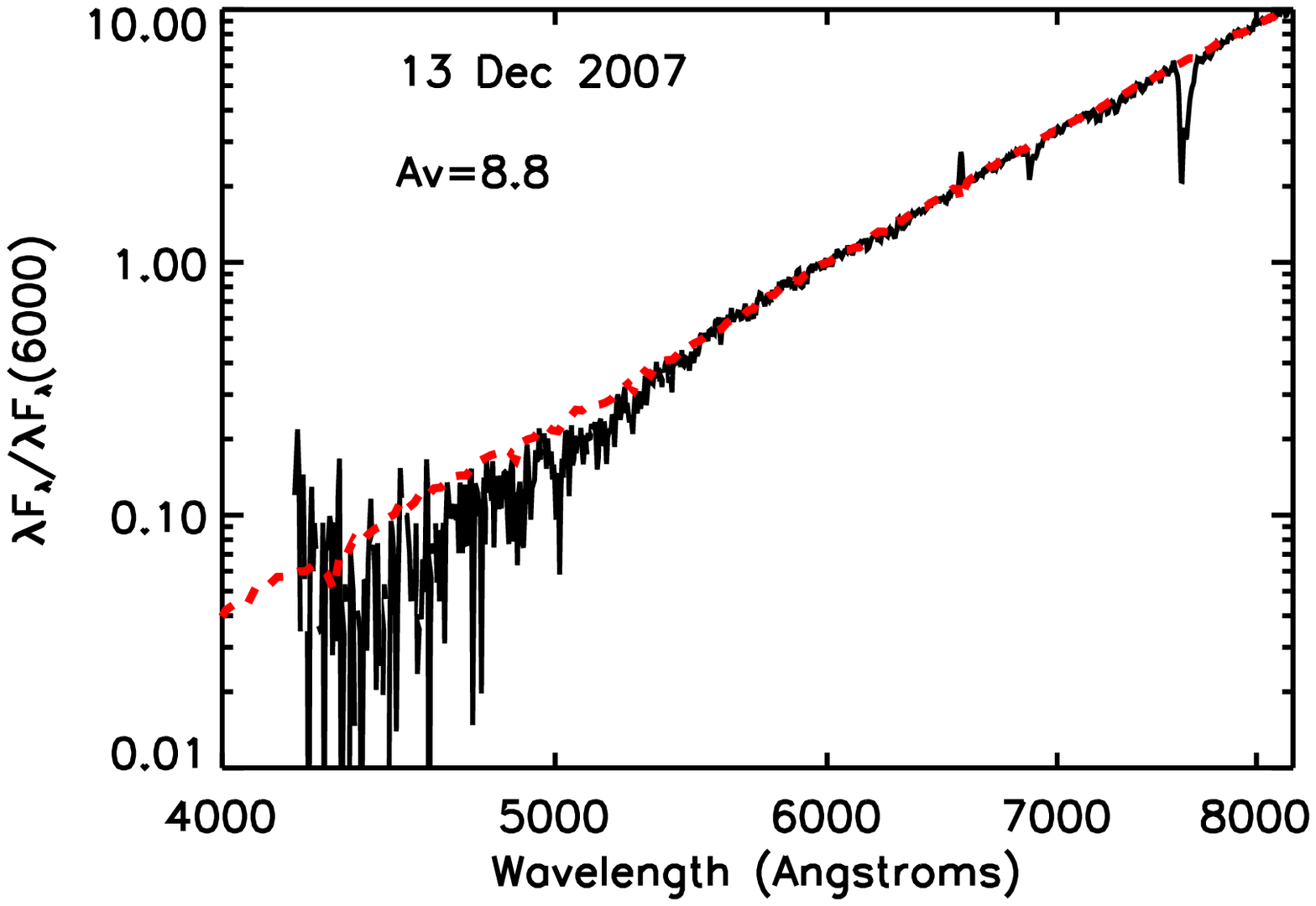}
\includegraphics[scale=.4]{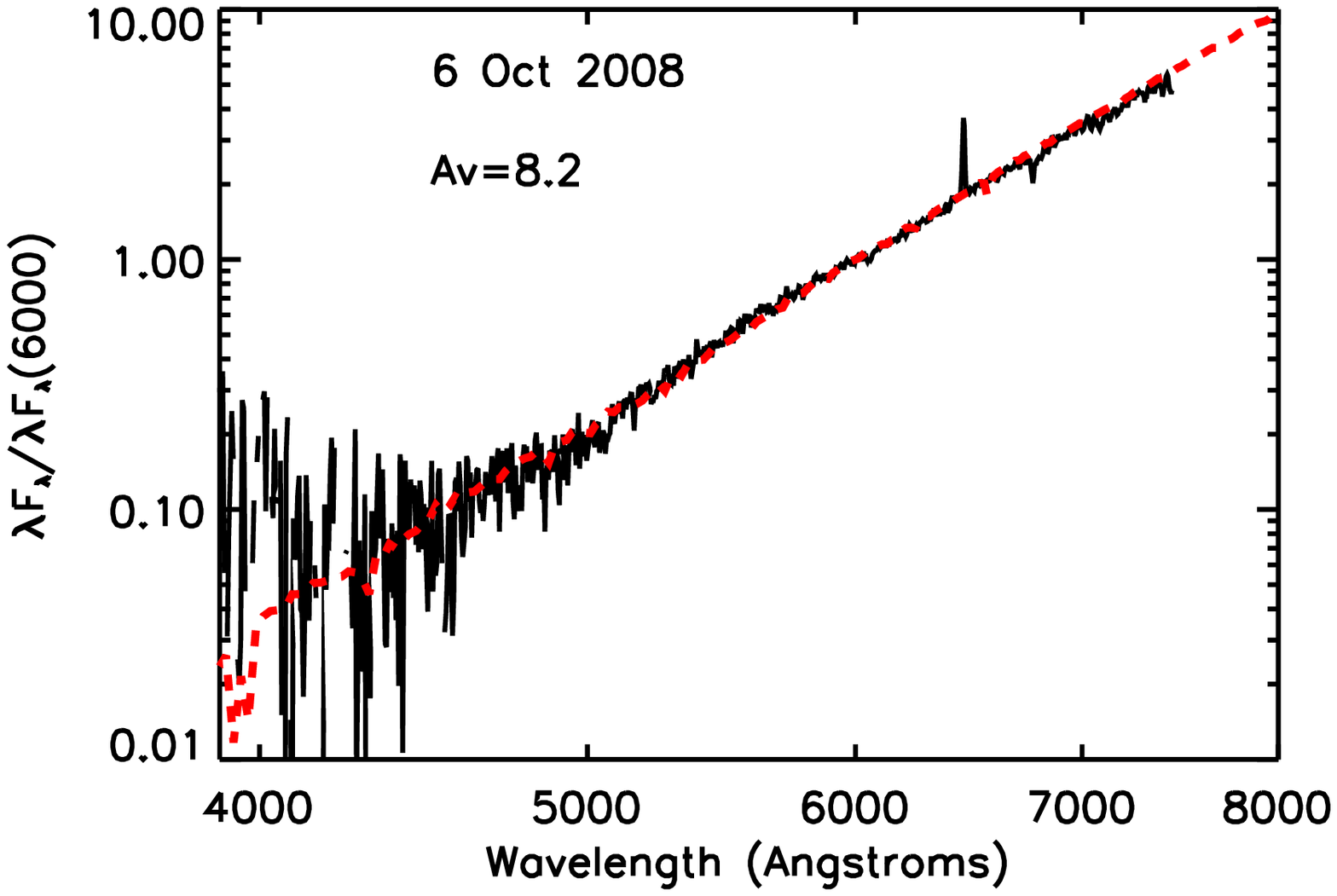}
\includegraphics[scale=.4]{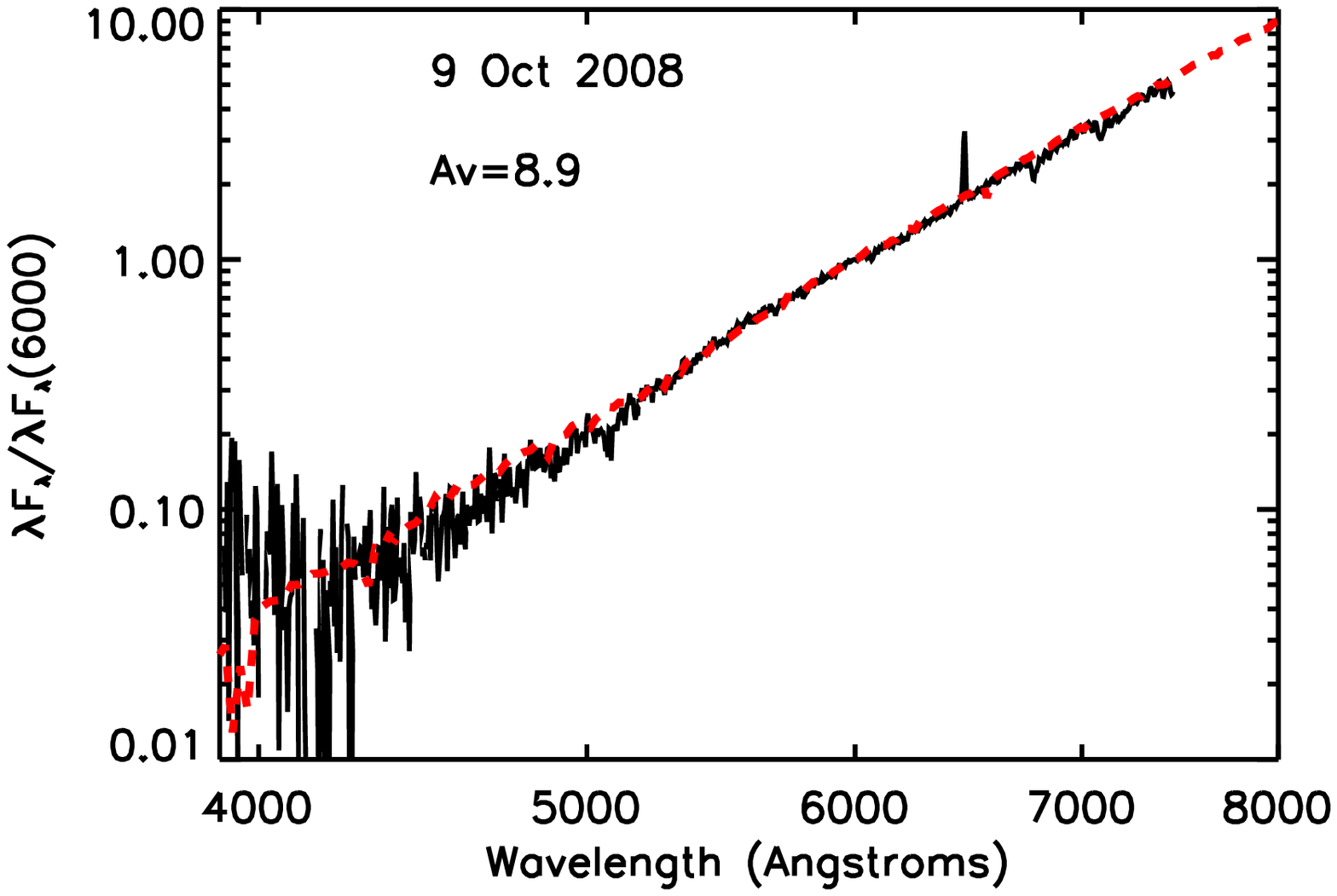}
\includegraphics[scale=.4]{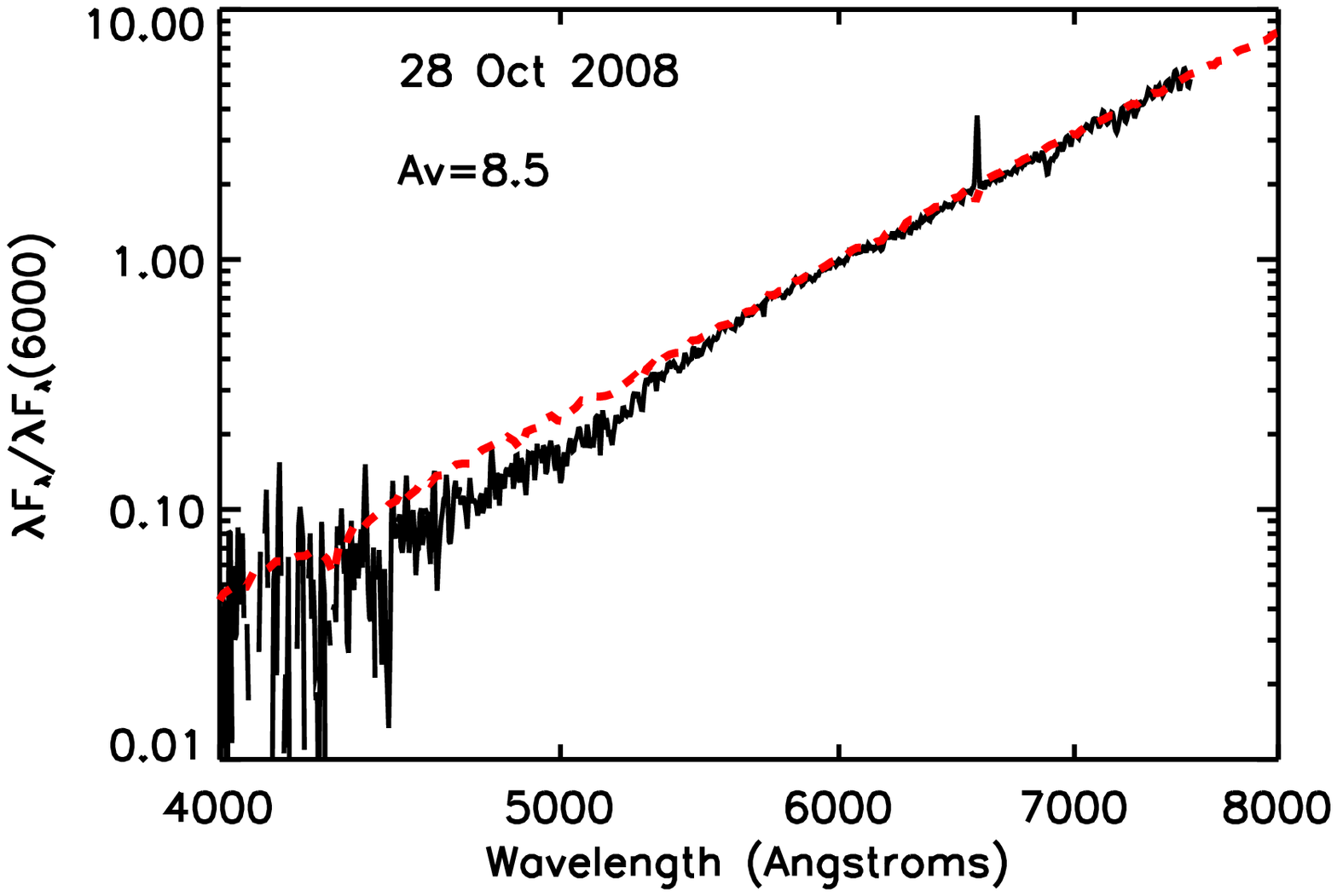}
\includegraphics[scale=.4]{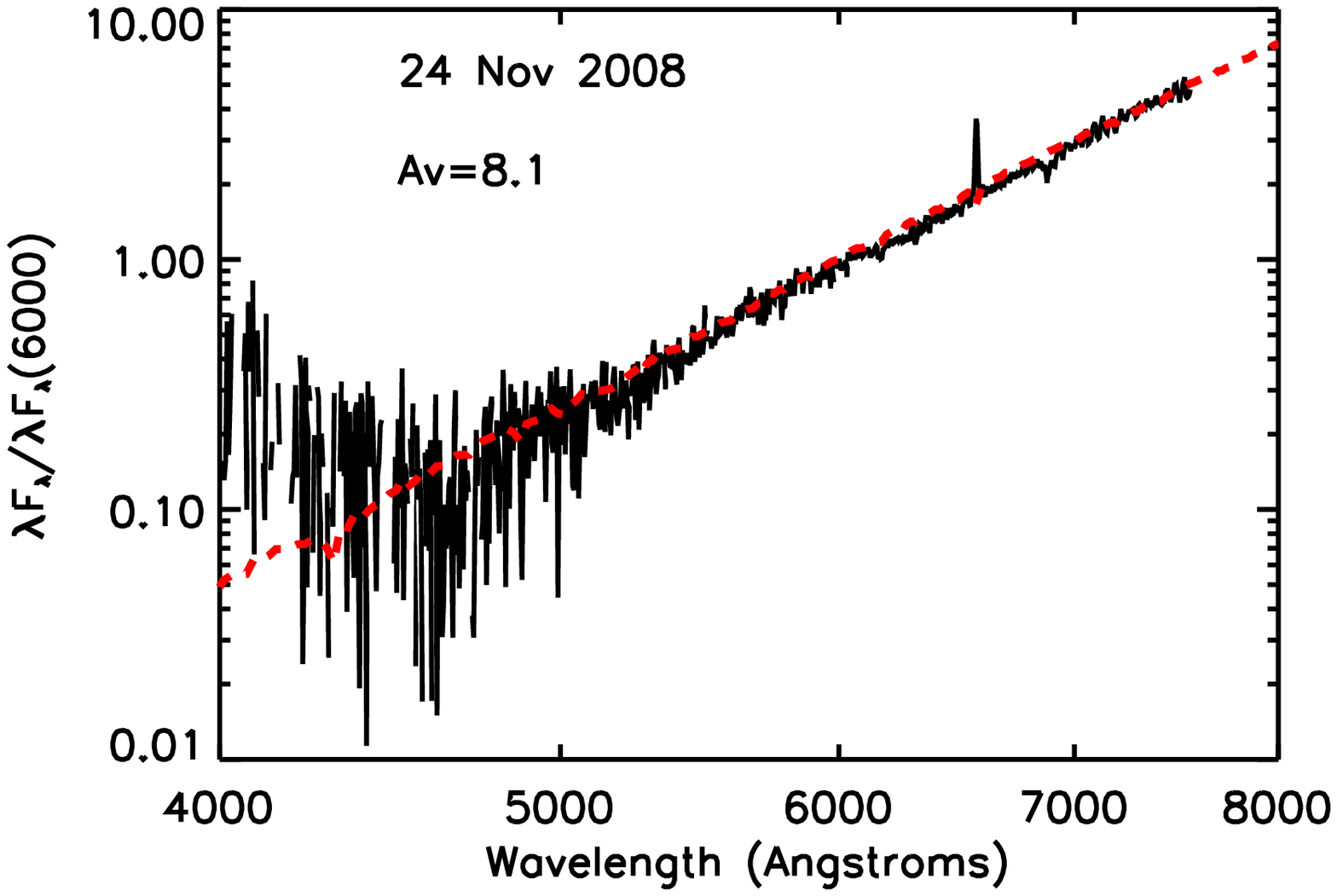}
\caption{Optical spectra of LRLL 31 normalized at 6000\AA. Red dashed line shows a reddened kurucz model that best matches the shape of the spectra.\label{opt_spec}}
\end{figure}

The stellar luminosity can be derived from the J band assuming it is dominated by stellar flux, as demonstrated in Figure~\ref{spectra}, once it has been dereddened.  We use the shapes of both the near-infrared and spectro-polarimetry spectra to measure the reddening of LRLL 31. The spectra can be dereddened until their shapes match that of a standard of the same spectral type. We ignore the K band spectrum since it could be modified by emission from dust, rather than just extinction. We use the R$_V$=5.5 extinction law from \citet{wei01} in both the optical and near-infrared, which is more appropriate for a star embedded in a molecular cloud than R$_V$=3.1, and provides a good fit in comparing LRLL 31 and the standard. For our standard we use the SpeX spectrum of HD 283572 in the near-infrared, and in the optical we used a Kurucz model (Teff=5750, log g=2.5) \citep{cas04}. The results are listed in Table~\ref{extinction}. Since the shapes of the standard and the spectral do not match perfectly due to effects such as imperfect telluric correction and scattered light within the instrument, there is a 0.5 mag uncertainty in the best fit extinction. We find that on average the extinction is A$_{V}\sim8.8$. This is lower than previous estimates \citep{muz09} because of our use of a different extinction law, which will convert the same amount of reddening to a smaller value of A$_{V}$. Occasionally the extinction appears to drop down to A$_{V}\sim8.2$, but these values are consistent with the average given the large error bars. Figures~\ref{spectra} and ~\ref{spectra2} show the observed near-infrared spectra compared with a reddened standard along with the observed JHK photometry and simultaneous 3.6,4.5\micron\ photometry where available. The agreement between the shape of the spectra and the photometry suggests that we are accurately measuring the shape of the SED. Figure~\ref{opt_spec} shows the optical spectra along with a reddened standard. Using the optical spectra we derive extinctions that are consistent with the near-infrared, including occasional apparent drops in A$_V$.

The reddening is combined with the J band photometry, a bolometric correction from \citet{ken95} and a distance of 320 pc \citep{luh03} to produce the stellar luminosity, which is listed in Table~\ref{luminosity}. The uncertainties are due to the the errors in the photometry and the extinction and are typically 0.6$L_{\odot}$. There is a range of luminosities, but given the error bars they are consistent with a constant luminosity of 4.3$L_{\odot}$. This luminosity corresponds to a radius of 2.1R$_{\odot}$ given a G6 (T$_{eff}$=5700) spectral type. If the low values of luminosity represent a real deviation from the mean, they may illustrate moments where the star is partially obscured by dust that is not taken into account in our estimate of the extinction. Overall the star appears to be underluminous relative to a G6 star at the age of IC 348 (2-3 Myr), based on the \citet{sie00} isochrones, although it is roughly consistent with the 3 Myr isochrone of \citet{pal99} which \citet{luh03} finds is more appropriate for the high mass members of IC 348. Based on the \citet{sie00} isochrones and the observed luminosity, LRLL 31 is roughly 1.5 $M_{\odot}$.

Recent monitoring in I-band (Baliber et al. in prep) shows small variations of $\sim3\%$ with a period of 3.4 days, which are consistent with the rotation of cool spots across the stellar surface suggesting that 3.4 days is the rotation period of the star. Despite the frequent monitoring of IC 348, this periodicity has not been observed \citep{coh04,lit05,cie06,nor06}. These previous observations were taken in 1998-2004 and found I=15.24 \citep{coh04} and I=15.36 with occasional dips down to I=15.5 \citep[see star 0130 in Figure 4 of][]{lit05}, while the most recent observations measured I=15.82 for LRLL 31. The near-infrared photometry shows a long term decrease from the 2MASS data (operated from 1997-2001) to our measurements in 2008 and 2009. This suggests that there may be a long term decrease in the stellar flux. Continued optical monitoring of this star will be useful in confirming the periodicity and dimming of LRLL 31. 

Our J-band photometry does not show a large change in the stellar flux, and some of the optical monitoring by Baliber et al. overlaps the first week of our Spitzer warm-mission monitoring and does not show a substantial decrease in the optical flux, which leads us to assume that the stellar flux is constant during our infrared monitoring. With the period information and the $v\sin i$ measurement from \citet{dah08} we can estimate the inclination of this star to be $i=38\pm7^{\circ}$, which includes both the uncertainty in the $v\sin i$ measurement and the radius as derived from the luminosity. This is the inclination to the stellar rotation axis, and it is possible for the stellar magnetosphere and the disk to be misaligned with this axis.


We find a change in the radial velocity based on three epochs of high resolution optical spectra. On 28 Nov 2006 the radial velocity was 12.82$\pm$0.42 km/sec, on 29 Feb 2008 it was 8.36$\pm$0.27 km/sec and on 03 Dec 2008 it was 15.7$\pm$1.0 km/sec. These changes could be due to the influence of a companion, or partial obscuration of the surface by cool spots or circumstellar material. If part of the stellar surface, such as the side rotating toward us, is partially blocked then the radial velocity will appear to become red-shifted. As the obscuring material rotates around the star the apparent radial velocity will change. The rotation of a spot across the surface of the star has been suggested to explain the radial velocity variations in AA Tau \citep{bou07} and the transitional disk T Cha \citep{sch09}. The rotational velocity of LRLL 31 has been measured to be 19.8 km/sec \citep{dah08}, which is consistent with the observed radial velocity variation, although the size of the spot required to produce such large radial velocity variations is bigger than typically seen around G-type stars. A change in the shape of the bisector of the cross correlation function used to derive the radial velocity would indicate the presence of obscuring source rotating in front of the star, but we do not have the signal to noise to test this theory. A clump of obscuring material, either in the disk or in the molecular cloud, would need to block almost half of the star in order to create the observed radial velocity variation. This would produce a drop in the optical flux of 0.75 mag, along with a substantial increase in the extinction, neither of which are observed. If these radial velocity measurement are confirmed then the most likely explanation is the presence of a companion somewhere in the system, although this does not prove that the companion is responsible for the observed infrared variability. In section 4.3 we place constraints on the mass and location of such a companion using the available data, although more radial velocity measurements are needed to confirm and fully characterize this companion.

\subsection{Dust Properties}
LRLL 31 was originally singled out based on its mid-infrared variability \citep{muz09}. In IRS spectra separated by one week the 5-8\micron\ flux dropped almost to the photosphere while the 8-40\micron\ flux increased. This behavior was unique and difficult to explain given typical sources of variability, such as the rotation of hot/cold spots across the surface of the star or extinction events. Modeling finds that the wavelength dependence, and timescale, of the variability is consistent with fluctuations in the scale height of the inner disk on a dynamical timescales, which is on the order of one week \citep{fla10}. As the inner disk grows it increases the short wavelength flux, while at the same time shadowing the outer disk, decreasing the long wavelength flux. The previous observations of the continuum emission were limited to probing material far from the star, at 1-10 AU, which may not be the source of the variability. The silicate emission may arise from closer to the star \citep[e.g.][]{kes07}, but this location is difficult to constrain without detailed radiative transfer models, especially given the transition disk nature of LRLL 31

Our 3-5\micron\ spectra give us information on the dust close to the star. By subtracting the stellar photosphere from the spectra we are left with only the emission from the dust. When the dust at the inner edge of the disk is optically thick, the emission should appear as a single temperature blackbody at the dust destruction temperature \citep{muz03,esp10}. The shape of the excess spectra will tell us the temperature of the hot dust which translates to the location of the inner edge of the dusty disk. To subtract the photosphere we must first normalize the standard to the LRLL 31 stellar component. We use the veiling at 2.15\micron\ to separate the stellar and excess emission of LRLL 31. 

\begin{figure}
\center
\includegraphics[scale=0.8]{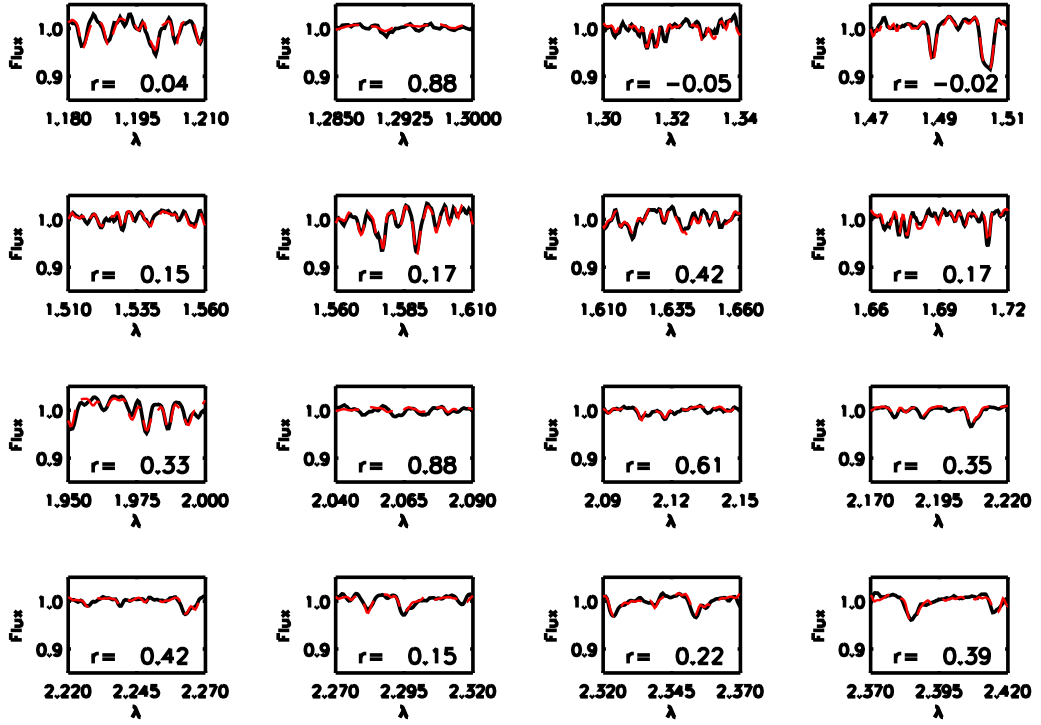}
\caption{Continuum normalized flux as a function of wavelength for the 16 bands used to measure the veiling for 08 Nov 2009. In each panel both the observed spectra (black line), along with the best fit veiled standard (red dashed line) are shown. The veiling increases from 0 in J band to a few tenths in K band. There can be a large spread in the veiling from band to band depending on the strength of the lines in the band and the continuum fit. \label{veiling}}
\end{figure}

We can derive the veiling based on the strength of photospheric lines in the K band relative to the G6 standard. Veiling is defined as $r=F/F_{phot}$ where F is an excess continuum flux that comes from the accretion disk; as r increases the strength of the lines relative to the continuum decreases. The value of r can be used to measure the amount of excess emission. We calibrate the veiling by adding a constant flux to our G6 standard in 16 small bands 0.05\micron\ wide from 0.8-2.5\micron\ and comparing the veiled standard to the observed spectra. The best-fit is determined independently in each band using the minimum chi-squared. These multiple measurements provide r as a function of wavelength and we use this linear fit to derive the veiling at 2.15\micron. There is substantial scatter between individual measurements, but we are able to use the line fit to estimate r with an uncertainty of $\pm$0.1. Figure~\ref{veiling} shows the measurements in our 16 bands for one day, demonstrating that we can obtain good fits across much of the spectrum, showing a trend of increasing veiling with increasing wavelength. The results are listed in Table~\ref{ir_excess}.

\begin{figure}
\includegraphics[scale=.3]{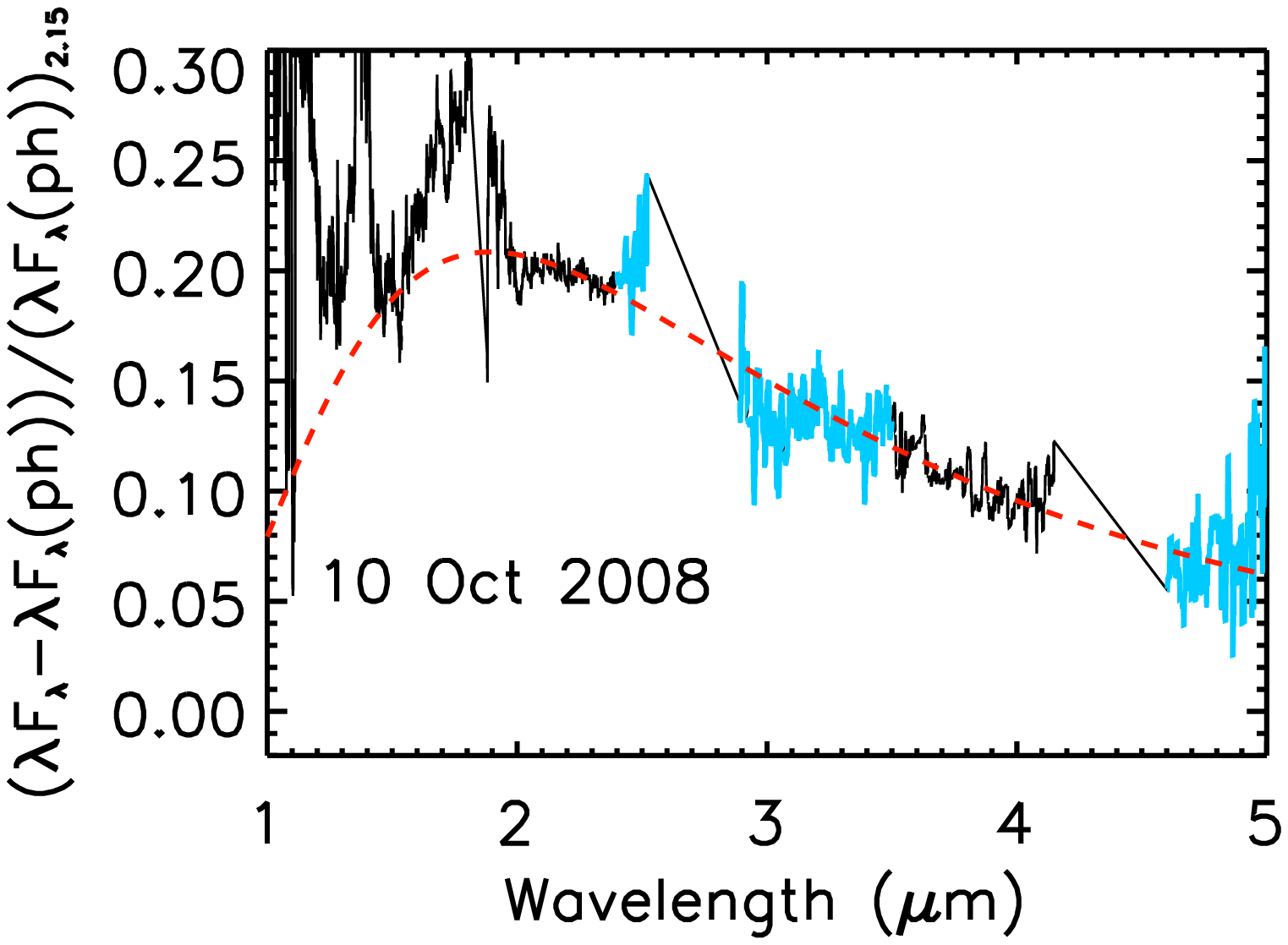}
\includegraphics[scale=.3]{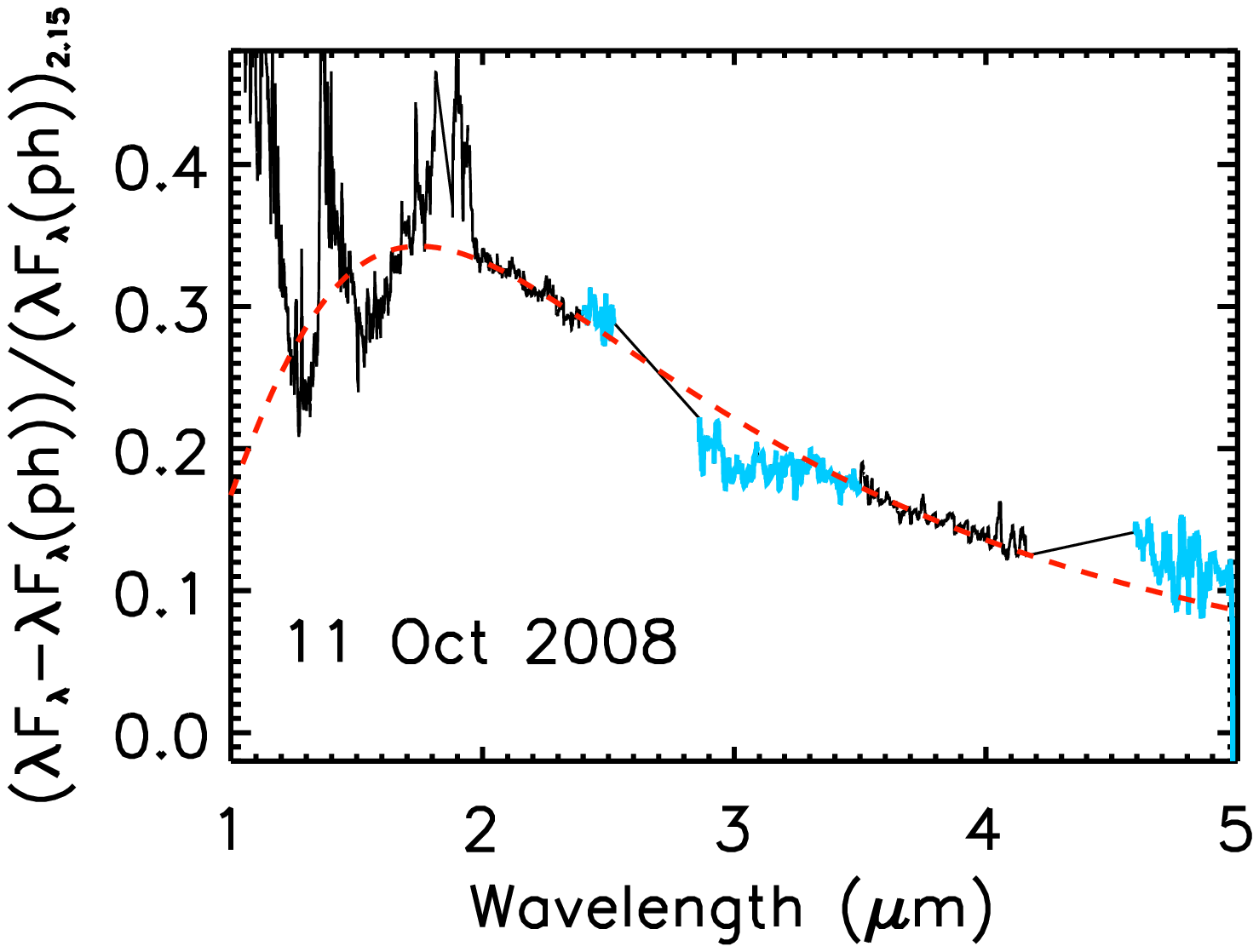}
\includegraphics[scale=.3]{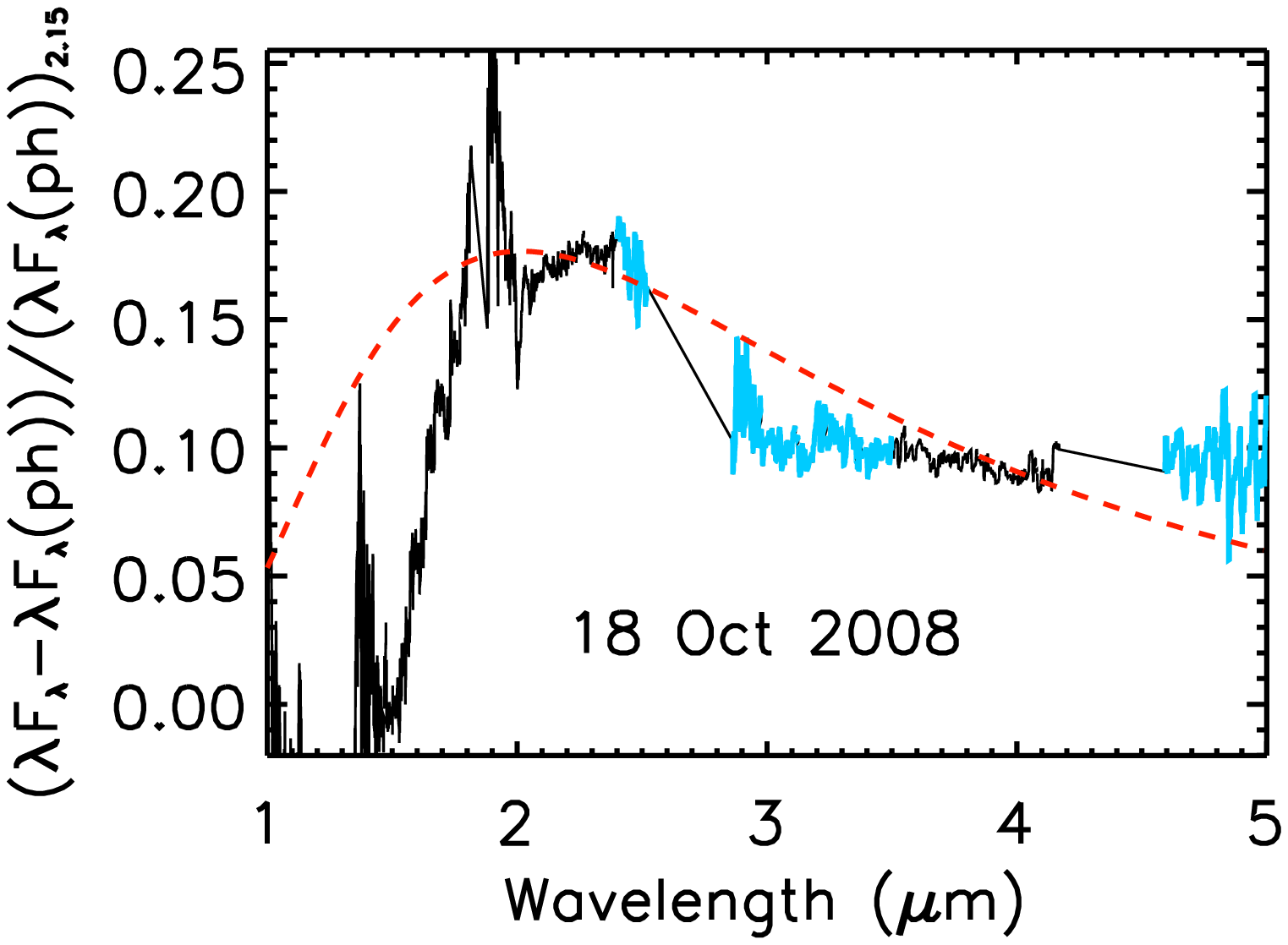}
\includegraphics[scale=.3]{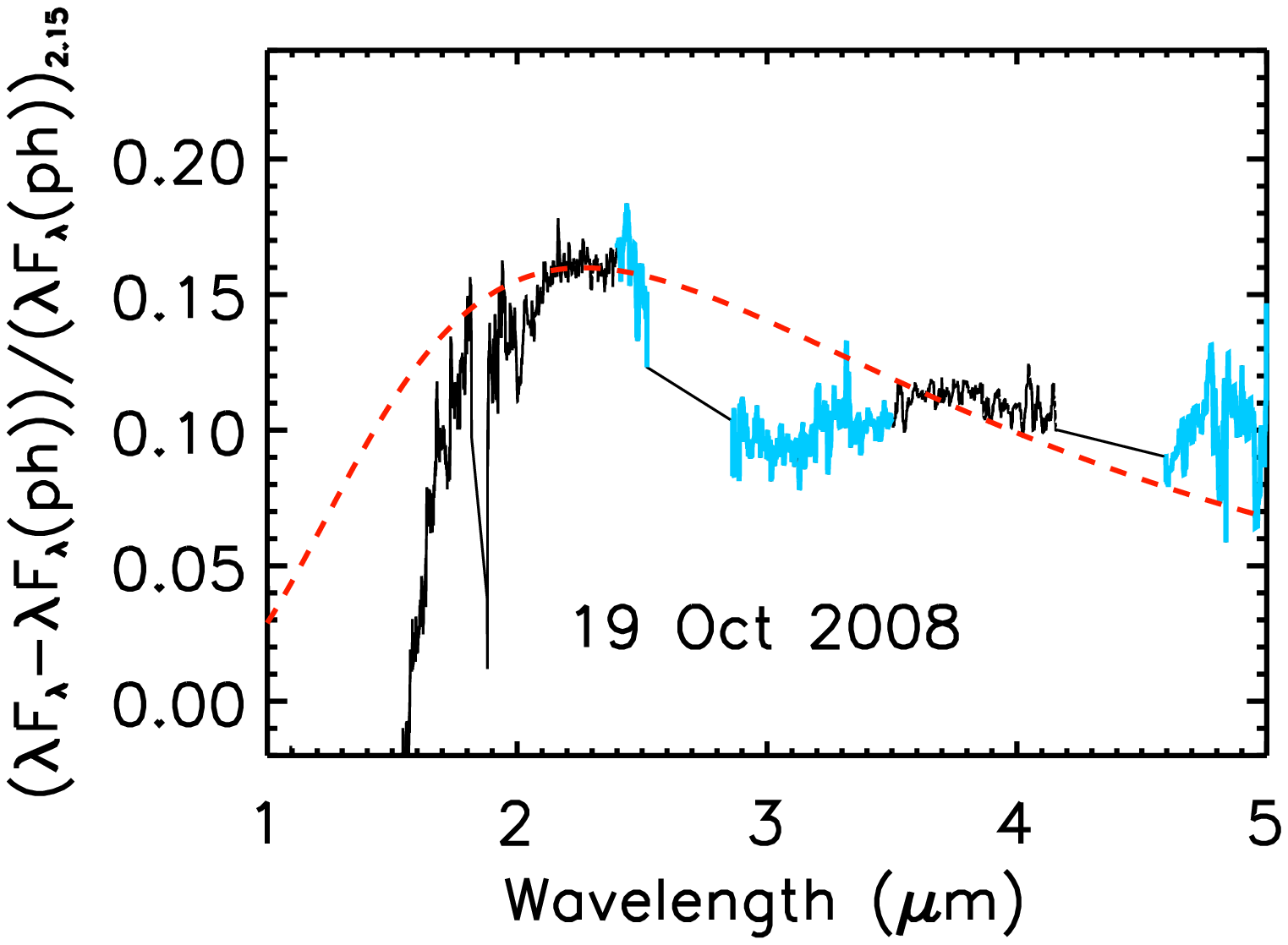}
\includegraphics[scale=.3]{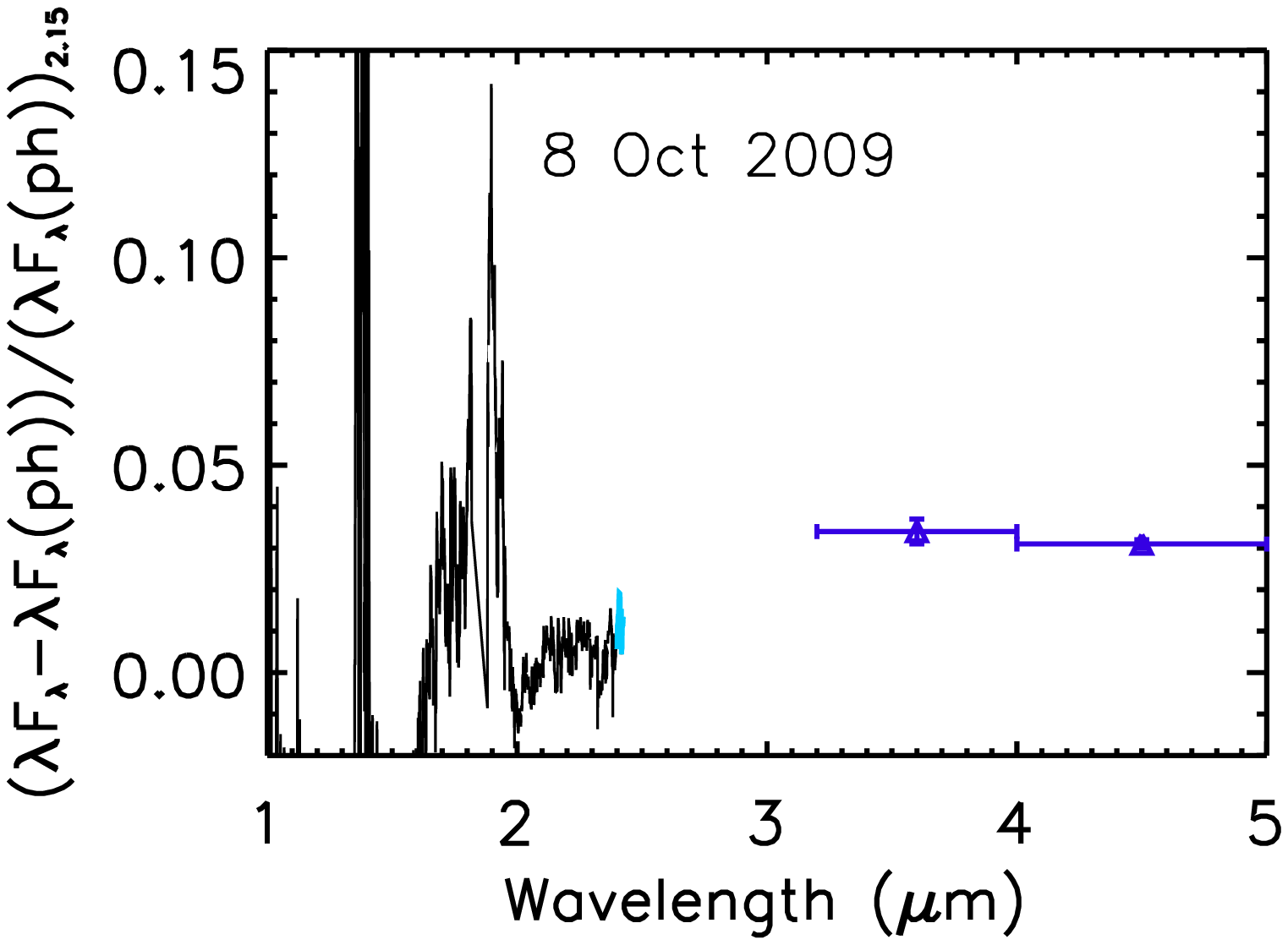}
\includegraphics[scale=.3]{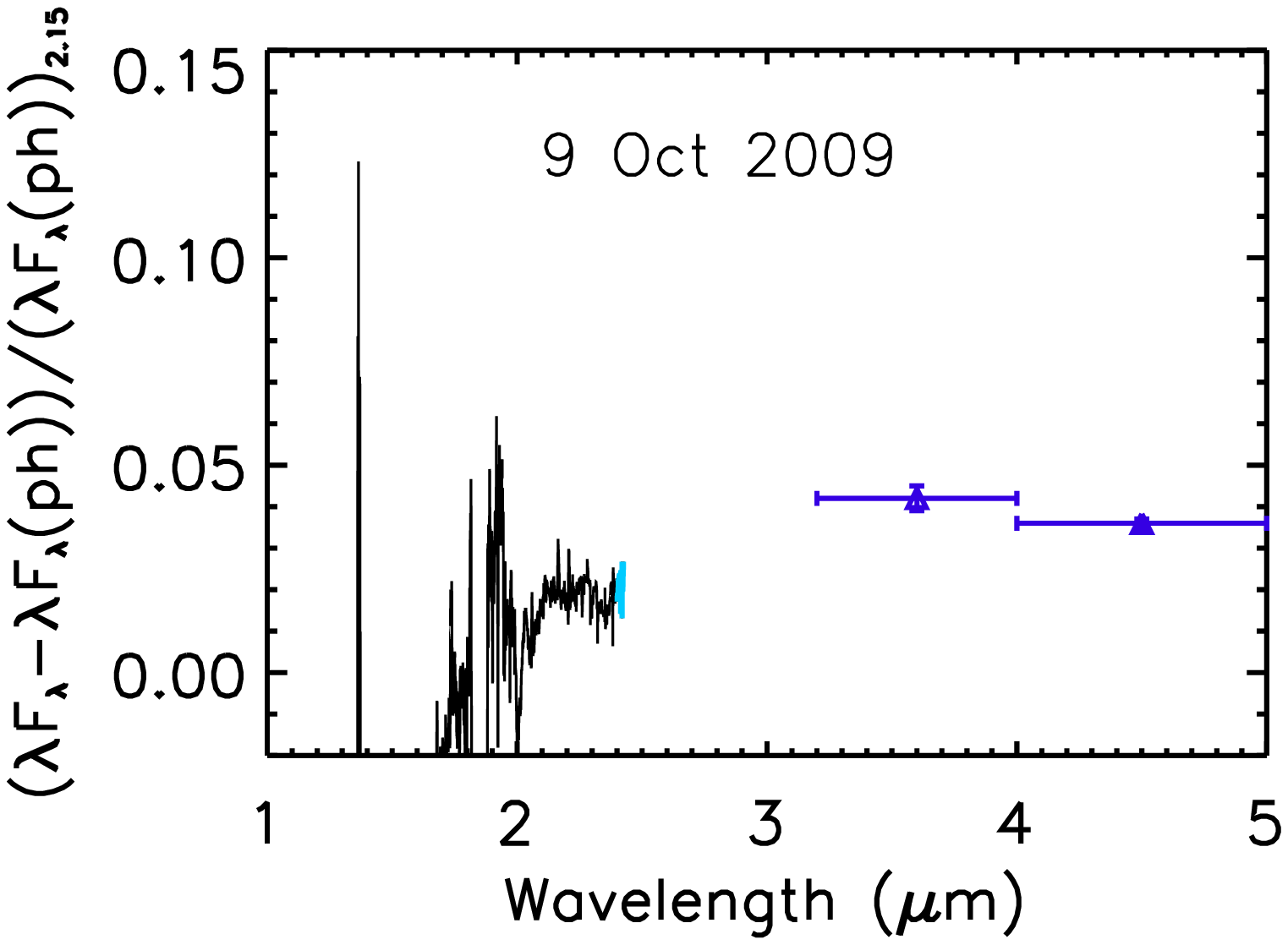}
\includegraphics[scale=.3]{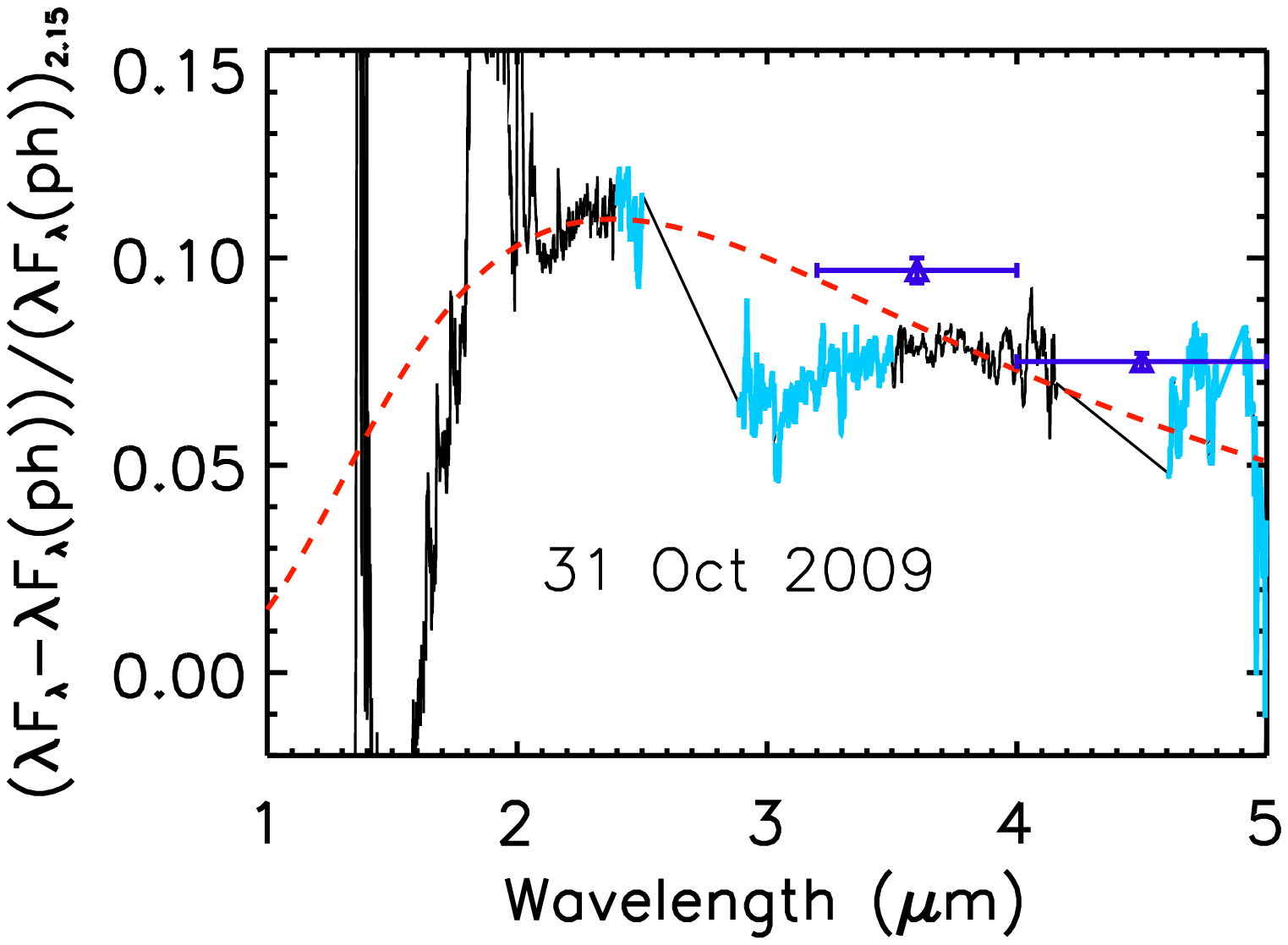}
\includegraphics[scale=.3]{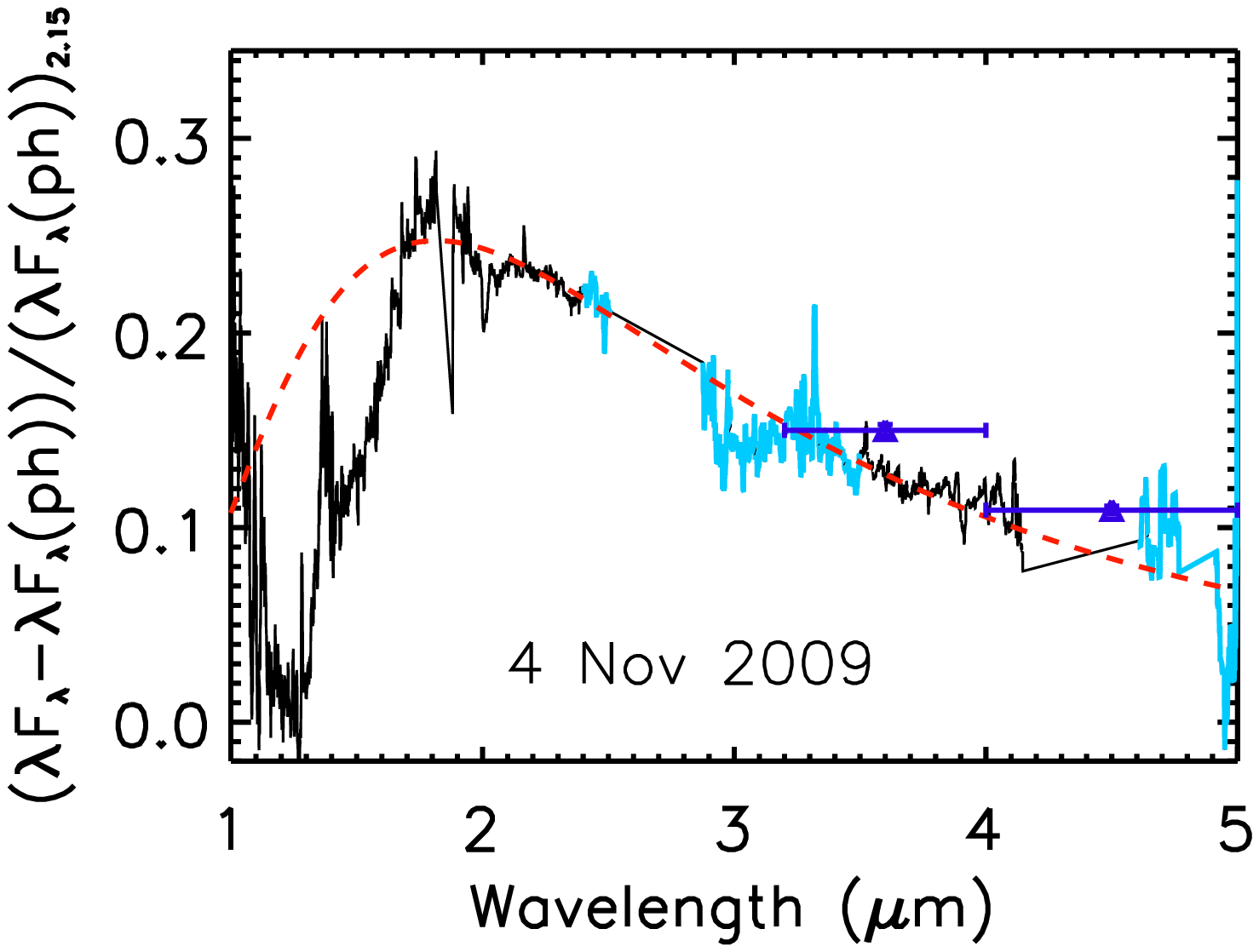}
\includegraphics[scale=.3]{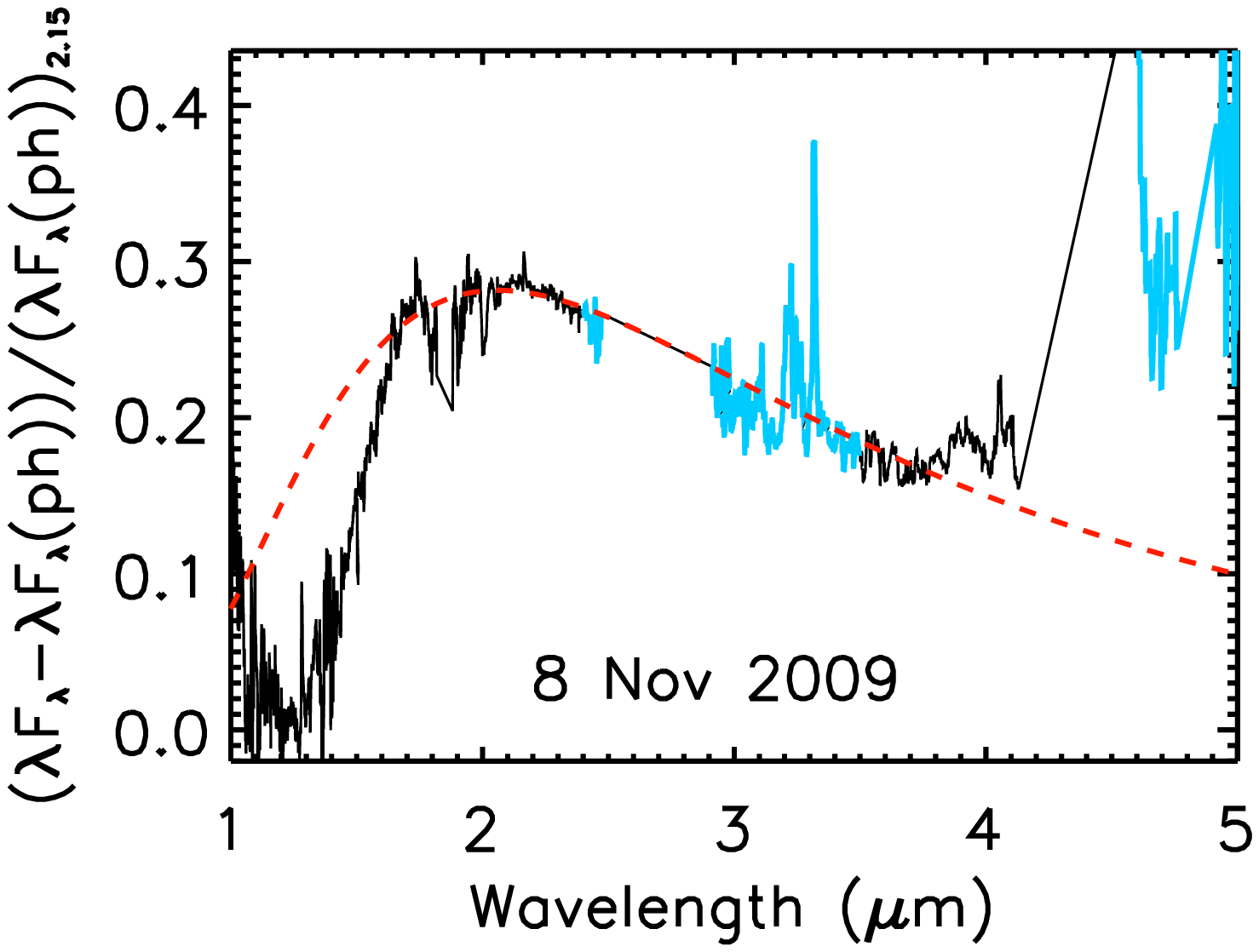}
\caption{The infrared excess. Each plot show the difference between LRLL 31 and the standard, normalized to the photospheric flux at 2.15\micron. On each plot the best fit is shown with a red dashed line and 3.6,4.5\micron\ photometry is included when available. Each spectra has been smoothed by a median filter 0.01\micron\ wide in order to reduce the noise in the continuum. The parts of the spectra marked in blue are strongly affected by the telluric correction. Note that the scale changes for each plot. The strength of the excess shows large variations from night to night, although the temperature stays roughly constant. \label{irexcess}}
\end{figure}

Once we have used the veiling to determine the excess emission at 2.15\micron\ we normalize the standard and LRLL 31 spectra at this wavelength. Since our G6 WTTS standard spectrum only extends to 2.3\micron\ we use a Kurucz model (Teff=5750, log g =2.5) \citep{cas04} to extend it out to 5\micron. The standard and LRLL 31 are subtracted yielding an excess spectrum whose flux is relative to the photospheric flux at 2.15\micron. These excess spectra are shown in Figure~\ref{irexcess}.

\begin{figure}
\includegraphics[scale=.3]{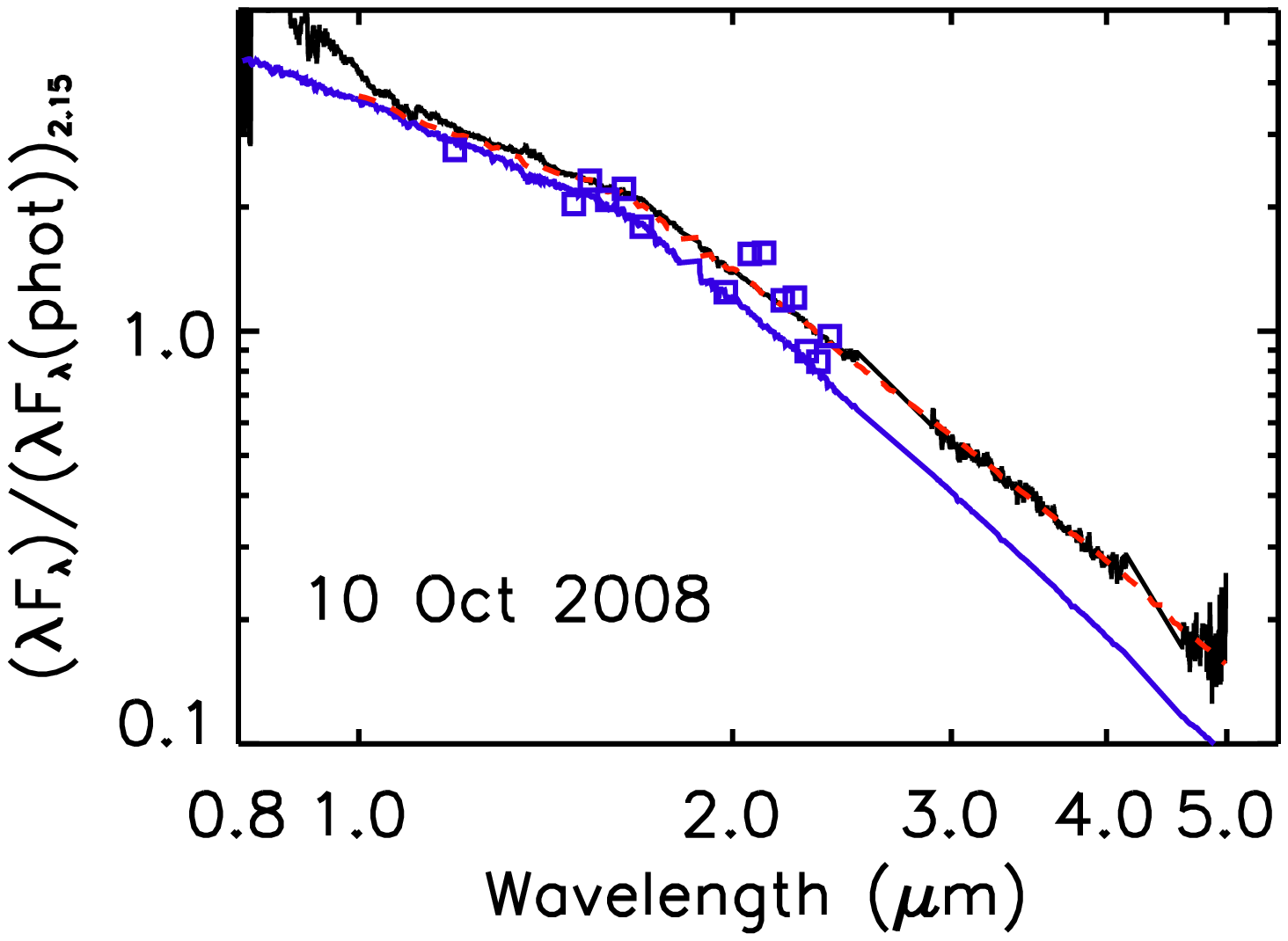}
\includegraphics[scale=.3]{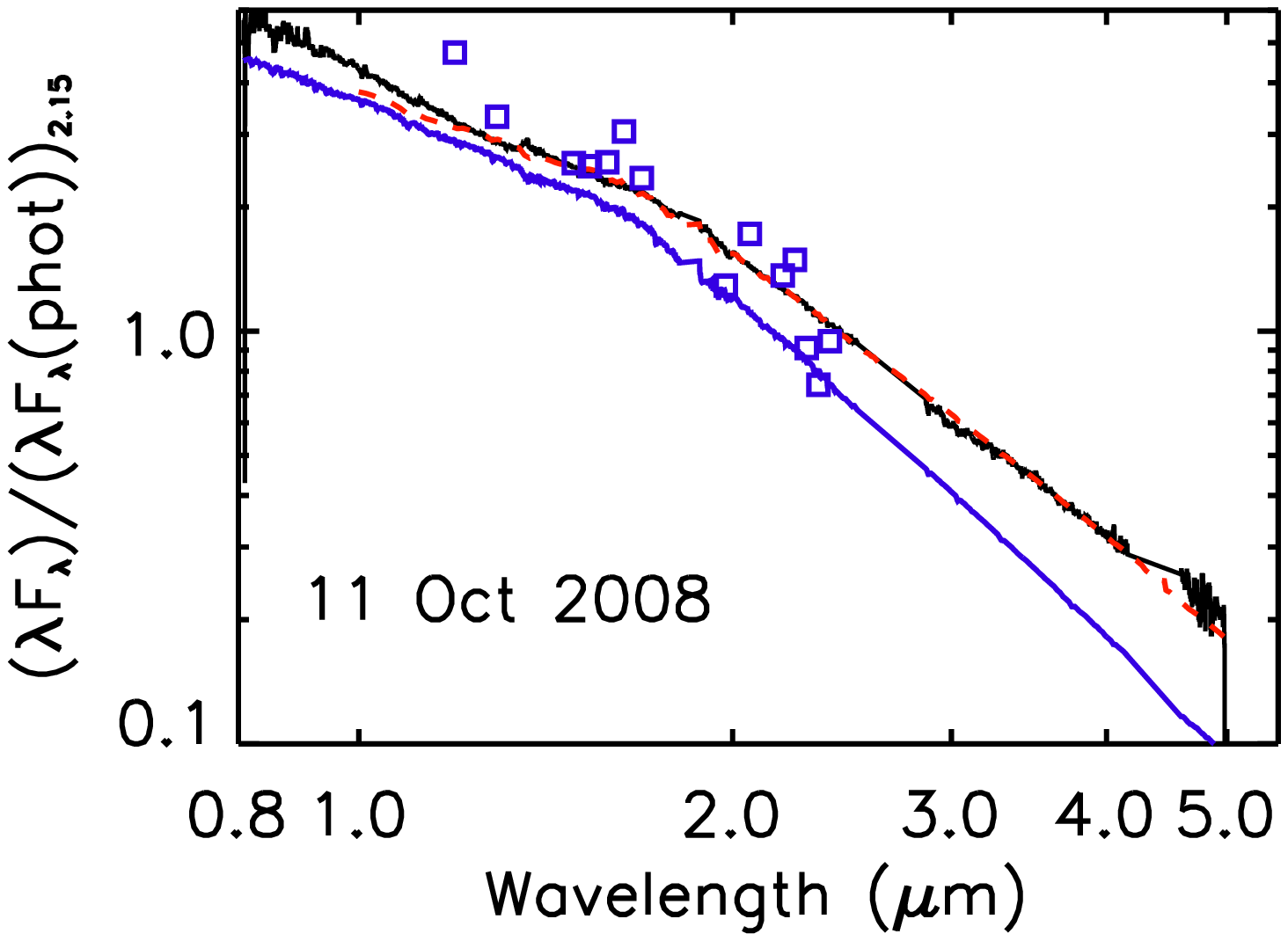}
\includegraphics[scale=.3]{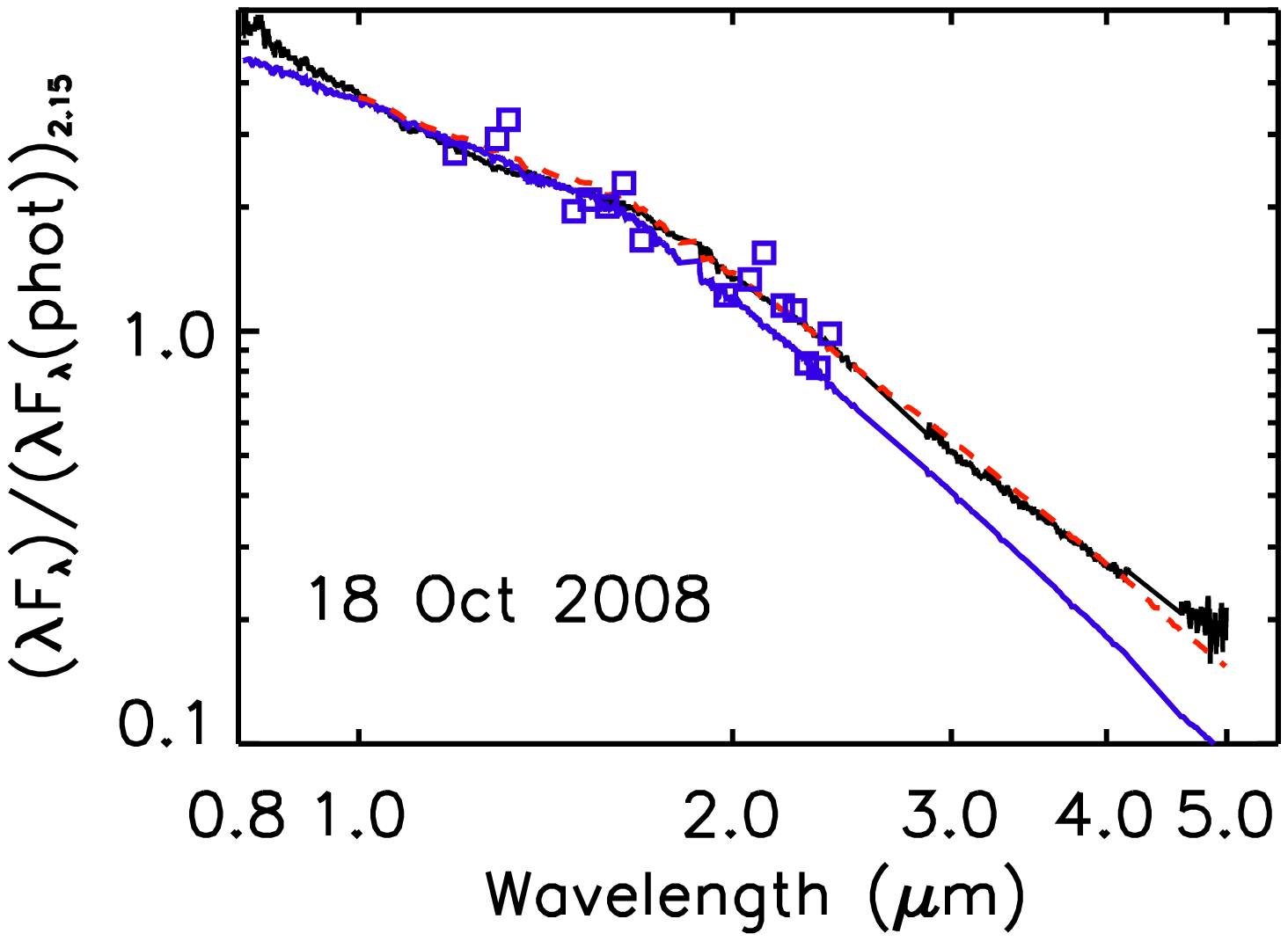}
\includegraphics[scale=.3]{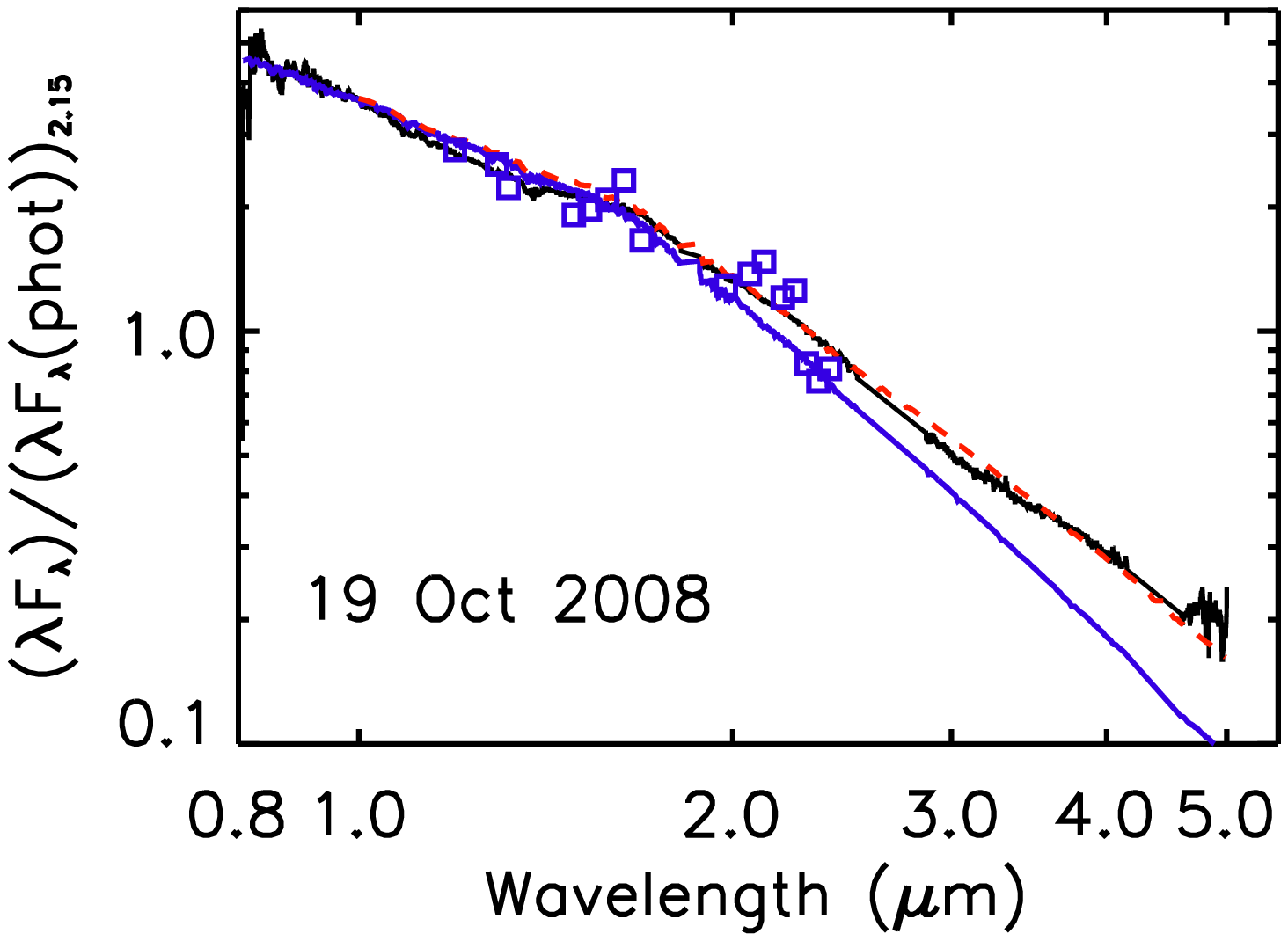}
\includegraphics[scale=.3]{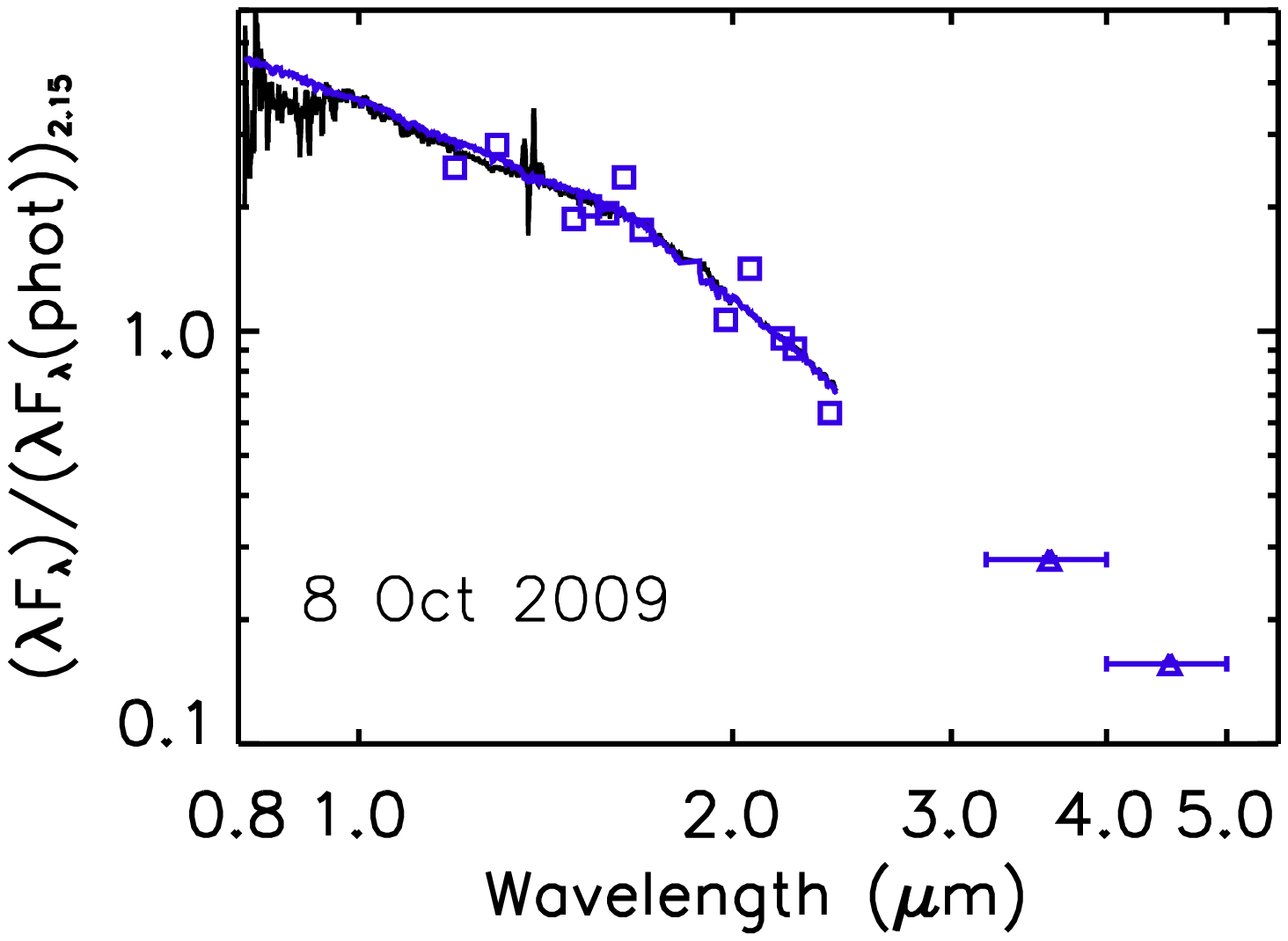}
\includegraphics[scale=.3]{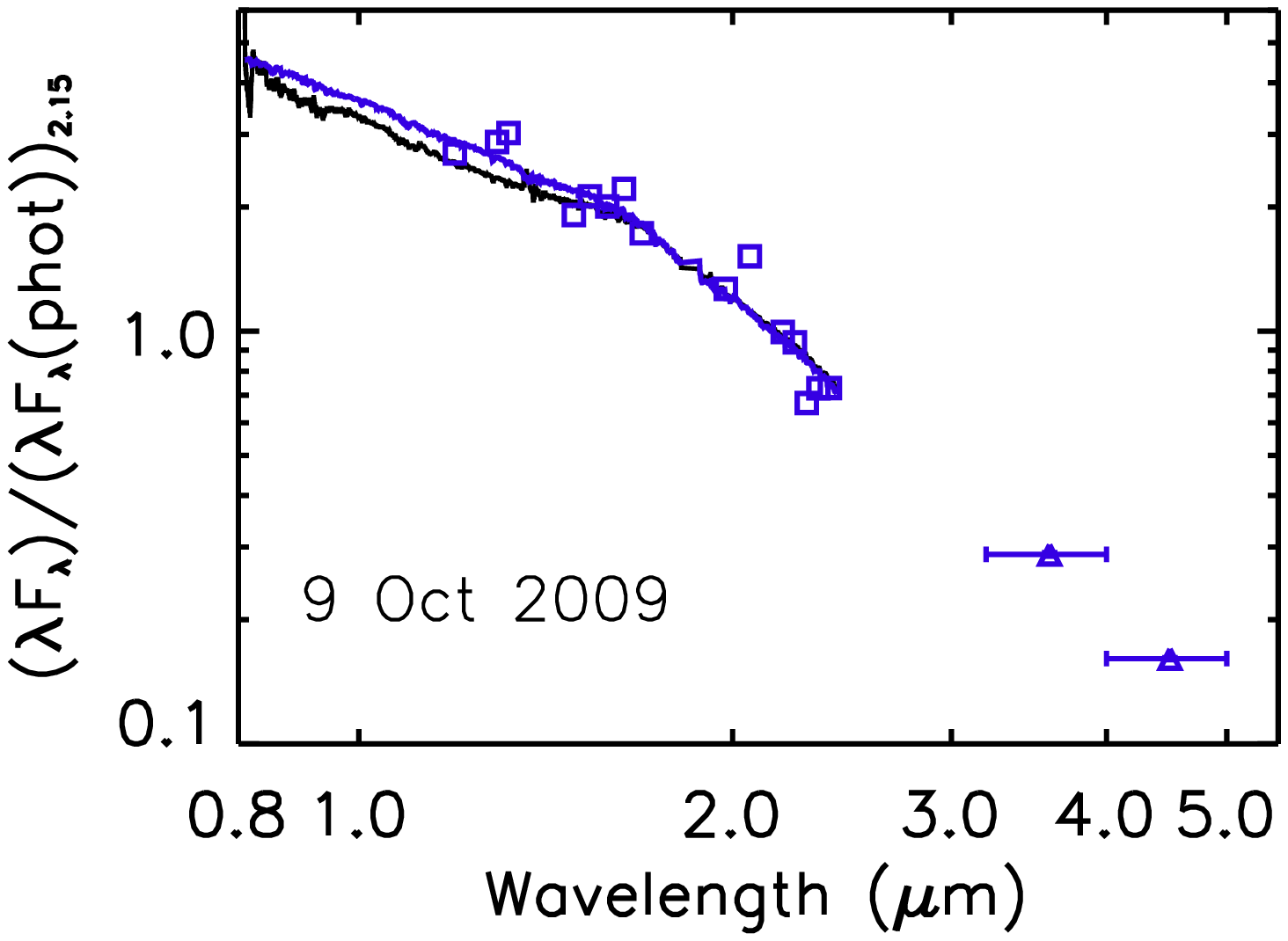}
\includegraphics[scale=.3]{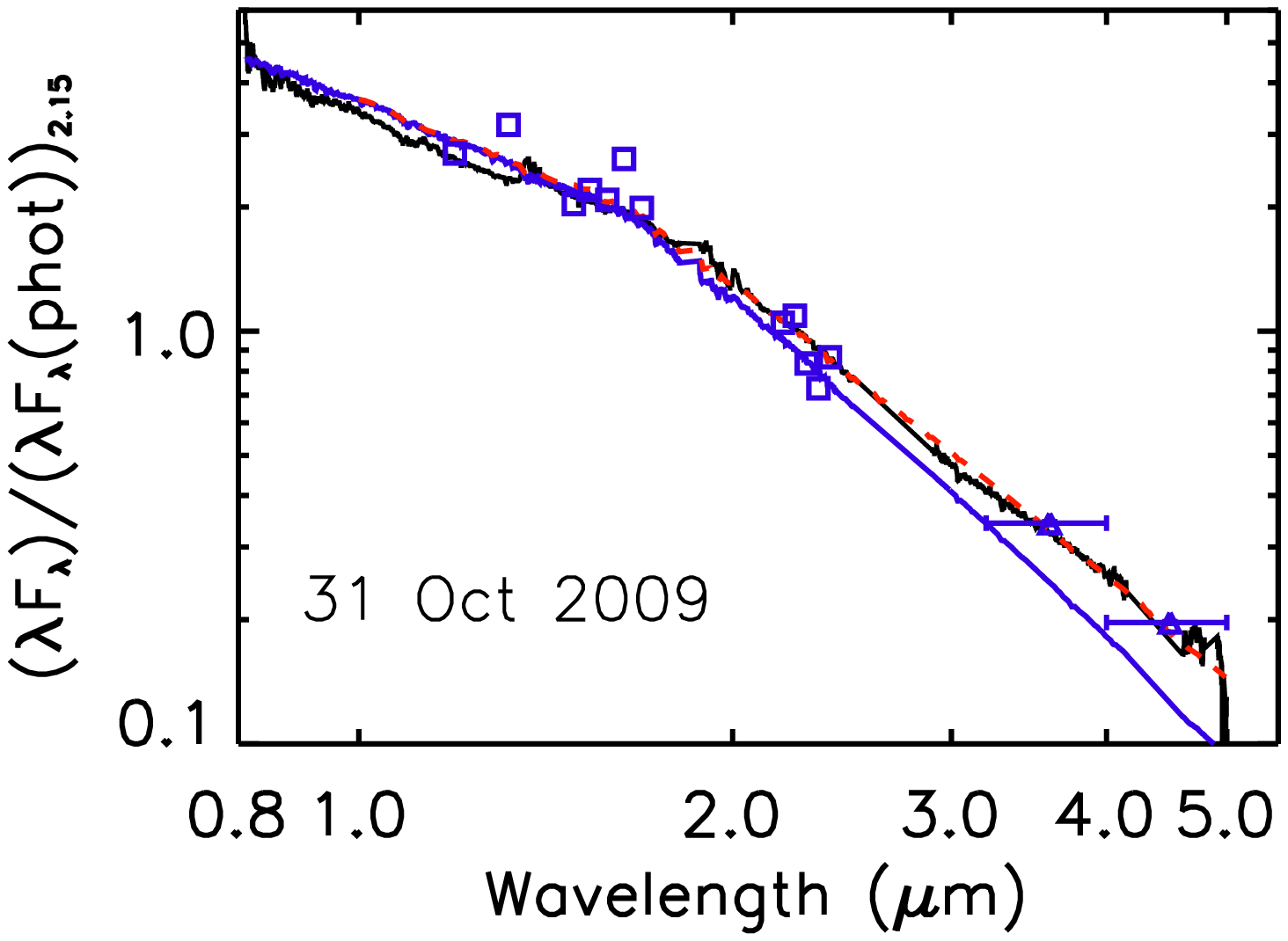}
\includegraphics[scale=.3]{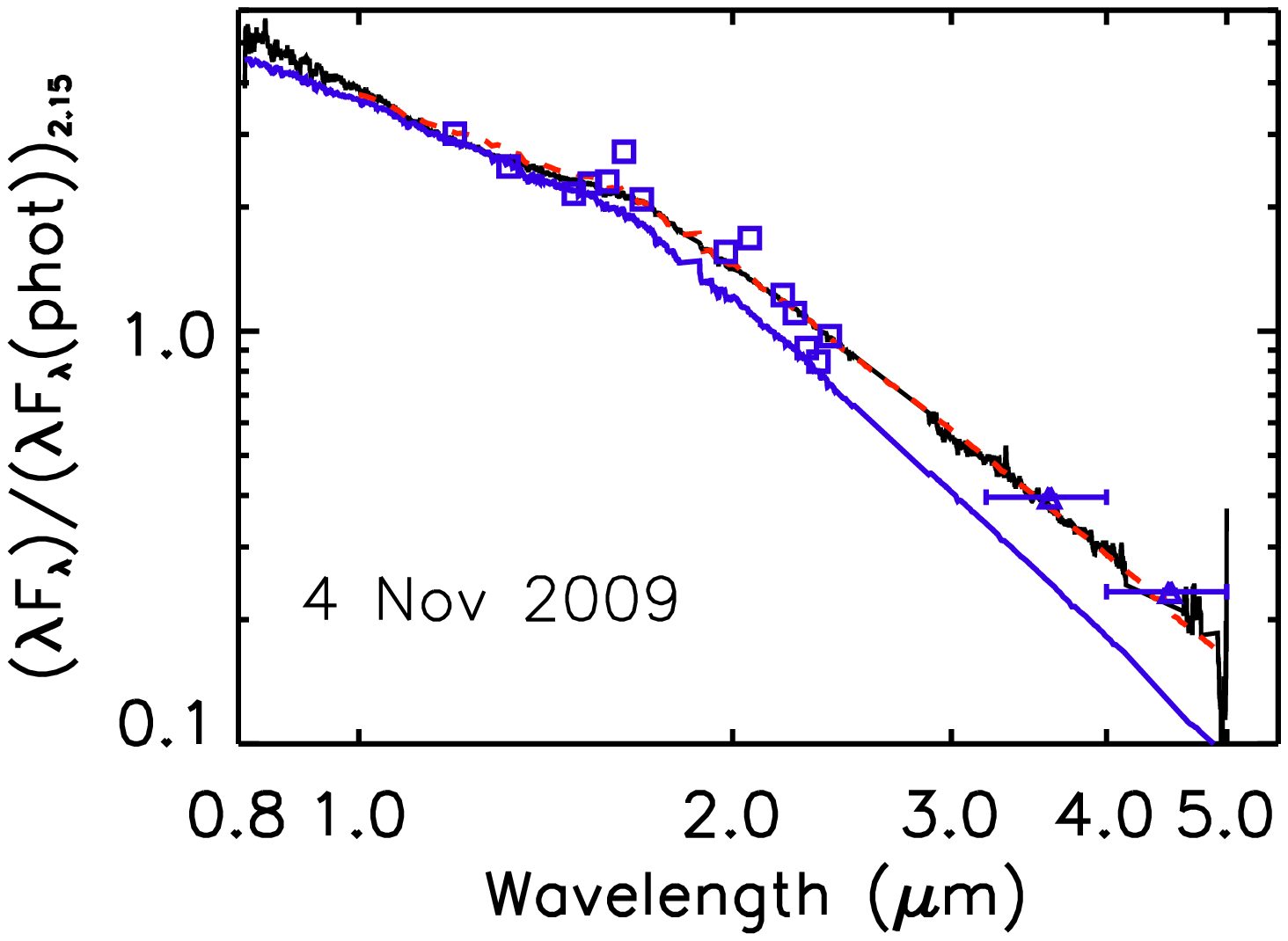}
\includegraphics[scale=.3]{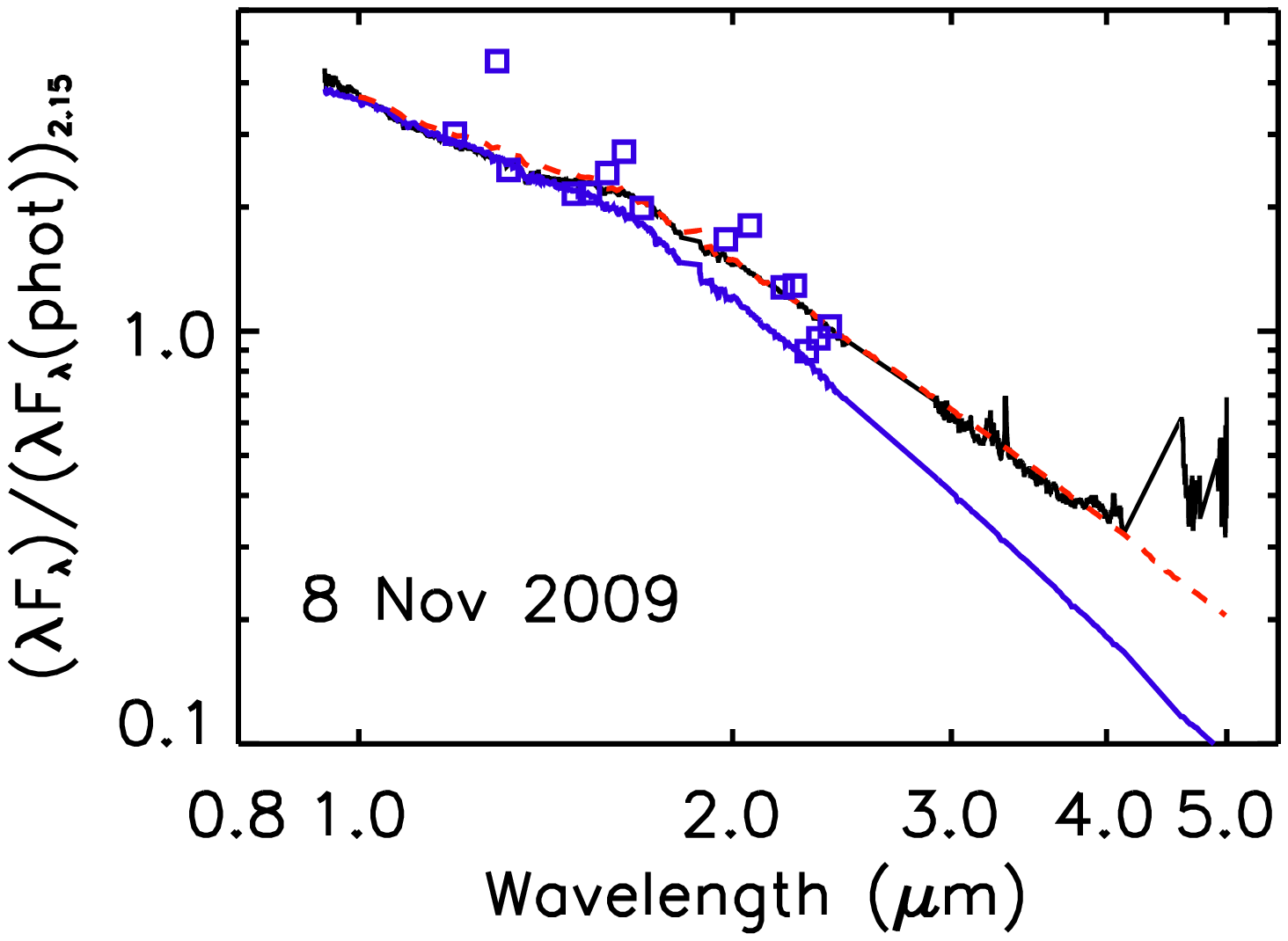}
\caption{The spectra for each night, along with a standard and the blackbody fit the the excess. Dark blue line is the standard, red dashed line is the standard plus the best fit blackbody. Blue squares are the veiling measurements, and blue triangles are the 3.5,4.5\micron\ photometry, where available. Spectra have been normalized by the photosphere at 2.15\micron\ and smoothed by a median filter 0.01\micron\ wide. \label{irexcess2}}
\end{figure}

To find the temperature of the hot dust we fit a single temperature blackbody to the excess spectrum. We artificially increased the uncertainties in the wavelength ranges $2.4<\lambda<3.5\micron$ and $\lambda>4.3\micron$ in order to account for systematic effects from an incomplete telluric correction, which affect some regions of the spectrum more than others. This is especially important for 08 Nov 2009 when the large airmass difference between the telluric standard and LRLL 31 (0.48) resulted in a poor correction of the telluric features. Typical uncertainties on the derived temperature are 250K, which is mostly due to the uncertainties in the veiling and in the extinction. \footnote{We derive the extinction assuming that there is very little excess in the J and H bands, which could potentially change the shape of the spectra.} As seen in Figure~\ref{irexcess} a blackbody is a good fit to the excess spectra. The full spectra, along with the veiling measurements as a function of wavelength are shown in Figure~\ref{irexcess2}. 

We can double check the shape of the excess spectra by normalizing both the standard and LRLL 31 at 1.15$\micron$, where we expect and measure zero veiling. This avoids using the veiling to normalize the spectra, since the large uncertainties in the veiling lead to large changes in the derived temperature. Veiling presents an extinction independent measure of the shape of the LRLL 31 spectra relative to the standard, which can be used in this case to confirm the shape of the spectra, rather than relying on them to define the shape. We find that the temperatures derived in this instance are consistent with our previous derivation within the uncertainties. The largest discrepancies are for the 31 Oct 2009, where the new temperature is 1900 K, and 11 Oct 2008, where the new temperature is 1620 K. Normalizing the spectra at 1.15\micron\ brings these two measurements into better agreement with the derived temperatures from the other five nights. 

The temperatures, listed in Table~\ref{ir_excess}, are consistent with each other, within the uncertainties, and the average temperature of 1830 K is consistent with the dust sublimation temperature \citep[e.g.][]{dul10}. This implies that the sublimation of dust sets the inner edge of the disk rather than some other process, such as the sculpting of the disk by a companion. Converting the dust temperature to a location in the disk depends on the properties of the dust, the luminosity of the star, and also on the covering fraction of the inner rim \citep{ise05}. If one of these dependencies was rapidly varying then the location of the inner rim could fluctuate while the temperature stays constant. Our measurements indicate that the accretion luminosity is much less than the stellar luminosity, the stellar luminosity does not vary significantly, and we do not expect the dust properties, such as grain size and crystallinity, to change on the observed timescales, suggesting that a constant dust temperature corresponds to a constant sublimation radius ($\sim0.15$AU).

While the average temperature of the hottest dust is roughly consistent with the dust sublimation temperature, it is still higher than is typically measured \citep{muz03,esp10}. This discrepancy may either be due to unique gas and dust properties that allow for higher than normal temperatures or systemic effects in our measurements that cause us to overestimate the dust temperature. Dust composed of irons and olivines at high gas densities has a higher sublimation temperature than dust particles more commonly seen around young stars \citep{pol94}. An inner disk composed mostly of this material would have an abnormally high temperature. While we cannot directly measure the dust and gas properties of the inner disk we can use the silicate feature and the accretion rate as proxies. The silicate feature is sensitive to dust properties  at radii of a few tenths of an AU to a few AU \citep{kes07} and the dust properties responsible for this feature generally do not change by orders of magnitude throughout the disk \citep{van04}. The shape of the silicate feature in LRLL 31 is consistent with typical ISM-like silicate dust with sub-micron sized grains, not a heavy mix of irons and olivines. Also the accretion rate, which is proportional to the gas density in a viscously evolving disk, is inconsistent with an extremely high gas density. The fact that this star is a transition disk would suggest that there is a decrease in the gas density rather than an increase. 

This suggests that there is a systematic effect in our measurements that lead to an overestimation of the dust temperature. Overestimating the veiling or the extinction, possibly due to an imperfect match between the standard star and the intrinsic photospheric emission of LRLL 31 will lead to a higher measured dust temperature. Given the difficulty in measuring the veiling at such low levels, this is the most likely source of the discrepancy. Our conclusions about the change in veiling should be unaffected by this systematic uncertainty since we use the same standard, accounting for changes in resolution, and derive the veiling using the same method on each night.

While the temperature of the dust is constant, the strength of the excess fluctuates throughout our observations. The variations in the strength of the excess occur on weekly to monthly timescales, but do not show daily fluctuations (Figure~\ref{ir_excess}). We integrate a blackbody at the derived temperature and strength for each night to determine the luminosity of the excess relative to the stellar luminosity, which is an estimate the covering fraction of the inner disk, since the flux from the disk is reradiated stellar flux. Our results are listed in Table~\ref{ir_excess}. For the nights of 8,9 Oct 2009 we assume the dust is at 1830 K, the average of the other seven nights, and fit the blackbody to the IRAC data to determine the strength of the excess. We find that the covering fraction varies by a factor of 8, from 1-8\%. 

In a typical T Tauri star, with a puffed inner rim, the covering fraction is 12\% \citep{dul01}, while typical debris disks have covering fractions of $<$0.1\% \citep[e.g.][]{pla09}. On average the covering fraction of the dust in LRLL 31 is below that expected for an optically thick puffed inner disk. In section 4 we discuss physical models that can produce changes in the infrared excess on short timescales, but here we consider what the small average covering fraction implies about the inner disk structure. The covering fraction for an optically thick disk is proportional to the ratio of disk height to radius, H/R \citep{dul10}, and if we assume that R stays constant, then we are observing a reduction in the scale height relative to a normal disk. This deviation in the average scale height from a typical disk could be due to (1) a dramatic decrease in the local surface density or (2) a decrease in the dust opacity, possibly due to the growth of the dust grains. Both of these effects would reduce the $\tau=1$ surface of the disk where stellar photons are absorbed, reducing the covering fraction. The accretion rate, which in a viscous disk is proportional to the surface density, is not unusually low for a T Tauri star, suggesting that enhanced grain growth and settling may be causing the decrease in the covering fraction, rather than a decrease in the surface density.  

This long term evolution of the grain size and settling toward the midplane may or may not be related to the rapid fluctuations seen in our data. The change in covering fraction implies a change in the scale height of a factor of 8 if the perturbation is axisymmetric. A model of the infrared fluctuations does not need to produce axisymmetric perturbations of the disk but does need to create features that are large enough to match the observations. If the change in covering fraction is due to a localized blob being launched from the disk and the disk always has a covering fraction of 1\%\ then the other 7\%\ can be due to a blob with a circular diameter of $\sim$50$^{\circ}$ as seen from the star.

\begin{figure}
\center
\includegraphics[scale=.7]{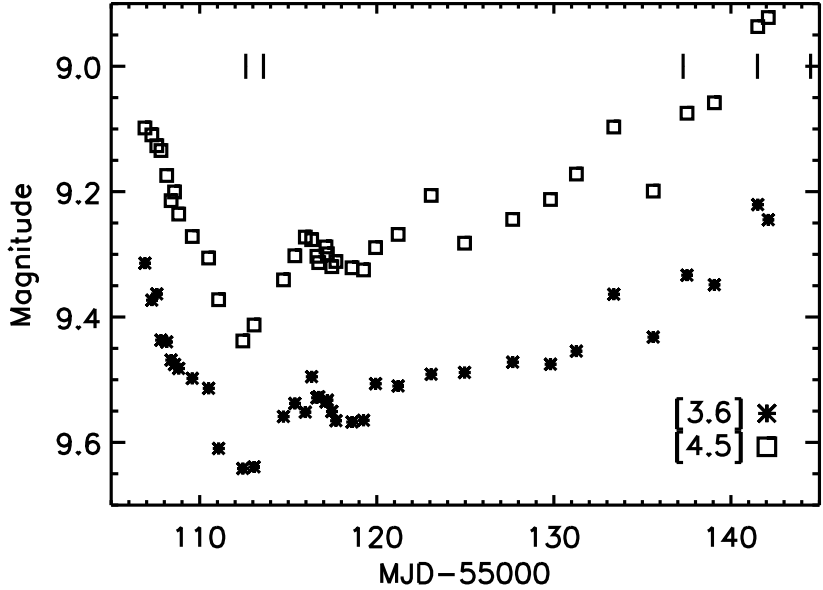}
\caption{3.6 and 4.5\micron\ light curves for LRLL 31. The data show a rapid drop in flux followed by a slow increase over the course of a few weeks. Uncertainties in the photometry are similar to the size of the points. Vertical lines mark the times when we obtained spectra. \label{wm_lc}}
\end{figure}

The 3.6 and 4.5\micron\ flux (Figure~\ref{wm_lc}), where we have more complete monitoring, shows a slow variation over the course of weeks, consistent with the change in the infrared excess measured from the spectra. The infrared flux is a mix of stellar and disk emission and our excess spectra taken during the warm mission monitoring suggest that most of the change in the infrared flux comes about from a change in the disk emission, not from variability in the stellar emission. Some of the optical monitoring of Baliber et al. overlaps with our infrared monitoring, and the stellar flux does not follow the same trend as the infrared data, supporting our interpretation that the dust is varying while the star is constant. 

\begin{figure}
\includegraphics{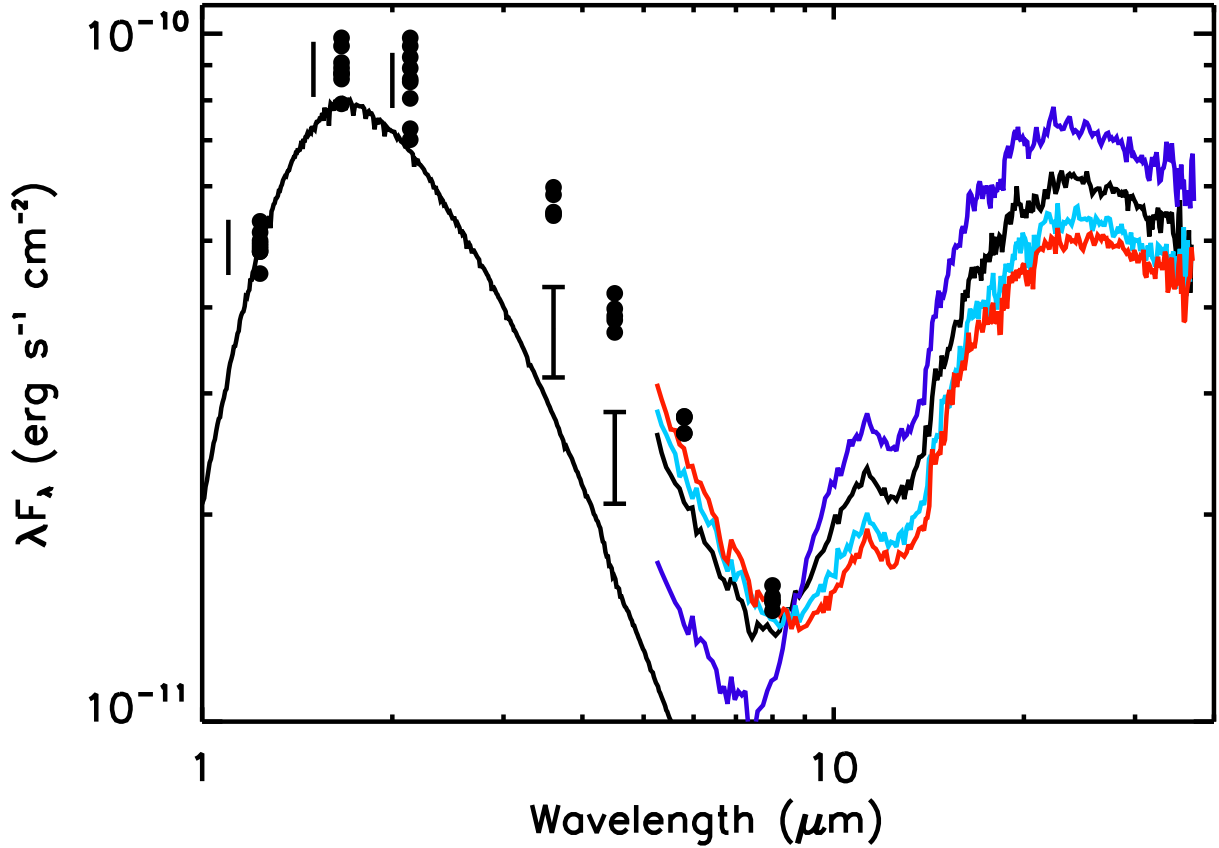}
\caption{Spectral energy distribution of LRLL 31 showing our early IRS data (rereduced), the IRAC-cryo data, the JHK photometry and the range of magnitudes seen in the IRAC warm mission data (error bars at 3.6,4.5\micron). A kurucz model photosphere, reddened by A$_V$=8.8 and scaled to the average J band  magnitude has been included for comparison. Vertical lines near the near-infrared photometry demonstrate the uncertainties in this data. Uncertainties in the mid-infrared photometry are similar to the size of the data points\label{sed}}
\end{figure}

In our mid-infrared observations the flux varies continuously (Figure~\ref{sed}). This suggests that the physical cause is not a single event, but a continuously occurring phenomenon. Unlike a FU Ori or EXor outburst, the variability in LRLL 31 does not occur once and slowly decay. Our IRS spectra separated by six months show that the wavelength dependence of the variability does not change. This suggests that the same mechanism dominates the variability over many years. It also supports our assumption that the variability we study in any one year has the same characteristics as in any other year. 

The extinction and polarization (Tables~\ref{extinction} and ~\ref{polarimetry}) can provide clues to the structure of the disk, assuming that they come from the disk. If the grains in the molecular cloud are aligned, they can create polarized flux from the star. Starlight that is parallel to the long axis of the grains will be more highly extincted than starlight that is parallel to the short axis of the grains. This differential extinction will lead to polarized light leaving the cloud, with the polarization being proportion to the extinction. Two other nearby cluster members were observed by SPOL and have previous polarimetry measurements. LRLL 169 (11" from LRLL 31) has a P=4.2\% and q=6.9$^{\circ}$ while LRLL 4 (3.5' from LRLL 31) has P=2.8\% and q=166.5$^{\circ}$, and the extinction is A$_V$=2.8 and 1.8 respectively \citep{luh03}. If the polarization is from the molecular cloud then the ratio of polarization to extinction should slightly decrease with extinction \citep{ger95}. For LRLL 169,4 and 31 the ratio is 1.5, 1.6 and 0.9 \%/mag respectively, which is consistent with the expected decrease.  The molecular cloud density needs to be highly variable in order to explain the observations since LRLL 169 is only 3500 AU away from LRLL 31 and shows a much smaller extinction and a different polarization angle. The high polarization/extinction could also be coming from the outer disk, if the disk is close to edge on and the system is moderately flared. Differential extinction can work with material in the outer disk in much the same way as it would operate in the molecular cloud. An edge-on disk is consistent with the inclination used by \citet{fla10} to fit the LRLL 31 SED with models of a warped disk. We can exclude the inner disk as the source of the extinction and polarization because of the lack of strong variability in these quantities, whereas we have observed continuous and large fluctuations of the inner disk.

\subsection{Gas Properties}
\begin{figure}
\includegraphics[scale=.5]{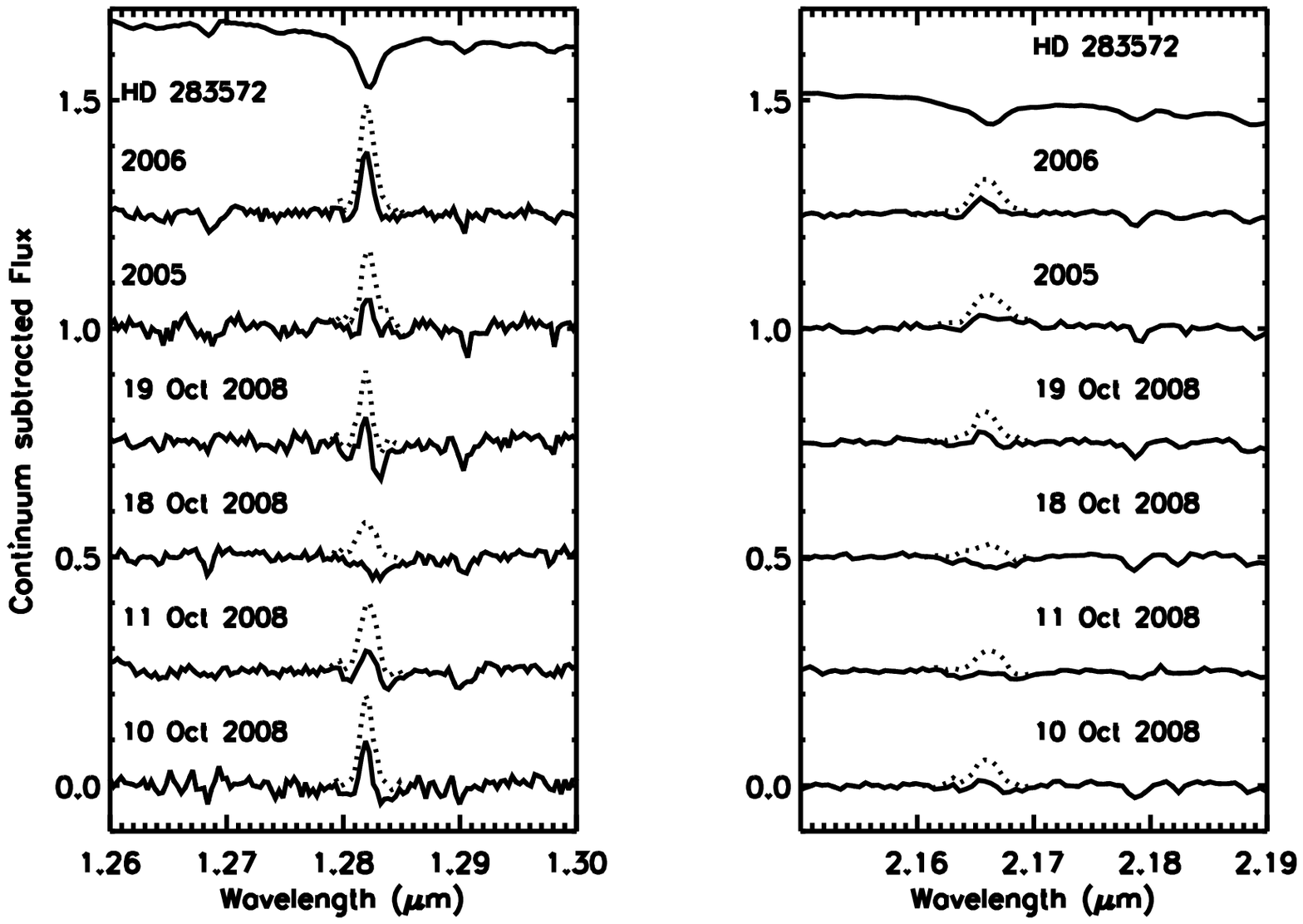}
\includegraphics[scale=.5]{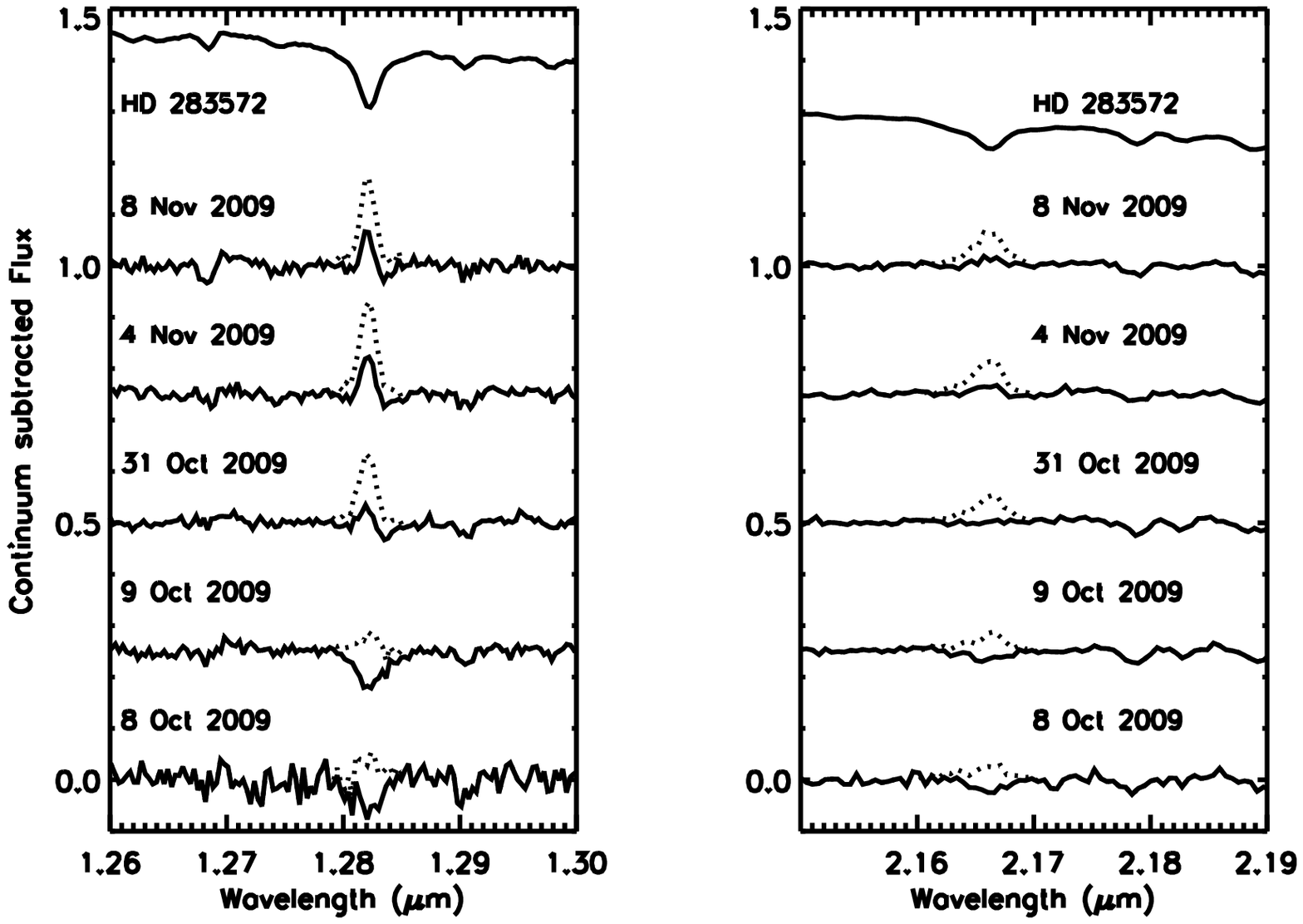}
\caption{Line measurements from 2005,2006, 2008 (left two panels) and 2009 (right two panels). Both the Pa$\beta$ line ($\lambda=1.282\micron$) and Br$\gamma$ ($\lambda=2.16\micron$) are shown for each day. The solid line is the observed spectrum, while the dashed line has had the photospheric absorption subtracted from the line. At the top of each panel is the G6 WTTS standard. There are large variations in the emission lines, mainly on timescales of weeks. \label{lines}}
\end{figure}

Next we consider the activity of the gas, which we trace through a combination of emission lines from the infrared and optical. Our observations of the Pa$\beta$ and Br$\gamma$ lines in the infrared let us measure the accretion rate onto the star directly \citep{muz98}. To derive the line fluxes, we first subtract the photospheric absorption as measured by the G6 WTTS standard. In this case we find that using a WTTS, rather than a dwarf or giant spectrum, does better at subtracting the photospheric absorption especially on the wings of the line. We normalize the absorption-subtracted spectrum to the continuum around the line. These spectra for each day are shown in Figure~\ref{lines}. The equivalent widths of the lines, as measured from the continuum normalized spectra (Figure~\ref{lines}), are listed in Table~\ref{nir_lines}. Veiling will change the strength of a line, especially for Br$\gamma$ where the veiling is significant, and in Table~\ref{nir_lines} we include the veiling corrected equivalent width (EW), which is derived by artificially veiling HD 283572 and subtracting this new standard from the observed spectrum. Using our nearly simultaneous measurements of J and K band photometry to estimate the continuum level, we derive the line flux. Since we use the observed K band photometry, which is a mix of stellar emission and veiling flux, we had to account for the veiling when measuring the EW, resulting in the corrected EW being lower than the uncorrected EW. The smaller EW, combined with the stellar plus veiling flux of the K band continuum produces an accurate estimate of the line flux. While others have found significant veiling near 1\micron\ in some stars \citep{fis08}, we measure little veiling in the J band, and assume it is negligible when deriving the flux of the Pa$\beta$ line. Assuming a distance of 320pc to the star, we can then derive line luminosities. These line luminosities are directly correlated with the accretion luminosity \citep{muz98} and we can derive the accretion rate using the formula:

\begin{equation}
L_{acc}=\frac{3G\dot{M}M_{*}}{5R_{*}}
\end{equation}

The measured line luminosities and accretion rates are listed in Table~\ref{accretion}. We cannot derive accurate accretion rates from the 2005 and 2006 spectra because we do not have simultaneous photometry, but the strong emission lines suggest that they were on the high end of the range of measured accretion rates. There is a large discrepancy between the accretion rate derived by Pa$\beta$ and Br$\gamma$ that is unexpected given that both measure the accretion rate onto the star. The uncertainties listed in Table~\ref{accretion} only include the uncertainty in the photometry and the equivalent width measurements and do not include the uncertainty in the conversion between line flux and accretion luminosity, or the uncertainties in the mass and radius of LRLL 31. The conversion uncertainty is closer to a factor of 2, much larger than the typical 10\% uncertainty, and can account for the discrepancy between the accretion rate derived by Pa$\beta$ versus Br$\gamma$. Unresolved absorption in the line may also explain some of the observed differences. Further conclusions drawn about the accretion rate are based on the Pa$\beta$ because it is the stronger and is less susceptible to veiling, which makes it a more reliable tracer of the accretion rate.

There are large variations in the Pa$\beta$ and Br$\gamma$ emission throughout our observations. The EW ranges from almost zero up to 4.5\AA. The accretion rates vary by a factor of 5. The infrared emission lines demonstrate that the accretion is changing constantly with the largest changes occurring on weekly, rather than daily, timescales. Accretion variability is common among T Tauri stars, and the daily level of variability seen in LRLL 31 is similar to that seen in many other stars \citep{har91,gul96}. The change seen in 2009, when the Pa$\beta$ line goes from absorption to emission is less common. 

Our measurements of the H$\alpha$ equivalent width (Table~\ref{haew}) from the optical spectra show emission at every epoch, implying that accretion is continuous in LRLL 31. The equivalent width is continuously changing with a range of a factor of 3.4, suggesting that the accretion rate is always varying as was found with the infrared hydrogen recombination lines. Similar to the dust emission, the variations in the accretion rate onto the star are not a single isolated event, and are always occurring with varying amplitude. Our high-resolution H$\alpha$ spectra are shown in Figure~\ref{halpha}. They show broad profiles, consistent with the presence of accretion. It appears as though the red side of the feature experiences the largest variations, while the blue side stays relatively constant. This may be due to partial obscuration of the accretion flow by the disk, or a change in the magnetospheric geometry.

\begin{figure}
\center
\includegraphics[scale=.5]{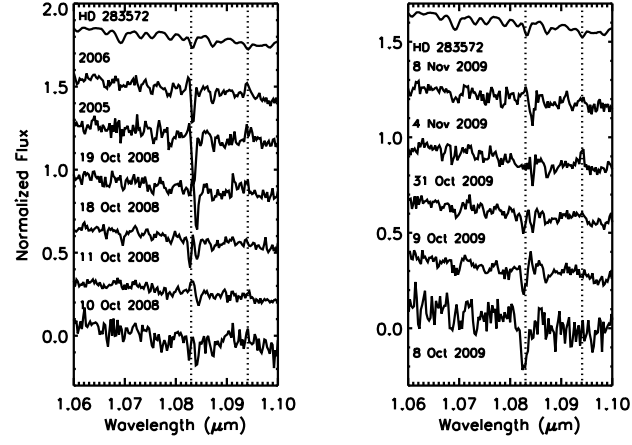}
\caption{ HeI ($\lambda=1.083\micron$) and Pa$\beta$ ($\lambda=1.0941\micron$) lines in LRLL 31. The HeI line shows large variations in both the blue and the red side in as little as one day. \label{hei} }
\end{figure}

\begin{figure}
\includegraphics[scale=.5]{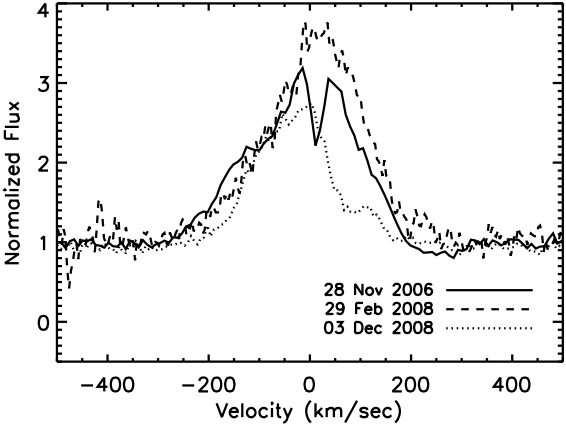}
\caption{High resolution H$\alpha$ measurements, normalized to the continuum. The line shows continuous variations in its shape and strength. \label{halpha}}
\end{figure}

The outflow of material, as traced by the blue-shifted emission/absorption in the HeI line (Figure~\ref{hei}), changes on daily to weekly timescales. In 2008 it shows large variations from night to night, while in 2009 the line slowly goes from showing very strong blue-shifted absorption to no blue-shifted features. Blue-shifted absorption in the HeI line is believed to come from a combination of stellar wind and disk winds \citep{fis08} and variability in the lines is sometimes seen \citep{edw06}. The red side of the HeI line is connected with the accretion flow, and the strength of the red-shifted absorption is broadly correlated with the strength of the Pa$\beta$ and Br$\gamma$ emission lines. When the Pa$\beta$ line is the strongest in 2005 and 2006, the red-shifted absorption is the deepest, and in 2009 the absorption increases with the accretion rate. This red-shifted absorption may result if we are looking down the accretion column, so the gas that is flowing onto the star absorbs the emission from the hot column of shocked gas striking the surface of the star. The variations in the relative strengths of the two sides of the line suggest that the two do not change contemporaneously. There are days with just red-shifted absorption, days with just blue-shifted absorption and days with both.

There is also evidence for correlated changes in the accretion rate and the infrared excess. In 2009 we were able to obtain near-infrared spectra during our Spitzer warm mission monitoring (vertical tick marks in Figure~\ref{wm_lc}). We observed two spectra during the minimum in the light curve and three spectra after the infrared excess had substantially increased. We find that while the infrared excess is at its minimum, the Pa$\beta$ line (8,9 Oct) is weaker than in any other epoch. As the infrared excess increases over the next few weeks the Pa$\beta$ line emission also increases. We have one observation of the H$\alpha$ line taken on 12 Oct 2009, close to the infrared minimum, and the equivalent width is lower than seen in other epochs. This suggests that the dust at the inner edge of the disk and the gas flowing onto the star are closely connected by the physical mechanism that causes this variability, which has been previously considered as a source of infrared variability \citep{car01}. Also, the blue-shifted absorption in the HeI line is strongest at the infrared minimum and slowly disappears as the excess gets stronger, suggesting that the dust and outflowing gas are also connected.

\section{Theoretical Implications}
Our observations of LRLL 31 target the inner disk, as shown by the variability timescales of a few days to a week and the temperature of the dust close to the sublimation temperature. Studies of circumstellar disks have gained a reasonably complete and consistent picture of the behavior of the accretion/protoplanetary disks at radii greater than 1 AU, but the zone within this radius presents many puzzles. Not only is it beyond our ability to resolve well observationally, but it is the venue for complex processes involving intense magnetic fields, gas flow, accretion luminosity, and extreme heating and sublimation of dust. The variability observed here can help to provide constraints on some of the physics occurring in this part of the disk. In this section, we consider possible physical causes for this behavior such as variable accretion, perturbations by a companion, winds, and the influence of magnetic fields. Schematic diagrams of the various models considered are shown in Figure~\ref{schematic}.

\subsection{Variable Accretion}
In a disk in hydrostatic equilibrium, the scale height of the dust is set partly by the luminosity illuminating its surface \citep{dul01}, which is a combination of stellar luminosity and accretion luminosity. A substantial change in either would lead to a variation in the scale height of the disk (Figure ~\ref{schematic}a). Our observations rule out any large changes in the star, but do indicate large variations in the accretion. Since the accretion luminosity is much smaller than the stellar luminosity, a much larger change in the accretion luminosity than the observed variation would be needed to produce the observed fluctuations in scale height. We can then rule out a variable illumination of the inner disk as the source of the variability.

The viscous nature of the disk also means that the accretion rate through the disk is directly tied to its surface density \citep{dal98}. If the surface density of the disk were to suddenly increase, possibly due to a large inflow of material from the outer disk, then the height of the $\tau=1$ surface would increase (Figure ~\ref{schematic}b).  Outgoing stellar photons will be absorbed by the disk below the height of the $\tau=1$ surface of the disk. A change in the height of this surface would change the amount of stellar light that is absorbed by the disk. This would lead to a change in the covering fraction of the inner disk and the shadowing of the outer disk, consistent with the behavior of this system. The timescale for a large mass of dust to travel through the inner disk is the viscous timescale, which is roughly 2000 years at 0.2 AU. This is much longer than the dust continuum variability timescales we observe. Also, if the magnetosphere radius where the stellar accretion flow is launched is significantly less than the dust sublimation radius (see below), then the long viscous timescale is also inconsistent with the observed correlation between dust and stellar accretion rate variations.

These arguments suggest that the variations are not caused directly by changes in the accretion rate. It is possible that the relationship between the infrared excess and accretion is not a causal one. The change in accretion rate may not directly lead to a variation in the covering fraction of the inner disk, but instead they may both be effects of another physical process. 

\subsection{Winds}
A highly variable wind could lead to a loss of material in the inner disk. This wind could carry both gas and dust out from the inner disk \citep{vin07,kon00} and this rapid removal of material would lead to a drop in both the infrared excess and the accretion rate, if gas were removed just before it began its free fall along the magnetic field lines (Figure ~\ref{schematic}c). Disk wind models usually start the wind with a velocity on the order of the sound speed, and then it rapidly accelerates to larger than the Keplerian velocity \citep{bla82}, which would imply a slower decrease in the disk emission than is observed. The strong blue-shifted absorption on the HeI line during the minimum of the infrared excess, and the rapid fluctuations in this component are consistent with a strong wind. If we assume that the inner disk is just barely optically thick before the wind starts to remove material with a gas to dust mass ratio of 100, and the wind removes the entire inner disk in only 5 days (the length of the drop in the infrared light curve) then the outflow rate is $2 \times10^{-5}M_{\odot}/yr$. The accretion rate required to refill the disk over the next 30 days is $3\times10^{-6}M_{\odot}/yr$. This accretion rate is a lower limit since the disk is not just barely optically thick but it is still much larger than is observed for LRLL 31. Also, filling the inner disk by accretion from the outer disk should occur on a viscous timescale, which is much longer than the timescale for the increase in the infrared flux.

The wind could be asymmetric about the plane of the disk, leading to a variation in the disk structure (Figure ~\ref{schematic}d). This can arise when a stellar magnetic field that is not symmetric about the equatorial plane launches the wind \citep{lov10}. If the mass flow in the wind is larger on one side of the disk than the other, then the asymmetric loss of angular momentum would lead to warping of the disk, which would then shadow the outer disk, leading to infrared variability \citep{fla10}. If the fluctuations in the wind were rapid enough, as is suggested by the fast changes in the blue-shifted side of the HeI line, then they could cause the observed variability. This model suggests that a large wind would be associated with more warping in the disk, and a larger infrared flux, which is inconsistent with our observations. As with the accretion flow, it is difficult to get a wind to explain the infrared variations directly, and the fluctuations seen in the blue-shifted side of the HeI line may be a result of the same physical process that changes the infrared excess and the accretion rate rather than the source of this behavior.

\subsection{Perturbation by a Companion}
A companion on an orbit that is misaligned with the disk (Figure ~\ref{schematic}e) can lead to a warped disk \citep{lar97}. The height of the warp will periodically vary as the companion passes out of the midplane of the disk and drags dust with it \citep{fra09}. If a companion were sitting inside the inner disk then it could create large perturbations while also periodically disturbing the accretion rate, assuming it did not get close enough to the inner disk to completely remove all of the dust \citep{art96}. It is also possible for a companion to be sitting outside of the inner disk, again assuming it isn't close enough to remove the inner disk. The deficit of flux around 10\micron\ relative to a typical T Tauri star indicates that the disk is missing material within a few AU of the star, i.e. it is a transition disk. Our near-infrared observations indicate that there is still an optically thick disk at the dust destruction radius, but beyond this point there may be a gap in the disk. If so, gas and dust would still flow through this gap as evidenced by the continuous accretion of gas onto the star. If the gap were caused by a companion clearing out material near its orbit, then the accretion rate through the gap would be expected to be periodic, and material would cross the gap at a speed much higher than the typical velocity through a full disk \citep{art96}. If the inner disk is narrow enough, then the shock heating as this material strikes the inner disk at its outer edge could lead to heating of the inner edge of the disk, which would change its scale height.

Our data allow us to provide some constraints on the position and mass of a companion. A companion larger than 1M$_{\odot}$ would be noticeable in our near-infrared spectra as anomalously strong photospheric lines in the K band. An equal mass binary would not produce anomalous line strengths, but would make the star appear over-luminous, which is not seen, although there is a large uncertainty in the luminosity. A smaller companion near the dust destruction radius can still lead to a large change in the structure of the dust, and would be consistent with the observed changes in the radial velocity. Figure~\ref{periodigram} shows our constraints on the mass and location of a companion in the disk. The upper and lower limits on the mass come from the near-infrared spectra and the RV data. The limits on the location of a companion come from the requirement that the companion cannot remove material from the dust destruction radius (taken to be 0.15 AU) or from the outer disk ($\sim$ 7 AU) \citep{art96}. The exact location of the inner disk, along with the limits on the location of a companion near this dust, depends on the dust composition, the luminosity of the central source and the assumption that the central source is a single star rather than a binary \citep{nag10}. Moving the inner disk in or out will change the absolute position of the boundary, but not the relative positions, which are at r/a$\sim0.3,1.7$ for the outer and inner limits respectively, where a is the binary separation and r is the location of the dust. The radial velocity data suggests that this system is a binary, and the presence of dust at certain locations within the disk limits the location of this companion.

Even if this system is a binary, the companion may not be responsible for the infrared variability. A companion in a circular orbit in the plane of the disk will not cause substantial mid-infrared variability \citep{nag10}. Making the companion misaligned with the disk could lead to periodic perturbations \citep{fra09}. Figure~\ref{periodigram} shows the periodigram based on the monitoring at 3.6 and 4.5\micron. We are most sensitive to periods in the range 2-25 days and no significant peak is seen indicating that there is no periodicity over these timescales. If a companion were located 0.04-0.2 AU from the central star and perturbed the disk on every orbit then we would see evidence of this in the IRAC data. This excluded region is marked in left-hand side of Figure~\ref{periodigram}. This data suggests that it is unlikely that there is a companion on an inclined orbit within the inner disk, but does allow an object further out in the disk. Based on the limits discussed above, if the inner disk is at 0.15 AU then the companion is restricted to outside of 0.4 AU where the orbital period is $\sim100$ days. The rapid decline in the flux at the beginning of our infrared monitoring is difficult to explain in the context of the slow variations expected from a companion this far from the star. It is possible that the inner disk is closer than 0.15 AU, pushing in the outer limit on the position of a companion inward resulting in a timescale more comparable to what is observed. The companion could also be on an eccentric orbit, which would allow it to rapidly disturb the disk during its high velocity periastron motion, while still having a long period.

\subsection{Magnetic Fields}
\begin{figure}
\includegraphics[scale=.5]{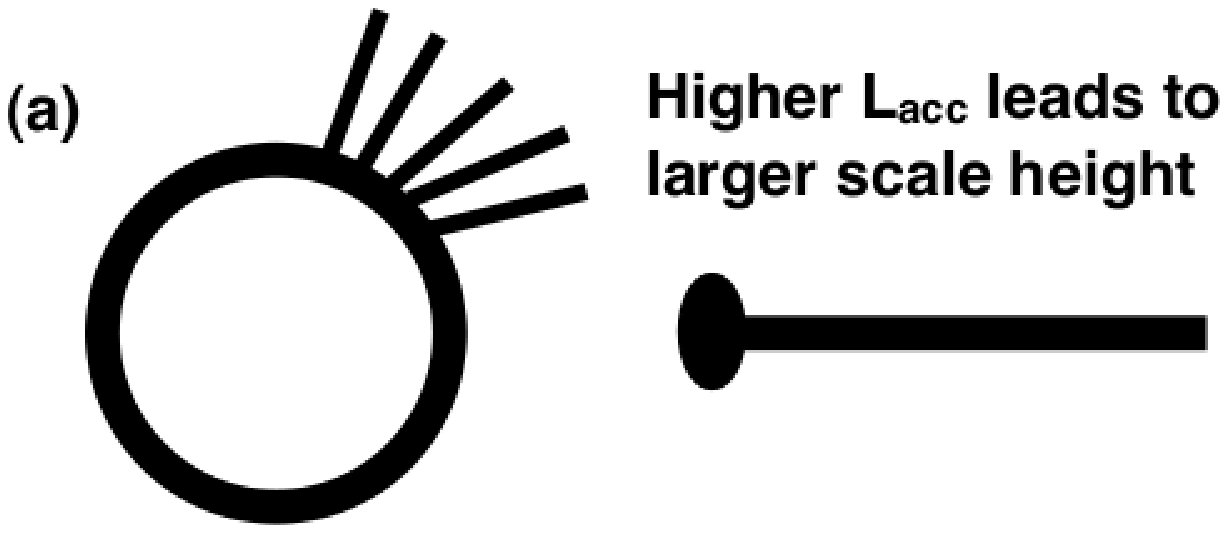}
\includegraphics[scale=.55]{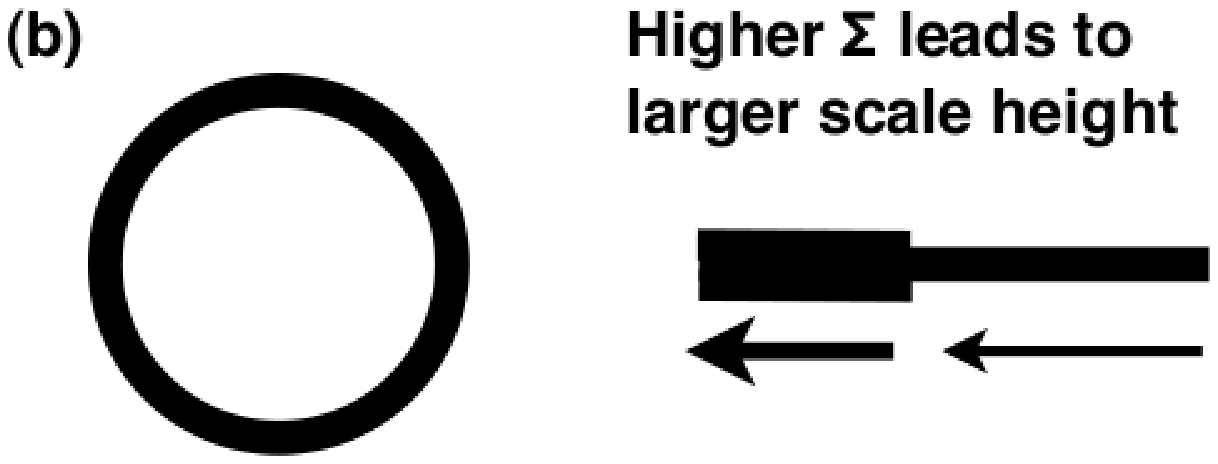}
\includegraphics[scale=.5]{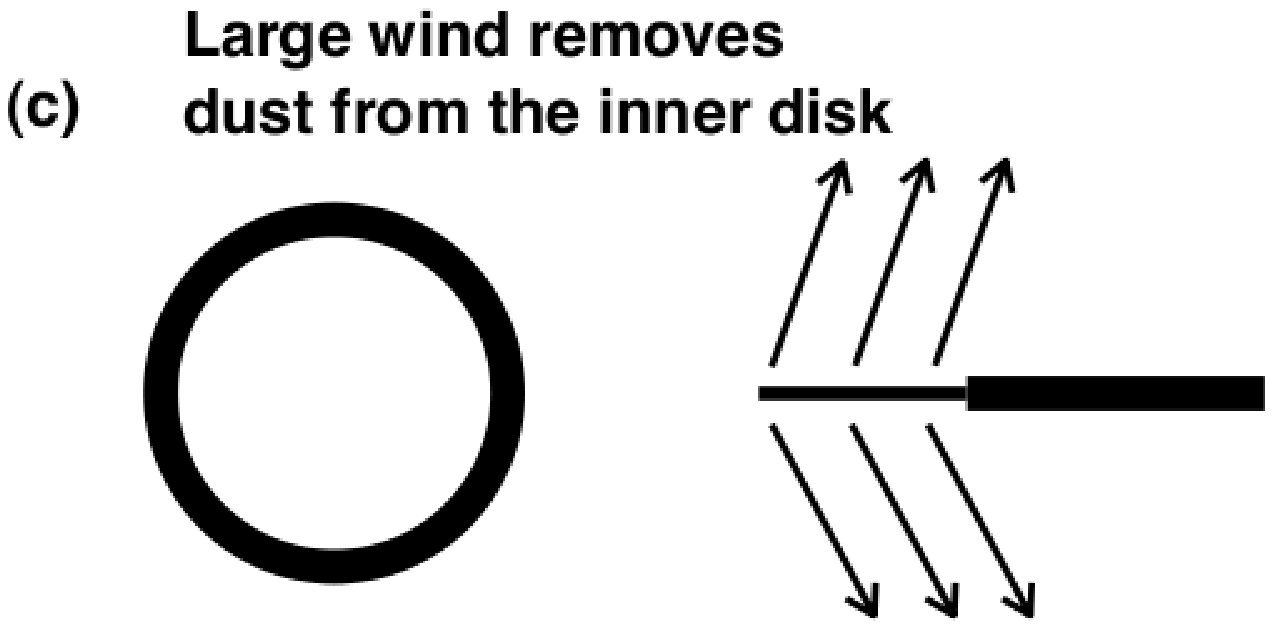}
\includegraphics[scale=.5]{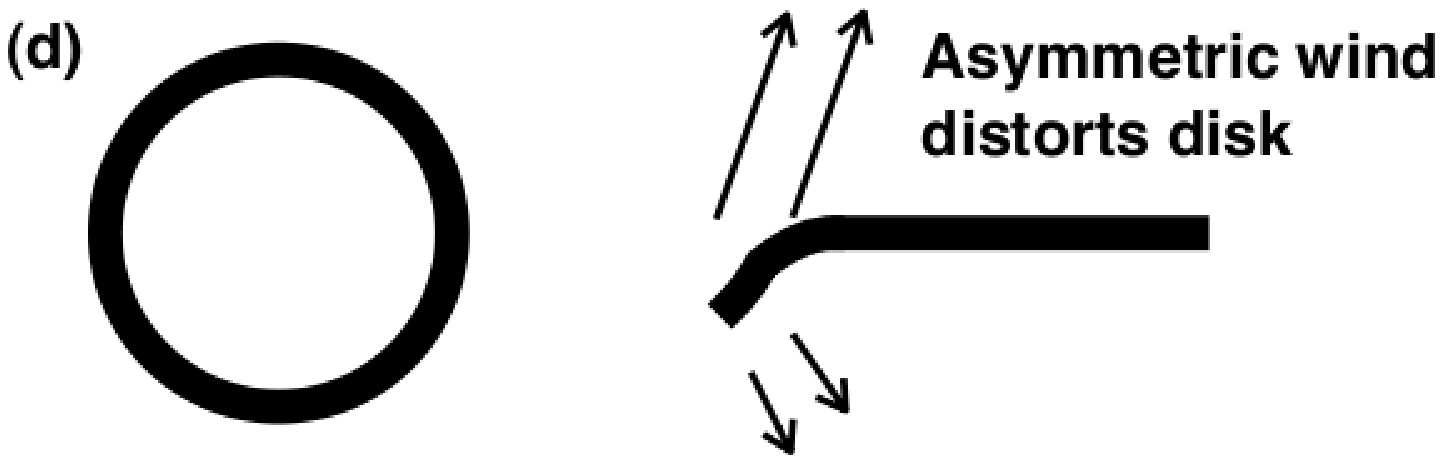}
\includegraphics[scale=.5]{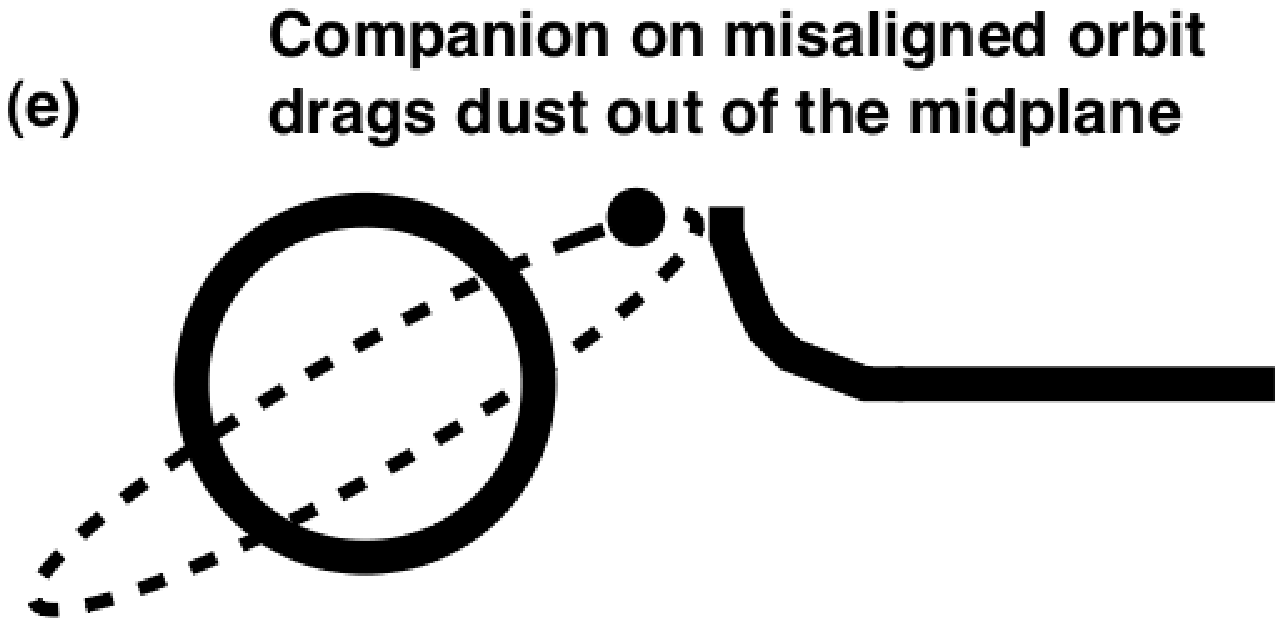}
\includegraphics[scale=.55]{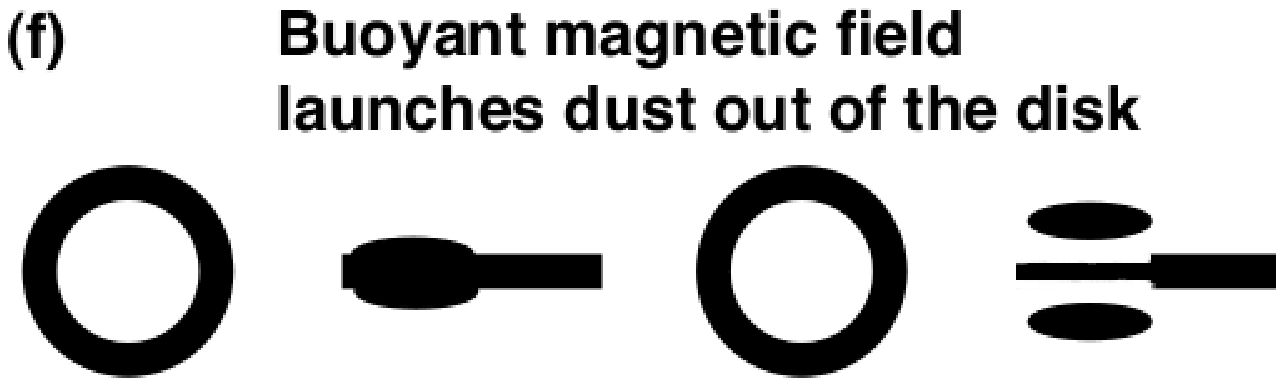}
\includegraphics[scale=.5]{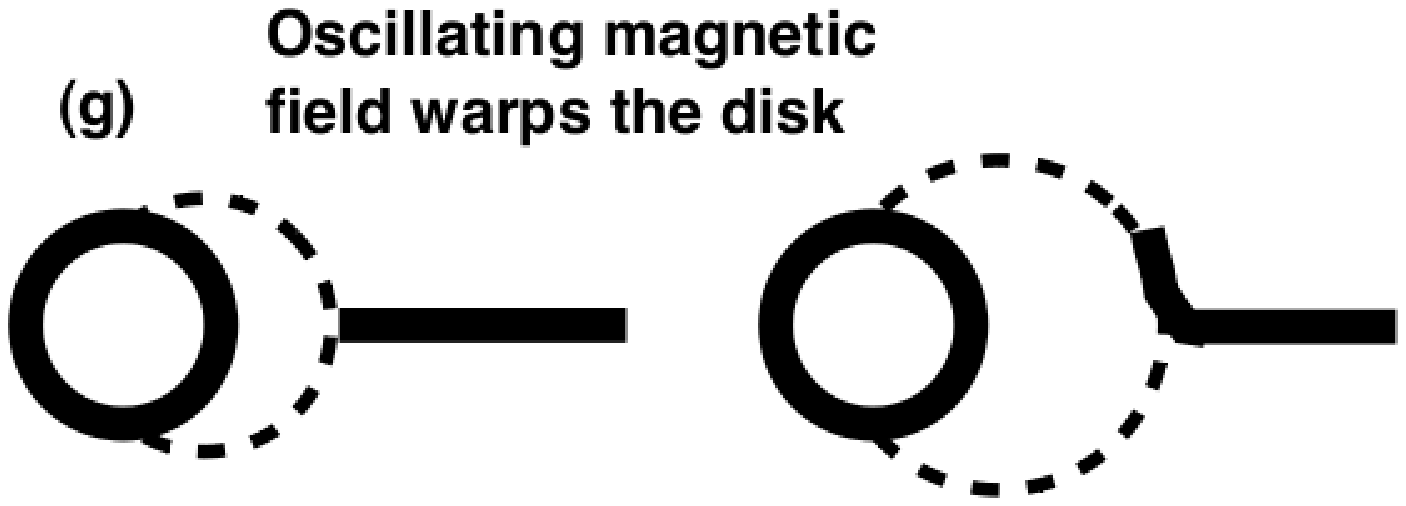}
\includegraphics[scale=.52]{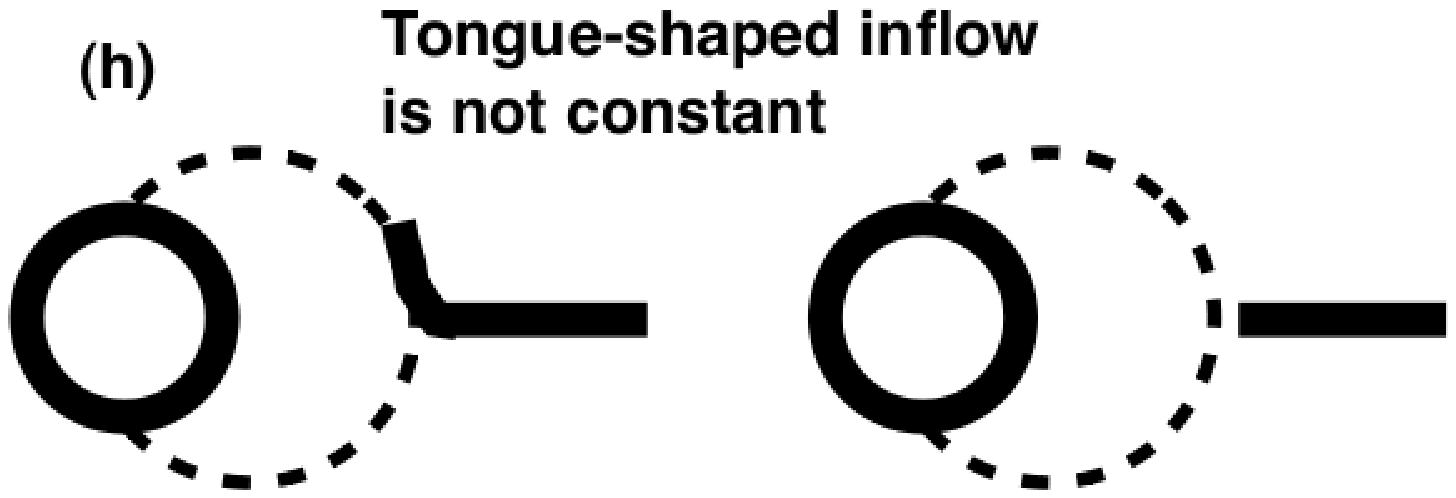}
\includegraphics[scale=.5]{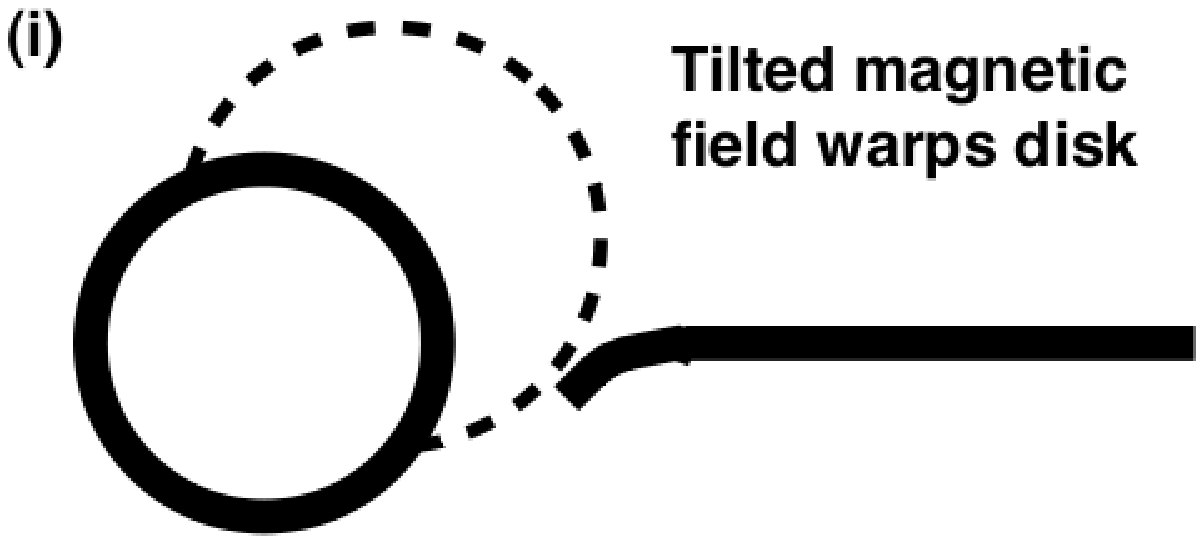}
\caption{Schematic diagrams of the various models considered for the variability of LRLL 31. \label{schematic}}
\end{figure}

\begin{figure}
\includegraphics[scale=.5]{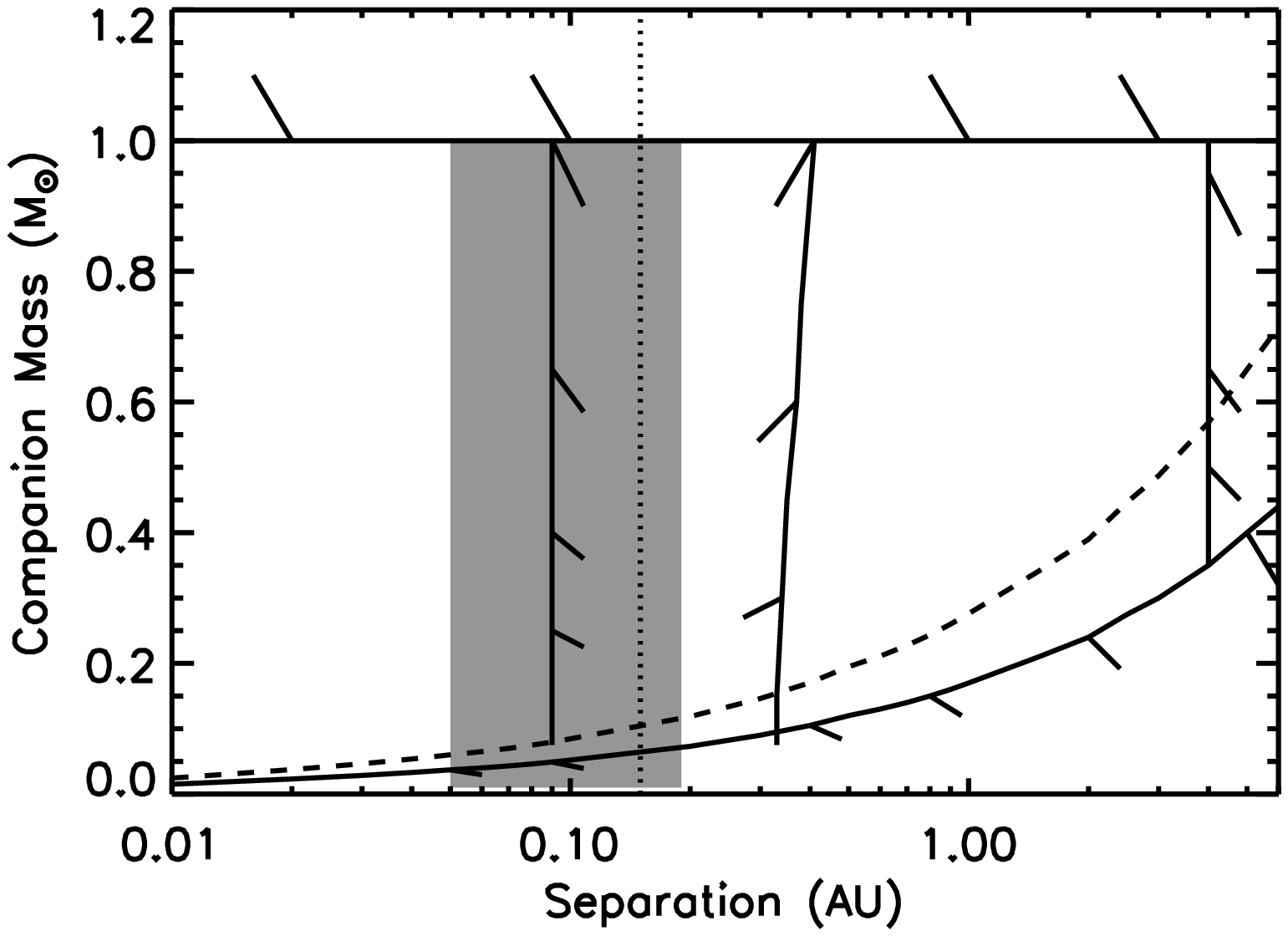}
\includegraphics[scale=.5]{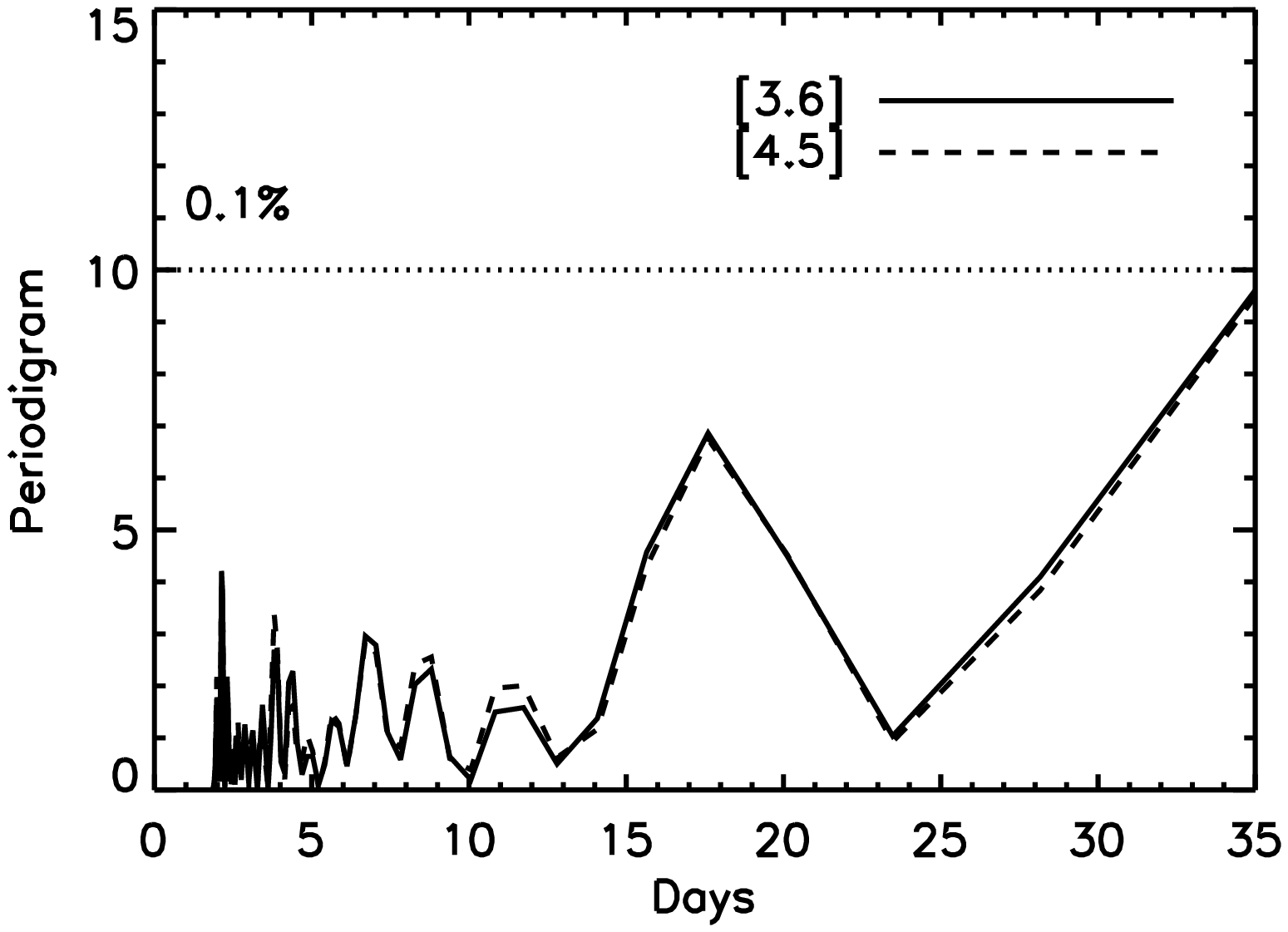}
\caption{(Left) Limits on the mass of a companion. Hashed areas show regions of mass and position of a companion that can be ruled out based on our data. The upper limit is from the near-infrared spectra and the lower limit is from our RV data (dashed line assumes i=38$^{\circ}$). The limits on position come from the dynamical clearing of the disk by a companion. Disk material exists at the dust sublimation radius ($\sim$0.15 AU, dotted line) and in the outer disk ($\sim$7 AU), and we can rule out a companion close to these locations. The lack of periodicity in the 3.6,4.5\micron\ photometry rules out a companion within the grey area perturbing material on every orbit. (Right) Lomb-Scargle Normalized Periodigram based on the 3.6 and 4.5\micron\ photometry. The dashed lined indicates a false alarm probability of 0.1\%. There are no peaks above this line, indicating that there is no significant periodic signal in the data. \label{periodigram}}
\end{figure}

A magnetic field threading a turbulent disk could lead to variable structure. Using magneto-hydrodynamic simulations \citet{tur10} find rapid variations in the disk due to the turbulence caused by the magneto-rotational instability (Figure ~\ref{schematic}f). They see sharp drops in the scale height of the disk by 30\% in an orbital timescale as dust is lifted out of the disk by a buoyant magnetic field. There is evidence for a small change in the silicate feature \citep{muz09}, which could be explained by the limiting of dust out of the midplane \citep{tur10}. The speed of the changes in this model is consistent with our observations, but the small amount of excess in the minimum of the 3.6,4.5\micron\ light curve would require that the entire inner disk have a scale height close to zero. Such global behavior is not seen in these models and the changes in the scale height are smaller than have been observed.\footnote{If the dips in the extinction that are sometimes seen in LRLL 31 are proven to be real, the large error bars on A$_V$ mean they are still marginally consistent with each other, and the extinction arises from a nearly edge-on disk then the model of \citet{tur10} would be a promising explanation for them.}

The interaction between the disk and stellar magnetosphere can be highly dynamic \citep{bou07b}. According to some models, where the magnetic field truncates the disk oscillates (Figure ~\ref{schematic}g) on timescales of $\sim25$days \citep{goo99}, which is similar to the timescale seen in the warm-mission photometry.  The oscillations can be very large (a factor of 3 change in the size) due to the expansion, opening and reconnection of the magnetosphere, although we do not see a large change in the location of the dusty disk. \citet{rom09} find large variations in both the accretion rate and the outflow rate associated with these magnetic field oscillations. The timescale between bursts is $\sim60$ days for rapidly rotating stars, which is consistent with our infrared light curve. \citet{kul08} present a set of three-dimensional simulations of magnetospheric accretion and they show that strong instabilities develop in the inward flow of material, breaking it up into a small number of tongue-shaped features extending from the disk, along the magnetic field lines, to the stellar surface, where they create hot spots (Figure~\ref{schematic}h). Variations in the number and geometry of the tongues occur on the dynamical timescale of the inner disk. If the tongues originate at or just beyond the sublimation radius and dust is carried along with the gas flow, they might provide a framework for explaining our observations. They would carry the dust out of the disk plane, potentially shadow the outer disk from the star when this occurs (as needed to explain the longer wavelength variations reported in \citet{fla10}), and would appear as a large increase in disk scale height, all occurring on a timescale compatible with that of the observed variations. 

Another possibility related to the interface of the stellar magnetic field and the disk is if the magnetic field is misaligned with respect to the plane of the disk (Figure ~\ref{schematic}i). As material flows onto the field lines it is pulled out of the midplane, preferentially on one hemisphere, creating a warp in the inner disk.  \citet{lai08} develop a model in which the magnetic field coupling between the star and disk can excite waves in the disk when the stellar rotation axis, stellar magnetic axis and disk axis are misaligned. This mechanism could set in place a disk geometry similar to the warps proposed by \citet{fla10} to explain the variability, assuming either the density of the inflowing material or the structure of the magnetic field rapidly varies in order to change the height of the warp.

These types of interaction require that the dust extend close to where the disk is truncated by the stellar magnetosphere. We can estimate where material stops flowing in through the disk and is loaded onto the magnetic field lines using the corotation radius, 0.05 AU based on the optical photometric period. Material needs to be within the corotation radius before it is loaded onto the magnetic field lines or else it will be flung out in a wind rather than accreted onto the star \citep{bou07b}, although the exact location where the magnetic field truncates the disk depends on the strength of the magnetic field and the accretion rate. Our observations of the infrared excess find that the dust extends to the sublimation point, but translating this information into a position requires detailed knowledge of the density and composition of the dust \citep{ise05,kam09}. Based on different assumptions about the dust properties and the maximum temperature of the dust results in a range of dust sublimation radii from 0.05 - 0.3 AU given the measured luminosity of LRLL31 \citep[see the review by][for details on various calculations of the location of the dust inner rim]{dul10}. The smallest radii correspond to large grains ($\sim$ a few \micron) at 1800 K, which because of their relative high efficiency in radiating away thermal emission compared to small grains (0.1\micron) can survive closer to the star before sublimating, while the largest radii correspond to small grains at 1500 K. We have no direct measurements of the grain properties of the inner disk, but the strong silicate emission feature at 10\micron\ is consistent with the presence of small amorphous grains rather than the large grains needed to get the disk down to 0.05 AU. Our measurements of the covering fraction for the inner disk find that it is consistently smaller than expected for a typical T Tauri star, which could be due to significant increase in the grain size for the inner disk versus the outer disk. Substantial grain growth in the inner disk would move the dust edge inward, making it more susceptible to oscillations of the stellar magnetosphere. We would expect to see variability from this process more often around systems with a slow rotation period, hence a much larger corotation radius and it would not preferentially occur around transition disks for which the type of infrared variability observed here is common \citep{esp11}. Without detailed information on the exact termination of the dusty disk, it is difficult to determine if oscillations in the stellar magnetosphere cause the observed variability.

\section{Conclusion}
We present results from a large synoptic study of the transition disk variable LRLL 31. The star itself stays relatively constant showing variability consistent with rotation of cool spots across its surface. We find large variations in the infrared excess and the accretion rate on timescales of weeks. The dust appears to stay at the dust destruction radius, while its scale height rapidly fluctuates by a factor of eight and the infrared flux changes by 0.3 mag. The accretion rate varies by a factor of 5, and appears to be correlated with the strength of the infrared excess. However, the change in accretion rate is unlikely to be the direct cause for the change in infrared excess. It is also unlikely that the observed variations are due to the influence of a companion within 0.4 AU, based on the lack of periodicity in our Spitzer warm mission monitoring and the presence of an optically thick inner disk. The source of the variability may be related to a companion outside of, but still close to, 0.4 AU or to the dynamic interface between the disk and the stellar magnetic field, assuming the dust extends close enough to the star. Our observations are able to limit the list of plausible physical models, although they cannot exactly define what is happening to the disk. Further constraints on the radial velocity variations, and the exact position of the inner disk will help to select between these models.

\acknowledgements
We would like to thank Paul Smith for obtaining the spectropolarimetric data and Joan Najita for useful discussion regarding the physical mechanism behind the variability. Our results are partly based on observations obtained at the Centro Astrono\'omico Hispano Alem\'an (CAHA) at Calar Alto, operated jointly by the Max-Planck Institut f\"ur Astronomie and the Instituto de Astrof\'{\i}sica de Andaluc\'{\i}a (CSIC). The observations were supported by OPTICON. OPTICON has received research funding from European Community's Sixth Framework Programme under contract number RII3-CT-001566. This work is based in part on observations made with the Spitzer Space Telescope, which is operated by the Jet Propulsion Laboratory, California Institute of Technology under a contract with NASA. This work was partially supported by contracts 1255094 and 960785 from JPL/Caltech to the University of Arizona.

\clearpage
\begin{deluxetable}{ccccc}
\tablewidth{0pt}
\tablecaption{Observing Log\label{obs_log}}
\tablehead{\colhead{Date}&\colhead{MJD}&\colhead{Wavelength}&\colhead{Resolution}&\colhead{Note}}
\startdata
\cutinhead{Spex}
29 Dec 2005 & 53734.2 & 0.8-5\micron\ & $\sim$1500 & \\
09 Nov 2006 & 54049.3 & 0.8-5\micron\ & $\sim$1500 & \\
10 Oct 2008 & 54750.5 & 0.8-5\micron\ & $\sim$1500 & \\
11 Oct 2008 & 54751.5 & 0.8-5\micron\ & $\sim$800 & \\
18 Oct 2008 & 54757.4 & 0.8-5\micron\ & $\sim$1500 & \\
19 Oct 2009 & 54758.4 & 0.8-5\micron\ & $\sim$1500 & \\
08 Oct 2009 & 55112.6 & 0.8-2.5\micron\ & $\sim$800 & \\
09 Oct 2009 & 55113.6 & 0.8-2.5\micron\ & $\sim$800 & \\
31 Oct 2009 & 55137.3 & 0.8-5\micron\ & $\sim$800 & \\
04 Nov 2009 & 55141.5 & 0.8-5\micron\ & $\sim$800 & \\
08 Nov 2009 & 55144.5 & 0.8-5\micron\ & $\sim$800 & \\
\cutinhead{Optical Spectra}
28 Nov 2006 & 54067 & 4700-9000\AA & $\sim$33,000 & Keck HIRES\\
29 Feb 2008 & 54526 & 6450-6650\AA & $\sim$30,000 & MMT Hectochelle\\
03 Dec 2008 & 54803 & 4700-9000\AA & $\sim$33,000 & Keck HIRES\\
13 Dec 2007 & 54447 & 4000-7500\AA & $\sim$600 & SPOL\\
16 Dec 2007 & 54450 & 4000-7500\AA & $\sim$600 & SPOL\\
06 Oct 2008 & 54745 & 4000-7500\AA & $\sim$600 & SPOL\\
09 Oct 2008 & 54748 & 4000-7500\AA & $\sim$600 & SPOL\\
28 Oct 2008 & 54767 & 4000-7500\AA & $\sim$600 & SPOL\\
24 Nov 2008 & 54794 & 4000-7500\AA & $\sim$600 & SPOL\\
12 Oct 2009 & 55116 & 6000-9500\AA & $\sim$3500 & CAFOS\\
\cutinhead{IRS}
09 Oct 2007 & 54382 & 5-40\micron & $\sim$600 & \\
16 Oct 2007 & 54389 & 5-40\micron & $\sim$600 & \\
24 Feb 2008 & 54520 & 5-40\micron & $\sim$600 &\\
02 Mar 2008 & 54527 & 5-40\micron & $\sim$600 & \\
\cutinhead{MIPS}
21 Feb 2004 & 53056 & 24\micron & photometry & \\
19 Sep 2004 & 53267 & 24\micron & photometry & \\
23 Sep 2007 & 53466 & 24\micron & photometry & \\
24 Sep 2007 & 54367 & 24\micron & photometry & \\
25 Sep 2007 & 54368 & 24\micron & photometry & \\
26 Sep 2007 & 54369 & 24\micron & photometry & \\
27 Sep 2007 & 54370 & 24\micron & photometry & \\
12 Mar 2008 & 54537 & 24\micron & photometry & \\
19 Mar 2008 & 54544 & 24\micron & photometry & \\
\cutinhead{IRAC}
11 Feb 2004 & 53046 & 3.6,4.5,5.8,8.0\micron & photometry & Cold-mission\\
08 Sep 2004 & 53257 & 3.6,4.5,5.8,8.0\micron & photometry & Cold-mission\\
19 Mar 2009 & 54910 & 3.6,4.5,5.8,8.0\micron & photometry & Cold-mission\\
20 Mar 2009 & 54911 & 3.6,4.5,5.8,8.0\micron & photometry & Cold-mission\\
21 Mar 2009 & 54912 & 3.6,4.5,5.8,8.0\micron & photometry & Cold-mission\\
22 Mar 2009 & 54913 & 3.6,4.5,5.8,8.0\micron & photometry & Cold-mission\\
23 Mar 2009 & 54914 & 3.6,4.5,5.8,8.0\micron & photometry & Cold-mission\\
03 Oct-07 Nov 2009 & 55107-55142 & 3.6,4.5\micron & photometry & Warm-mission\\
\enddata
\end{deluxetable}

\begin{deluxetable}{cccc}
\tablewidth{0pt}
\tablecaption{Photometry\label{photometry}}
\tablehead{\colhead{Date}&\colhead{J}&\colhead{H}&\colhead{K}}
\startdata
2MASS & 12.09 & 10.54 & 9.69\\
10 Oct 2008 & 12.19 & 10.71 & 9.95\\
11 Oct 2008 & 12.23 & 10.81 & 10.08\\
18 Oct 2008 & 12.26 & 10.81 & 10.07\\
19 Oct 2008 & 12.25 & 10.79 & 9.99\\
08 Oct 2009 & 12.34 & 10.92 & 10.29\\
09 Oct 2009 & 12.22 & - & 10.25\\
31 Oct 2009 & 12.26 & 10.83 & 10.14\\
04 Nov 2009 & 12.24 & 10.77 & 10.03\\
08 Nov 2008 & 12.15 & 10.68 & 9.92\\
\enddata
\tablecomments{Typical uncertainties in the photometry, dominated by the absolute uncertainty in the flux measurements, are 0.1 mag.}
\end{deluxetable}

\begin{deluxetable}{ccccc}
\tablewidth{0pt}
\tablecaption{IRAC photometry\label{irac-cryo}}
\tablehead{\colhead{Date}&\colhead{[3.6]}&\colhead{[4.5]}&\colhead{[5.8]}&\colhead{[8.0]}}
\startdata
19 Mar 2009 & 9.058 $\pm$ 0.002 & 8.781 $\pm$ 0.003 & 8.404 $\pm$ 0.006 & 7.997 $\pm$ 0.03\\
20 Mar 2009 & 9.064 $\pm$ 0.002 & 8.721 $\pm$ 0.003 & 8.403 $\pm$ 0.006 & 8.033 $\pm$ 0.03\\
21 Mar 2009 & 8.995 $\pm$ 0.002 & 8.736 $\pm$ 0.003 & 8.347 $\pm$ 0.006 & 8.002 $\pm$ 0.03\\
22 Mar 2009 & 8.968 $\pm$ 0.002 & 8.639 $\pm$ 0.003 & 8.343 $\pm$ 0.006 & 7.941 $\pm$ 0.03\\
23 Mar 2009 & 9.071 $\pm$ 0.002 & 8.695 $\pm$ 0.002 & 8.342 $\pm$ 0.006 & 7.981 $\pm$ 0.03\\
\enddata
\end{deluxetable}

\begin{deluxetable}{ccc}
\tablewidth{0pt}
\tablecaption{Extinction\label{extinction}}
\tablehead{\colhead{Date}&\colhead{A$_{V}$}&\colhead{Wavelength Range}}
\startdata
13 Dec 2007 & 8.8 & Optical\\
06 Oct 2008 & 9.2 & Optical\\
09 Oct 2008 & 8.9 & Optical\\
10 Oct 2008 & 9.5 & NIR\\
11 Oct 2008 & 8.2 & NIR\\
18 Oct 2008 & 9.2 & NIR\\
19 Oct 2008 & 9.2 & NIR\\
28 Oct 2008 & 8.5 & Optical\\
24 Nov 2008 & 8.1 & Optical\\
08 Oct 2009 & 8.8 & NIR\\
09 Oct 2009 & 8.3 & NIR\\
31 Oct 2009 & 8.9 & NIR\\
04 Nov 2009 & 8.7 & NIR\\
08 Nov 2009 & 8.8 & NIR\\
\enddata
\tablecomments{The uncertainty in the extinction is 0.5 mag.}
\end{deluxetable}

\begin{deluxetable}{cc}
\tablewidth{0pt}
\tablecaption{Stellar Luminosity\label{luminosity}}
\tablehead{\colhead{Date}&\colhead{Stellar Luminosity (L/L$_{\odot}$)}}
\startdata
10 Oct 2008 & 5.3 $\pm$ 0.8\\
11 Oct 2008 & 3.6 $\pm$ 0.6\\
18 Oct 2008 & 4.6 $\pm$ 0.7\\
19 Oct 2008 & 4.6 $\pm$ 0.7\\
08 Oct 2009 & 3.8 $\pm$ 0.6\\
09 Oct 2009 & 3.8 $\pm$ 0.6\\
31 Oct 2009 & 4.2 $\pm$ 0.7\\
04 Nov 2008 & 4.1 $\pm$ 0.6\\
08 Nov 2009 & 4.6 $\pm$ 0.7\\
\enddata
\tablecomments{Errors include the 0.1 mag uncertainty in the photometry and the 0.5 mag uncertainty in A$_V$.}
\end{deluxetable}

\begin{deluxetable}{cccc}
\tablewidth{0pt}
\tablecaption{Infrared Excess\label{ir_excess}}
\tablehead{\colhead{Date}&\colhead{Veiling}&\colhead{Dust Temperature}&\colhead{Covering Fraction}}
\startdata
29 Dec 2005 & 0.33 $\pm$0.1& - & -\\
09 Nov 2006 & 0.20 $\pm$0.1 & - & \\
10 Oct 2008 & 0.20 $\pm$0.1 & 1940 $\pm$ 250 & 0.039\\
11 Oct 2008 & 0.32 $\pm$0.1 & 2100 $\pm$ 250 & 0.084\\
18 Oct 2008 & 0.17 $\pm$0.1 & 1820 $\pm$ 200 & 0.037\\
19 Oct 2008 & 0.16 $\pm$0.1 & 1620 $\pm$ 200 & 0.031\\
08 Oct 2009 & -0.05 $\pm$0.1 & - & 0.010\tablenotemark{a}\\
09 Oct 2009 & 0.02 $\pm$0.1 & - & 0.012\tablenotemark{a}\\
31 Oct 2009 & 0.10 $\pm$0.1 & 1540 $\pm$ 250 & 0.025\\
04 Nov 2009 & 0.23 $\pm$0.1 & 2020 $\pm$ 300 & 0.048\\
08 Nov 2009 & 0.29 $\pm$0.1 & 1780 $\pm$ 250 & 0.054\\
\enddata
\tablenotetext{a}{Derived assuming a dust temperature of 1830 K and fitting to the IRAC photometry}
\end{deluxetable}

\begin{deluxetable}{ccc}
\tablewidth{0pt}
\tablecaption{Polarimetry\label{polarimetry}}
\tablehead{\colhead{Date}&\colhead{P(\%)}&\colhead{q($^o$)}}
\startdata
13 Dec 2007 & 8.44 $\pm$ 0.14 & 146.6 $\pm$ 0.5\\
16 Dec 2007 & 7.66 $\pm$ 0.25 & 144.0 $\pm$ 0.9\\
06 Oct 2008 & 8.34 $\pm$ 0.12 & 145.7 $\pm$ 0.4\\
09 Oct 2008 & 7.70 $\pm$ 0.07 & 144.9 $\pm$ 0.3\\
28 Oct 2008 & 7.84 $\pm$ 0.11 & 144.4 $\pm$ 0.4\\
24 Nov 2008 & 7.77 $\pm$ 0.23 & 143.2 $\pm$ 0.8\\
\enddata
\end{deluxetable}

\begin{deluxetable}{ccccc}
\tablewidth{0pt}
\tablecaption{Emission Line Tracers of Accretion Rate\label{nir_lines}}
\tablehead{\colhead{Date}&\colhead{Pa$\beta$ EW ($\AA$)}&\colhead{Br$\gamma$ EW ($\AA$)}&\colhead{Veiling Corrected Br$\gamma$ EW ($\AA$)}&\colhead{Veiling}}
\startdata
29 Dec 2005 & -3.36 $\pm$ 0.08 (0.50) & -3.05 $\pm$ 0.09 (0.84) & -2.56 $\pm$ 0.09 (0.71) & 0.33 $\pm$ 0.1 \\
09 Nov 2006 & -4.29 $\pm$ 0.04 (0.37) & -2.55 $\pm$ 0.02 (0.18) & -2.08 $\pm$ 0.02 (0.15) & 0.20 $\pm$ 0.1 \\
10 Oct 2008 & -2.74 $\pm$ 0.07 (0.49) & -1.52 $\pm$ 0.04 (0.20) & -1.20 $\pm$ 0.04 (0.17) & 0.20 $\pm$ 0.1\\
11 Oct 2008 & -2.69 $\pm$ 0.07 (0.45) & -1.27 $\pm$ 0.05 (0.22) & -0.80 $\pm$ 0.05 (0.16) & 0.32 $\pm$ 0.1\\
18 Oct 2008 & -1.54 $\pm$ 0.04 (0.12) & -0.86 $\pm$ 0.02 (0.07) & -0.59 $\pm$ 0.02 (0.05) & 0.17 $\pm$ 0.1\\
19 Oct 2008 & -1.78 $\pm$ 0.06 (0.30) & -2.11 $\pm$ 0.03 (0.21) & -1.84 $\pm$ 0.03 (0.19) & 0.16 $\pm$ 0.1\\
08 Oct 2009 & -1.12 $\pm$ 0.12 (0.31) & -1.16 $\pm$ 0.04 (0.14) & -1.16 $\pm$ 0.04 (0.14) & -0.05 $\pm$ 0.1\\
09 Oct 2009 & -0.74 $\pm$ 0.05 (0.08) & -1.39 $\pm$ 0.03 (0.11) & -1.36 $\pm$ 0.03 (0.11) & 0.02 $\pm$ 0.1\\
31 Oct 2009 & -2.56 $\pm$ 0.04 (0.22) & -2.04 $\pm$ 0.02 (0.14) & -1.87 $\pm$ 0.02 (0.13) & 0.10 $\pm$ 0.1\\
04 Nov 2009 & -3.31 $\pm$ 0.05 (0.35) & -2.29 $\pm$ 0.03 (0.21) & -1.94 $\pm$ 0.03 (0.18) & 0.23 $\pm$ 0.1\\
08 Nov 2009 & -3.08 $\pm$ 0.05 (0.32) & -2.30 $\pm$ 0.03 (0.18) & -1.87 $\pm$ 0.03 (0.16) &  0.29 $\pm$ 0.1\\
\enddata
\tablecomments{Uncertainties in parenthesis include the uncertainties in the continuum}
\end{deluxetable}

\begin{deluxetable}{ccccc}
\tablewidth{0pt}
\tablecaption{Accretion Rates\label{accretion}}
\tablehead{\colhead{Date}&\colhead{Flux ($10^{-14}$ erg s$^{-1}$ cm$^{-2}$)}&\colhead{$\log$(L/L$_{\odot}$)}&\colhead{$\log$(L$_{acc}$/L$_{\odot}$)}&\colhead{$\dot{M}$ ($10^{-8} M_{\odot}yr^{-1}$)}}
\startdata
\cutinhead{Pa$\beta$}
10 Oct 2008 & 16.6 $\pm$ 2.7 & -3.28 $\pm$ 0.07 & -0.59 $\pm$ 0.08 & 1.62 $\pm$ 0.13\\
11 Oct 2008 & 11.2 $\pm$ 1.8 & -3.45 $\pm$ 0.07 & -0.78 $\pm$ 0.08 & 1.05 $\pm$ 0.08\\
18 Oct 2008 & 8.08 $\pm$ 1.3 & -3.59 $\pm$ 0.07 & -0.94 $\pm$ 0.08 & 0.73 $\pm$ 0.06\\
19 Oct 2008 & 9.43 $\pm$ 1.6 & -3.52 $\pm$ 0.07 & -0.86 $\pm$ 0.08 & 0.87 $\pm$ 0.07\\
08 Oct 2009 & 4.93 $\pm$ 1.2 & -3.81 $\pm$ 0.11 & -1.19 $\pm$ 0.13 & 0.41 $\pm$ 0.05\\
09 Oct 2009 & 3.21 $\pm$ 0.6 & -3.99 $\pm$ 0.08 & -1.40 $\pm$ 0.09 & 0.25 $\pm$ 0.02\\
31 Oct 2009 & 12.5 $\pm$ 2.0 & -3.40 $\pm$ 0.07 & -0.73 $\pm$ 0.08 & 1.18 $\pm$ 0.09\\
04 Nov 2009 & 15.5 $\pm$ 2.5 & -3.31 $\pm$ 0.07 & -0.62 $\pm$ 0.08 & 1.52 $\pm$ 0.12\\
08 Nov 2009 & 16.1 $\pm$ 2.6 & -3.29 $\pm$ 0.07 & -0.60 $\pm$ 0.08 & 1.59 $\pm$ 0.13\\
\cutinhead{Br$\gamma$}
10 Oct 2008 & 3.10 $\pm$ 0.38 & -4.01 $\pm$ 0.05 & -0.62 $\pm$ 0.06 & 1.52 $\pm$ 0.09\\
11 Oct 2008 & 1.61 $\pm$ 0.29 & -4.29 $\pm$ 0.08 & -0.98 $\pm$ 0.10 & 0.66 $\pm$ 0.07\\
18 Oct 2008 & 1.32 $\pm$ 0.18 & -4.38 $\pm$ 0.06 & -1.09 $\pm$ 0.08 & 0.51 $\pm$ 0.04\\
19 Oct 2008 & 4.45 $\pm$ 0.48 & -3.85 $\pm$ 0.05 & -0.42 $\pm$ 0.06 & 2.40 $\pm$ 0.14\\
08 Oct 2009 & 2.04 $\pm$ 0.26 & -4.19 $\pm$ 0.06 & -0.85 $\pm$ 0.08 & 0.89 $\pm$ 0.07\\
09 Oct 2009 & 2.35 $\pm$ 0.26 & -4.13 $\pm$ 0.05 & -0.77 $\pm$ 0.06 & 1.09 $\pm$ 0.06\\
31 Oct 2009 & 3.81 $\pm$ 0.40 & -3.92 $\pm$ 0.05 & -0.51 $\pm$ 0.06 & 2.19 $\pm$ 0.12\\
04 Nov 2009 & 4.28 $\pm$ 0.46 & -3.87 $\pm$ 0.05 & -0.45 $\pm$ 0.06 & 2.76 $\pm$ 0.13\\
08 Nov 2009 & 4.63 $\pm$ 0.50 & -3.83 $\pm$ 0.05 & -0.40 $\pm$ 0.06 & 3.32 $\pm$ 0.15\\
\enddata
\tablecomments{Uncertainties include the uncertainty in the spectra and the J and K photometry, but do not include the uncertainty in the conversion from line flux to accretion luminosity. The Br$\gamma$ line fluxes are derived with veiling of the line taken into account.}
\end{deluxetable}

\begin{deluxetable}{cc}
\tablewidth{0pt}
\tablecaption{H$\alpha$ EW\label{haew}}
\tablehead{\colhead{Date}&\colhead{H$\alpha$ EW(\AA)}}
\startdata
28 Nov 2006 & 11.5\\
13 Dec 2007 & 4.86\\
29 Feb 2008 & 12.0\\
06 Oct 2008 & 12.06\\
09 Oct 2008 & 10.32\\
28 Oct 2008 & 10.77\\
24 Nov 2008 & 13.56\\
03 Dec 2008 & 8.12\\
12 Oct 2009 & 4.02\\
\enddata
\end{deluxetable}

\begin{deluxetable}{ccc}
\tablewidth{0pt}
\tablecaption{IRAC Warm Mission Photometry\label{irac_wm}}
\tablehead{\colhead{MJD-55000}&\colhead{[3.6]}&\colhead{[4.5]}}
\startdata
106.91 & 9.314 & 9.098\\
107.30 & 9.373 & 9.109\\
107.57 & 9.363 & 9.127\\
107.81 & 9.437 & 9.134\\
108.13 & 9.439 & 9.174\\
108.38 & 9.469 & 9.214\\
108.58 & 9.476 & 9.200\\
108.81 & 9.482 & 9.236\\
109.58 & 9.498 & 9.271\\
110.50 & 9.514 & 9.306\\
111.07 & 9.609 & 9.372\\
112.43 & 9.642 & 9.438\\
113.07 & 9.639 & 9.413\\
114.73 & 9.559 & 9.341\\
115.37 & 9.538 & 9.302\\
115.97 & 9.552 & 9.272\\
116.32 & 9.495 & 9.277\\
116.62 & 9.529 & 9.303\\
116.72 & 9.527 & 9.313\\
117.13 & 9.535 & 9.288\\
117.21 & 9.533 & 9.299\\
117.46 & 9.551 & 9.319\\
117.69 & 9.565 & 9.311\\
118.60 & 9.568 & 9.321\\
119.25 & 9.565 & 9.325\\
119.93 & 9.506 & 9.289\\
121.21 & 9.510 & 9.268\\
123.07 & 9.491 & 9.206\\
124.96 & 9.488 & 9.282\\
127.68 & 9.472 & 9.244\\
129.81 & 9.475 & 9.212\\
131.27 & 9.454 & 9.171\\
133.38 & 9.363 & 9.097\\
135.61 & 9.432 & 9.199\\
137.52 & 9.333 & 9.075\\
139.07 & 9.349 & 9.058\\
141.52 & 9.221 & 8.937\\
142.13 & 9.245 & 8.922\\
\enddata
\tablecomments{Uncertainties are 0.017 and 0.014 mag at [3.6] and [4.5] respectively based on the rms fluctuations of non-variable stars in the field}
\end{deluxetable}

\clearpage

\end{document}